\documentclass[10pt]{article}
\usepackage[numbers,sort&compress]{natbib}
\usepackage{scistyle}
\usepackage{projectmacros}
\usepackage{placeins}

\title{Long-horizon autoformalization of a core theorem underlying MIP*=RE}

\contribnote{equal}{These authors contributed equally.}
\paperauthor{1,2,equal}{Sirui Lu}
\paperauthor{3,equal}{Ruixuan Deng}
\paperauthor{4}{Yanqiao Zhu}
\paperauthor{3}{Zhengfeng Ji}
\affiliation{1}{Max-Planck-Institut f\"ur Quantenoptik,
  Hans-Kopfermann-Stra\ss{}e 1, D-85748 Garching, Germany}
\affiliation{2}{Munich Center for Quantum Science and Technology (MCQST),
  Schellingstra\ss{}e 4, D-80799 Munich, Germany}
\affiliation{3}{Department of Computer Science and Technology,
  Tsinghua University, Beijing, China}
\affiliation{4}{Department of Computer Science, University of California,
  Los Angeles, Los Angeles, California, USA}
\correspondence{Zhengfeng Ji}{jizhengfeng@tsinghua.edu.cn}
\metadata[Repository]{\url{https://github.com/LionSR/MIPStarRE},
  \url{https://github.com/Dengnifer/MIPStarRE-QPBT}}

\renewcommand{\suppappendix}[1]{%
  \clearpage
  \refstepcounter{suppappendix}%
  \setcounter{subsection}{0}%
  \renewcommand{\thesubsection}{\Alph{suppappendix}.\arabic{subsection}}%
  \renewcommand{\thesubsubsection}{\thesubsection.\arabic{subsubsection}}%
  \renewcommand{\theHsubsection}{\Alph{suppappendix}.\arabic{subsection}}%
  \renewcommand{\theHsubsubsection}{\theHsubsection.\arabic{subsubsection}}%
  \section*{\Alph{suppappendix}\hspace{0.75em}#1}%
  \addcontentsline{toc}{section}{\protect\numberline{\Alph{suppappendix}}#1}}
\renewcommand{\beginsupplement}{%
  \clearpage
  \floatplacement{table}{H}%
  \floatplacement{figure}{htbp}%
  \setcounter{equation}{0}\renewcommand{\theequation}{S\arabic{equation}}%
  \setcounter{figure}{0}\renewcommand{\thefigure}{S\arabic{figure}}%
  \setcounter{table}{0}\renewcommand{\thetable}{S\arabic{table}}%
  \renewcommand{\theHequation}{S\arabic{equation}}%
  \renewcommand{\theHfigure}{S\arabic{figure}}%
  \renewcommand{\theHtable}{S\arabic{table}}%
  \renewcommand{\tablename}{Table}%
  \renewcommand{\figurename}{Figure}%
  \begingroup\setlength{\parskip}{0pt}%
  {\raggedright\LARGE\sffamily\bfseries\color{scidark}Supplementary Information\par}%
  \vspace{7pt}%
  {\color{sciaccent}\rule{\linewidth}{1.1pt}}%
  \endgroup\par\vspace{11pt}}


\begin{document}

\begin{abstract}
Landmark mathematical formalizations have taken specialist teams years to
complete.
We present \sysname{}, a system that coordinates AI proving agents under human
supervision to address statement drift and proof composition in long-horizon
formalization.
Drawing on software-engineering principles and practices, it uses a shared
blueprint to guide nested planning, proving, and review loops.
Using \sysname{}, we completed a machine-checked Lean~4 proof of the quantum
soundness of the classical low individual-degree test, a core theorem underlying
$\MIPstarRE$, in 63 days.
The final library contains \NumLines{} lines of Lean code, all generated by
agents.
Beyond providing a verified foundation for quantum complexity, the formalization
corrects side conditions and intermediate error bounds in the published proof
and improves the error exponent of the main theorem.
The development record shows what it takes to trust proofs written by AI agents
and provides reusable tools for verifying research mathematics at scale.
\end{abstract}

\maketitle

\section{Introduction}\label{sec:introduction}

Gaps in mathematical proofs can go unnoticed for years, even after peer review.
Formalization translates informal arguments into proofs that can be checked by a
proof assistant such as Lean, Isabelle, or Rocq.
Large human-led formalization projects have achieved landmark results,
including the Four-Color Theorem~\citep{Gonthier2008Formal}, the
Feit--Thompson Odd Order Theorem (over 150,000 lines of proof scripts,
including supporting libraries)~\citep{Gonthier2013Machinechecked}, the
Kepler Conjecture (over 500,000 lines of proof
scripts)~\citep{Hales2017Formal}, and the Liquid Tensor
Experiment~\citep{Scholze2022Liquid,CommelinTopaz2024Abstraction}.
These achievements required years of manual work and coordination.
Artificial-intelligence systems have now advanced automated theorem proving and
autoformalization to the point where machine-checked proofs can be produced on
a far greater scale~\citep{Polu2020Generative,Lample2022HyperTree,
Zheng2022MiniF2F,
Wu2022Autoformalization,Trinh2024Solving,Ren2025DeepSeekproverV2,
Hubert2026Olympiadlevel}.
Recent results show that cutting-edge models can formalize
research-level mathematics~\citep{Rammal2026Formalizing,Tsoukalas2026Advancing,
OpenAI2026Ten}.
Concurrent work on LeanMarathon~\citep{Zhang2026LeanMarathon} and
Theo~\citep{SoltaniMoakhar2026Beyond} also studies agent coordination for
Lean autoformalization.

In long-horizon autoformalization, agents must assemble results from a large,
evolving codebase into a complete proof.
Successive agent calls can introduce duplicate constructions, competing
representations, and layers of intermediate results that still depend on
unproved steps.
As these additions accumulate, later agents and the supervising mathematician
can have difficulty determining how the partial results fit together.
The project can then stall despite its growing size; if its mathematical
structure cannot be recovered, the formalization may have to be restarted.
Long developments therefore require repeated consolidation of useful results
and an explicit account of what remains to be proved.

\begin{figure}
  \centering
  \newcommand{\agentchip}[1]{\textcolor{cbOrange!85!black}{#1}}
  \adjustbox{max width=\linewidth,center}{%
  \begin{tikzpicture}[
      >={Stealth[length=4pt,width=3.2pt]},
      every node/.style={font=\small},
      box/.style={draw=black!50, rounded corners=2pt, fill=white,
                  line width=0.6pt, minimum height=0.75cm, align=center,
                  inner sep=4pt},
      agent/.style={box, draw=cbOrange!85!black, fill=cbOrange!12},
      focus/.style={box, draw=agentaccent, fill=agentaccent!25, line width=1.1pt},
      session/.style={box, draw=sessionaccent!80, fill=sessionaccent!12,
                      line width=1.1pt},
      human/.style={box, humanbox, text width=2.7cm},
      check/.style={box, densely dotted, draw=black!70},
      arrow/.style={->, draw=black!70, line width=0.9pt},
      retry/.style={arrow, densely dashed, rounded corners=2pt},
      note/.style={text=black!75, align=center, inner sep=2pt},
      rowlabel/.style={anchor=east, align=right, inner sep=1pt},
      sub/.style={text=black!60},
      gate/.pic={\fill[cbMagenta] (-0.08,-0.26) rectangle (-0.02,0.26);
                 \fill[cbMagenta] (0.02,-0.26) rectangle (0.08,0.26);},
      x=1cm, y=1cm
    ]

    \begin{pgfonlayer}{background}
      \fill[agentaccent!9] (6.55,-0.375) -- (9.45,-0.375)
        -- (12.95,-1.525) -- (3.05,-1.525) -- cycle;
      \fill[sessionaccent!10] (3.05,-2.275) -- (5.75,-2.275)
        -- (12.95,-3.825) -- (3.05,-3.825) -- cycle;
    \end{pgfonlayer}

    \node[rowlabel] at (2.75,1.15) {\textbf{Human decisions}};
    \node[human, minimum width=2.9cm] (obj) at (4.50,1.15) {Set objectives};
    \node[human, minimum width=4.4cm, text width=4.2cm] (audit) at (11.50,1.15)
      {Audit the final statement};
    \begin{pgfonlayer}{background}
      \node[draw=black!30, rounded corners=3pt, inner sep=5pt,
            fit=(obj)(audit)] {};
    \end{pgfonlayer}
    \node[font=\small\bfseries] at (15.00,-1.62) {Proof-gap protocol};
    \node[human, minimum height=1.35cm] (gap) at (15.00,-2.72)
      {Decide whether to correct the paper, repair the proof,
       or close the note};

    \node[rowlabel] at (2.75,0)
      {\textbf{(a)~Task planning}\\[-1pt]
       \agentchip{Orchestrator} \textcolor{black!60}{handles}\\[-2pt]
       \textcolor{black!60}{one task at a time}};
    \node[box,   minimum width=2.9cm] (todo) at (4.50,0) {Todo list};
    \node[focus, minimum width=2.9cm] (task) at (8.00,0) {One task};
    \node[box,   minimum width=2.9cm] (next) at (11.50,0) {Next task};
    \draw[arrow] (todo) -- (task);
    \draw[arrow] (task) -- (next);
    \draw[arrow] (next) -- (audit);
    \draw[arrow] (obj) -- (todo);

    \node[rowlabel] at (2.75,-1.90)
      {\textbf{(b)~Review loop}\\[-1pt]
       \agentchip{Review agents} \textcolor{black!60}{check each}\\[-2pt]
       \textcolor{black!60}{pull request before it merges}};
    \node[session, minimum width=2.7cm] (work) at (4.40,-1.90)
      {Work session};
    \node[agent, minimum width=3.3cm] (review) at (8.00,-1.90)
      {Compare with paper};
    \node[box, minimum width=2.7cm] (merge) at (11.60,-1.90) {Merge};
    \draw[arrow] (work) -- (review);
    \draw[arrow] (review) -- (merge);
    \pic at (9.95,-1.90) {gate};
    \draw[retry] (merge.north) -- (merge.north |- next.south);
    \node[note, anchor=east, align=right] at (11.35,-0.95)
      {Merged change: follow-up\\issues become new tasks};
    \draw[retry] (review.south) -- (8.00,-2.75) -- (4.40,-2.75)
      -- (work.south);
    \node[note] at (6.20,-2.52) {Findings send it back};
    \node[diamond, fill=cbRed, inner sep=2.0pt] at (13.17,-1.90) {};
    \draw[retry] (9.30,-2.275) -- (9.30,-2.75) -- (gap.west |- 0,-2.75);
    \node[note] at (11.60,-2.52) {Gap note};
    \draw[retry] (gap.west |- 0,-3.15) -- (4.25,-3.15) -- (4.25,-3.825);
    \node[note] at (9.00,-3.42) {Approved correction returns to the task};

    \node[rowlabel] at (2.75,-4.20)
      {\textbf{(c)~Work session}\\[-1pt]
       \agentchip{Task agents} \textcolor{black!60}{edit the proof}\\[-2pt]
       \textcolor{black!60}{until it compiles}};
    \node[agent, minimum width=2.4cm] (resume) at (4.25,-4.20)
      {Read task record};
    \node[agent, minimum width=2.1cm] (edit) at (6.90,-4.20) {Edit proof};
    \node[check, minimum width=2.4cm] (lean) at (9.55,-4.20)
      {Lean compiler};
    \node[box, minimum width=1.8cm, draw=green!60!black, fill=green!12,
          text=green!35!black] (commit) at (12.05,-4.20) {Commit};
    \draw[arrow] (resume) -- (edit);
    \draw[arrow] (edit) -- (lean);
    \draw[arrow] (lean) -- (commit);
    \pic at (10.95,-4.20) {gate};
    \draw[retry, draw=cbOrange!85!black] (lean.south) -- (9.55,-4.95)
      -- (6.90,-4.95) -- (edit.south);
    \node[note, text=cbOrange!70!black] at (8.00,-5.22)
      {Compiler error: repair and retry};
    \node[note, text width=2.8cm] at (11.95,-5.10)
      {A compiling proof\\still needs review};

    \draw[black!25] (0,-5.70) -- (16.50,-5.70);
    \node[rowlabel] at (2.75,-6.45)
      {\textbf{(d)~Check growth}\\[-1pt]
       \textcolor{black!60}{Review findings become}\\[-2pt]
       \textcolor{black!60}{new automated checks}};
    \node[box, minimum width=2.7cm] (failure) at (4.40,-6.45)
      {Observed failure};
    \node[agent, minimum width=3.3cm] (newcheck) at (8.00,-6.45)
      {Reviewed new check};
    \node[check, minimum width=2.7cm] (ci) at (11.60,-6.45) {CI checks};
    \draw[arrow] (failure) -- (newcheck);
    \draw[arrow] (newcheck) -- (ci);
    \node[note] at (8.00,-7.15)
      {Run on relevant changes; report findings or block a change};

    \node[humanbox, rounded corners=1pt, minimum width=0.38cm,
          minimum height=0.22cm, inner sep=0pt] (lga) at (3.20,-7.75) {};
    \node[anchor=west, inner sep=2pt] (lgat) at (lga.east) {Humans};
    \node[draw=cbOrange!85!black, fill=cbOrange!12, rounded corners=1pt,
          minimum width=0.38cm, minimum height=0.22cm, inner sep=0pt,
          right=10pt of lgat] (lgb) {};
    \node[anchor=west, inner sep=2pt] (lgbt) at (lgb.east) {AI agents};
    \node[draw=black!70, densely dotted, fill=white, rounded corners=1pt,
          minimum width=0.38cm, minimum height=0.22cm, inner sep=0pt,
          right=10pt of lgbt] (lgc) {};
    \node[anchor=west, inner sep=2pt] at (lgc.east)
      {Automated check (Lean compiler, CI)};
    \node[diamond, fill=cbRed, inner sep=1.8pt] (lgd) at (3.20,-8.25) {};
    \node[anchor=west, inner sep=2pt] (lgdt) at (lgd.east)
      {New failure pattern, observed at review};
    \fill[cbMagenta] ([xshift=12pt,yshift=-3pt]lgdt.east)
      rectangle ++(0.05,0.22);
    \fill[cbMagenta] ([xshift=15pt,yshift=-3pt]lgdt.east)
      rectangle ++(0.05,0.22);
    \node[anchor=west, inner sep=2pt] at ([xshift=20pt]lgdt.east)
      {Blocking gate where a check is deployed};
  \end{tikzpicture}}
  \caption{\textbf{Three nested scales of work and one cross-scale process.}
    Rows (a)--(c) read from the largest scale to the smallest:
    (a)~task planning, (b)~the review loop, and (c)~the work session, inside
    which the proof loop repeats until the proof compiles.
    The \textcolor{agentaccent}{highlighted task} expands into the review
    row, and the \textcolor{sessionaccent}{highlighted session} expands
    into the work-session row.
    In (a), the orchestrator takes one task at a time from the todo list;
    the follow-up issues of a merged change become new tasks.
    In (b), review agents compare each pull request with the paper:
    findings send it back to the work session, and a passing request merges
    through a blocking gate.
    In (c), task agents read the task record and edit the proof; a failing
    compiler check sends the edit back, as the dashed retry arrow shows, and
    a passing one reaches the green, kernel-checked commit, which still has to pass
    review.
    Human decisions bracket the sequence: humans set the objectives that
    seed the todo list and audit the final statement; when the proof and the
    paper disagree, a gap note goes to the proof-gap protocol, where humans
    decide whether to correct the paper, repair the proof, or close the note,
    and an approved correction returns to the task.
    The bottom row, (d)~check growth, is a separate process: failure
    patterns observed at review (red diamond) inform new checks, which are
    introduced through reviewed pull requests and run by continuous
    integration (CI) on relevant changes, where they may report findings
    before becoming blocking gates (magenta marks).
    Grey boxes are human decisions, orange boxes are AI agents, and dotted
    boxes are automated checks.}\label{fig:workflow-vE2}
\end{figure}
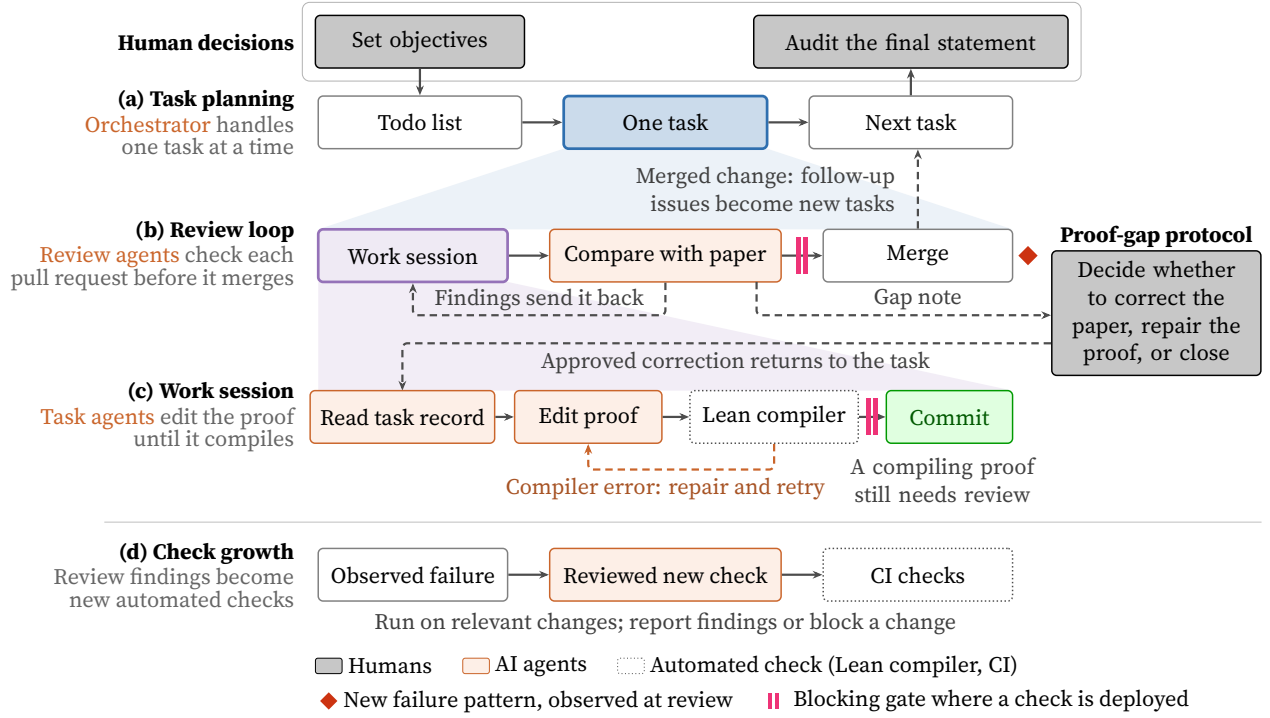

We address this problem with \sysname, an autoformalization system inspired
by software-engineering principles.
It coordinates agents through shared repositories, continuous integration, and
code review~\citep{Jimenez2024SWEbench,Yang2024SWEagent}.
The shared repository holds the paper, the Lean codebase, proof-gap notes, and
an interactive blueprint~\citep{Massot2021Leanblueprint}: a dependency graph
and mathematical specification that maps each informal claim in the paper to
its formal Lean declaration.
\sysname connects task planning, proving, and review through nested loops
(\cref{fig:workflow-vE2}), using the failures found during the work to improve
subsequent checks.
Human supervisors set milestones, resolve discrepancies between the paper and
the formalization, and inspect the final definitions.

\sysname coordinates formalization as an iterative process towards a fixed
point: a complete formalization in which all dependencies compile, all
statements express the intended mathematics accurately, and no open proof
obligations remain.
We call a claim that must be proved to complete an argument a \emph{proof
obligation}, for example a bound required by the next lemma; an obligation
remains open until it has been proved from the available assumptions.
In each iteration, the multi-agent system audits the evolving repository against
the paper and the blueprint, retains verified and reusable proofs, removes
abandoned proof attempts, and logs newly exposed obligations as issues.
Suspected discrepancies with the paper are recorded for human adjudication,
while missing intermediate proofs become explicit tasks for subsequent sessions.
This separation preserves valid partial results without treating their temporary
assumptions as established mathematics, and each subsequent pass audits the
updated statements and the remaining obligations required to reach the target
theorem.

We used \sysname to formalize a central component of the $\MIPstarRE$
theorem~\citep{Ji2021Mip}.
This theorem establishes that interactive proofs with two entangled provers can
decide every recursively enumerable language and refutes Connes' embedding
conjecture~\citep{Connes1976Classification}.
The component we formalize is the quantum soundness of the classical
low individual-degree test (LIDT)~\citep{jnvwy_ldt}, in which a referee checks
whether two non-communicating provers answer consistently with evaluations of a
low-individual-degree polynomial.
Quantum soundness states that when the provers win with high probability,
global measurements with polynomial outcomes describe their strategies.
This test has a history that makes it a natural target for formalization.
A proof gap in the analysis of the low-degree test
(LDT)~\citep{Vidick2016Threeplayer} affected the subsequent two-player
NP-hardness result~\citep{NatarajanVidick2018TwoPlayer}, quantum games
PCP~\citep{Natarajan2018Lowdegree},
$\mathrm{NEEXP}\subseteq\mathrm{MIP}^*$~\citep{Natarajan2019NEEXP}, and the
original proof of $\MIPstarRE$~\citep{Ji2021Mip,Vidick2020It}.
The LIDT is a weaker variant introduced in~\citet{jnvwy_ldt} to replace the LDT
and recover the latter two results.
The proof of $\MIPstarRE$ therefore depends on the quantum soundness of the
LIDT.

Our formalization is complete, contains no unproven placeholders
(\code{sorry}), and uses only the standard axioms of classical mathematics.
The resulting library contains \NumLines{} lines of Lean across \NumFiles{}
files (\cref{app:theorem}).
During this process, \sysname uncovered and repaired two side conditions in
the published theorem statement and several errors in the intermediate error
budget, and in a subsequent application established a
sharper final error bound that improves the error exponent of the main theorem.
We also provide a public development history recording the failures, repairs,
and changes to verification alongside the verified theorem, supporting library,
and reusable coordination tools for subsequent formalization projects.
By reducing the formalization of a hundred-page proof from years of specialist
labour to weeks of supervised agent work, this work opened the way to a fully
verified proof of $\MIPstarRE$, which \citet{Vidick2026Mip} has since completed
on top of our library.

\section{Results}\label{sec:results}

We present two linked results.
First, we introduce \sysname{}, a system that coordinates coding agents to
translate long informal mathematical proofs into verified Lean code, and
demonstrate its generality on a second theorem (\cref{sec:formalflow}).
Second, we use \sysname{} to formalize the quantum soundness theorem for the low
individual-degree test (LIDT)~\citep{jnvwy_ldt}, a central result in the proof
of $\MIPstarRE$~\citep{Ji2021Mip}, and establish several corrections to the
published proof (\cref{sec:formal-LIDT}).
The argument uses quantum information, non-commutative polynomial identities,
Naimark dilations, semidefinite programming (SDP) duality, spectral graph
expansion, and inductive pasting of local low-degree approximations.
Formalizing it demanded both mathematical work and coordination across the
codebase.
We formalized foundations absent from Mathlib, including state-dependent
distances between quantum measurements, finite-dimensional SDP duality with
complementary slackness, and bipartite Naimark dilations.
We also tracked error propagation across dozens of inductive stages and combined
thousands of intermediate lemmas.
In turn, the composition failures and statement drift this work exposed guided
the design of \sysname{}.

\subsection{\texorpdfstring{The \sysname{} autoformalization system}{%
  The FormalFlow autoformalization system}}
\label{sec:formalflow}

\sysname{} coordinates proving agents on long mathematical formalizations
through nested feedback loops around a shared GitHub repository
(\cref{fig:workflow-vE2}).
The repository holds the paper, blueprint, Lean code, open tasks, and review
history, providing a shared record of the intended proof and its current state.
\sysname{} supplies the orchestration scripts, review prompts, and automated
checks that coordinate work on this record.

\sysname{} addresses two related obstacles to long formalizations: composition
and statement drift.
Each agent has a finite context window and works on only part of the codebase,
while work may pass between different models, harnesses, and human users.
The resulting components must nevertheless use compatible definitions,
assumptions, and bounds: \emph{composition} requires each lemma to supply the
mathematical objects and bounds assumed by the next lemma.
When a proof is difficult, \emph{statement drift} occurs if a formal statement
or definition changes in a way that leaves the intended claim unproved, for
example through added assumptions or a weakened conclusion, causing further
issues for composition.

To manage these obstacles, the system connects three scales of work.
Task planning selects open proof obligations whose prerequisites are available.
Within each task, review checks proposed changes against the paper and
blueprint and returns newly exposed obligations to the task list.
Within a proving session, an agent edits Lean code using compiler feedback.
Human supervisors set objectives, resolve mathematical discrepancies, and audit
the assembled theorem and its definitions.
We developed these procedures during the LIDT formalization as agents
encountered missing dependencies, altered statements, and recurring errors.
\hyperref[sec:methods]{Methods} describes the procedures; the dated development
and repair records appear in \cref{app:trajectory,app:composition,app:checks}.

\paragraph*{Proof dependencies and progress.}
\phantomsection\label{sec:local-progress}
We began by asking agents to translate the sections of the paper into Lean.
The initial skeleton contained one file per section, definitions and theorem
statements, and placeholder proofs marked with \code{sorry}.
A blueprint linked the paper's claims to Lean declarations and recorded their
dependencies.
Its status markers distinguished declarations that were not yet fully
formalized (\emph{not ready} or \emph{stated}) from those whose statements and
proofs were recorded as complete (\emph{fully formalized}).
Agents worked on individual proof tasks, using compiler feedback to fill the
placeholders; a task was treated as finished when its target theorem compiled.
We initially expected these local results to supply the components needed to
assemble the main theorem.

Local completion did not ensure that these components would fit together.
Replacing a required construction with an assumption that it exists produces a
conditional theorem, leaving its callers to supply the construction; the
individual proof can compile even though the missing step has only moved
elsewhere in the argument.
On 29 April, only the main soundness theorem retained an explicit \code{sorry},
yet 114 of the 283 target declarations in the blueprint remained not fully
formalized (74 \emph{not ready} and 40 \emph{stated}) because their statements,
proofs, or dependencies were still unsettled (\cref{sec:blueprint-progress}).
The main theorem remained the only explicit placeholder for 335 commits,
through 8 May.
The placeholder count recorded explicit unfinished proofs, while blueprint
status also recorded unresolved statements and dependencies.
\Cref{fig:metrics-evolution} shows how these measures diverged as local proofs
accumulated and were reconciled with the requirements of the main theorem.

\begin{figure}
  \centering
  \pgfplotstableread[col sep=comma]{data/formalization_metrics_daily.csv}\MetricsDaily
  \begin{tikzpicture}[
      every node/.style={font=\scriptsize},
      phase/.style={font=\scriptsize, text=black!80, inner sep=1pt},
      ev/.style={font=\scriptsize, align=center, inner sep=1pt},
      pstem/.style={-, thin, draw=black!55},
      panel/.style={font=\bfseries\scriptsize, inner sep=1pt, anchor=north west},
    ]
    \pgfplotsset{
      evo/.style={
        scale only axis, width=13.6cm,
        xmin=0, xmax=109,
        axis line style={draw=cbInk, line width=0.45pt},
        tick style={draw=cbInk, line width=0.45pt},
        tick label style={font=\scriptsize},
        label style={font=\scriptsize},
        ylabel style={at={(axis description cs:-0.045,0.5)}, anchor=south,
                      rotate=0, yshift=0pt},
        xtick={0,25,55,86,109}, xticklabels={,,,,},
        extra x ticks={13,38,77}, extra x tick labels={},
        extra x tick style={grid=major,
          grid style={densely dashed, draw=black!35, line width=0.4pt},
          tick style={draw=none}},
        every axis plot/.append style={line width=0.8pt},
        clip=false,
        legend style={font=\scriptsize, draw=none, fill=none,
                      legend cell align=left},
      },
    }

    \begin{axis}[evo, name=ribbon, height=2.5cm, axis lines=none,
                 ymin=0, ymax=4.2, extra x ticks={}]
      \fill[black!7] (axis cs:0,0)  rectangle (axis cs:13,1);
      \fill[black!3] (axis cs:13,0) rectangle (axis cs:38,1);
      \fill[black!7] (axis cs:38,0) rectangle (axis cs:77,1);
      \fill[black!3] (axis cs:77,0) rectangle (axis cs:109,1);
      \draw[black!55, thin] (axis cs:0,0) rectangle (axis cs:109,1);
      \pgfplotsinvokeforeach{13,38,77}{%
        \draw[black!45, thin] (axis cs:#1,0) -- (axis cs:#1,1);}
      \node[phase] at (axis cs:6.5,0.5)  {Scaffolding};
      \node[phase] at (axis cs:25.5,0.5) {Blueprint construction};
      \node[phase] at (axis cs:57.5,0.5) {Proof filling};
      \node[phase] at (axis cs:93,0.5)   {Consolidation};
      \pgfplotsinvokeforeach{13,49,55,62,70,77}{%
        \fill[black!60] (axis cs:#1,1) circle (1.1pt);}
      \draw[pstem] (axis cs:13,1) -- (axis cs:13,1.75);
      \node[ev, anchor=south] at (axis cs:13,1.8)
        {LIDT target adopted\\{\itshape(20 Mar)}};
      \draw[pstem] (axis cs:49,1) -- (axis cs:49,1.75);
      \node[ev, anchor=south] at (axis cs:49,1.8)
        {Review against the paper\\in place {\itshape(late Apr)}};
      \draw[pstem] (axis cs:77,1) -- (axis cs:77,1.75);
      \node[ev, anchor=south] at (axis cs:77,1.8)
        {Last \texttt{sorry} closed\\{\itshape(23 May)}};
      \pgfplotsinvokeforeach{55,62,70}{%
        \draw[pstem] (axis cs:#1,1) -- (axis cs:#1,1.3);}
      \draw[pstem] (axis cs:55,1.3) -- (axis cs:70,1.3);
      \draw[pstem, densely dotted] (axis cs:62.5,1.3) -- (axis cs:62.5,3.35);
      \node[ev, anchor=south] at (axis cs:62.5,3.4)
        {Automated checks tightened {\itshape(1--16 May)}};
    \end{axis}

    \begin{axis}[evo, name=size, at={(ribbon.south west)}, anchor=north west,
                 yshift=-0.25cm, height=2.6cm, ymin=0, ymax=150,
                 ytick={0,50,100,150}, ylabel={k lines of Lean},
                 axis y line*=left]
      \addplot[draw=black!45, thin, fill=black!8] coordinates
        {(1,1) (7,1) (18,7) (34,24.5) (49,71) (63,127.3) (67,129)
         (67.5,121.2) (76,141.8) (82,126.9) (95,122.9) (109,126.4)}
        \closedcycle;
      \node[panel] at (rel axis cs:0,1) {a};
    \end{axis}
    \begin{axis}[evo, at={(size.south west)}, anchor=south west, height=2.6cm,
                 ymin=0, ymax=750, ytick={0,300,600}, axis y line*=right,
                 axis x line=none, extra x ticks={},
                 ylabel={Blueprint targets},
                 ylabel style={at={(axis description cs:1.075,0.5)},
                               anchor=north}]
      \addplot[draw=cbInk!60, densely dotted] table[x=day,y=targets] {\MetricsDaily};
      \node[font=\scriptsize, text=cbInk!70, anchor=south east]
        at (axis cs:108,600) {598 targets at the end};
    \end{axis}

    \begin{axis}[evo, name=comp, at={(size.south west)}, anchor=north west,
                 yshift=-0.3cm, height=3.0cm, ymin=0, ymax=750,
                 ytick={0,300,600}, ylabel={Declarations},
                 legend pos=north west, legend columns=1]
      \addplot[draw=cbBlue] table[x=day,y=formalized] {\MetricsDaily};
      \addlegendentry{fully formalized}
      \addplot[draw=cbOrange] table[x=day,y=not_ready] {\MetricsDaily};
      \addlegendentry{not ready}
      \addplot[draw=cbMagenta, dashed] table[x=day,y=no_leanok] {\MetricsDaily};
      \addlegendentry{no \texttt{leanok} marker}
      \node[panel] at (rel axis cs:0,1) {b};
    \end{axis}

    \begin{axis}[evo, name=sorry, at={(comp.south west)}, anchor=north west,
                 yshift=-0.3cm, height=2.3cm, ymin=0, ymax=170,
                 ytick={0,50,100,150}, ylabel={\texttt{sorry} count}]
      \addplot[draw=cbTeal] table[x=day,y=sorry] {\MetricsDaily};
      \node[font=\scriptsize, text=cbTeal!70!black, anchor=south west]
        at (axis cs:26,140) {149 (1 Apr)};
      \node[font=\scriptsize, text=cbTeal!70!black, anchor=south]
        at (axis cs:53,12) {1 (29 Apr)};
      \node[font=\scriptsize, text=cbTeal!70!black, anchor=south west]
        at (axis cs:78,8) {0 from 23 May};
      \node[panel] at (rel axis cs:0,1) {c};
    \end{axis}

    \begin{axis}[evo, name=gaps, at={(sorry.south west)}, anchor=north west,
                 yshift=-0.3cm, height=2.0cm, ymin=0, ymax=15,
                 ytick={0,5,10,15}, ylabel={Open gap notes},
                 xticklabels={7 Mar 2026,1 Apr,1 May,1 Jun,24 Jun}]
      \addplot[const plot, draw=cbRed!85!black, fill=cbRed!12]
        table[x=day,y=open_gaps] {\MetricsDaily} \closedcycle;
      \node[font=\scriptsize, text=cbRed!70!black, anchor=west, align=left]
        at (axis cs:79,11.0) {25 notes opened 29 Apr to 12 Jun,\\ all closed by 23 Jun};
      \node[panel] at (rel axis cs:0,1) {d};
    \end{axis}
  \end{tikzpicture}
  \caption{\textbf{Development of the formalization, 7 March to 24 June 2026.}
    The ribbon marks the phases of the work and selected changes to review and
    verification.
    \textbf{\textsf{a}}, Lean code size (grey area, left axis) and the number
    of blueprint-linked target declarations (dotted line, right axis).
    \textbf{\textsf{b}}, declarations marked \emph{fully formalized} or
    \emph{not ready}, and declarations without a \code{leanok} completion
    marker.
    \textbf{\textsf{c}}, explicit \code{sorry} placeholders in
    \code{MIPStarRE/LDT} (populated on 24 March).
    \textbf{\textsf{d}}, open notes documenting suspected discrepancies with
    the paper.
    Series show the last value per day.
    The blueprint's targets and status markers changed as statements were
    revised and missing obligations were exposed;
    \cref{sec:blueprint-progress} explains these changes, and
    \cref{fig:gap-lifecycle} traces individual notes.}
  \label{fig:metrics-evolution}
\end{figure}

\paragraph*{Statement review and repair.}
To reconcile local proofs with the global argument, we ran repository-wide
audit-and-repair passes (\cref{app:repair-prompt}) that required agents to
document departures from the intended statements and record the missing proof
obligations as tasks for later sessions.
Agents could retain conditional lemmas as intermediate helpers while those
obligations remained open, while removing abandoned proof attempts and unused
helper structures (\mippr{1440}) and consolidating duplicate constructions
across files.
A declaration representing a theorem from the paper had to recover the intended
statement, even when this reintroduced a tracked \code{sorry}.
The orchestrator and issue-tracking workflows used these records to schedule
repair tasks and re-audit updated statements until the formalization reached a
fixed point in which all dependencies compiled, statements matched the intended
mathematics, and no open proof obligations remained.

Repairing the rounding lemma required both a statement change and a new proof.
The paper requires projectors close to an input measurement on its original
Hilbert space, but an early declaration admitted the identity projector on an
unrelated one-dimensional space.
This proof compiled yet could not supply the approximation required by the next
stage.
Repair first restored the relation to the input measurement and exposed the
missing proof; later sessions constructed the required projectors on that
space.

The Laplacian identity exposed a different problem with definitions.
The paper equates an edge-sum operator with the matrix expression
$M^{-1}I - K$, but an early draft defined the operator directly as that
expression.
The equality then held by definition; the repair had to prove it from the
edge-sum formula.
In the self-improvement lemma (\cref{fig:outline}), an early declaration
assumed the existence and consistency properties of the sub-measurement it was
meant to construct using semidefinite programming.
Repair required restoring the statement, assembling the intermediate argument,
and proving the remaining SDP duality step.
\Cref{app:trajectory,app:composition} give the statements, detection records,
and completed repairs.

Some discrepancies instead required changing the mathematics stated in the
paper.
The proof-gap protocol recorded the paper's claim, the obstruction, a proposed
correction, and the consequences for dependent results.
Human supervisors decided whether to correct the paper or repair the
formalization, and agents implemented the resolution.
This process distinguished an approved mathematical correction from a temporary
assumption whose proof remained open.

Restoring faithful statements during the May repairs raised the \code{sorry}
count from one to a peak of ten and revoked blueprint completion markers on
dependent nodes (\cref{fig:metrics-evolution}b,c).
These changes exposed the work required to complete the argument; the last
placeholder closed on 23 May after the main theorem incorporated the corrected
positive-sampling condition $0 < k$ (\cref{app:trajectory}).

\paragraph*{Review and automated checks.}
\phantomsection\label{sec:review}
Recurring failures led us to update review instructions and introduce
continuous-integration (CI) checks that run automatically on proposed changes.
For example, the proof-debt scanner flags names associated with unproved
intermediate assumptions in declarations linked to paper results, while a
separate scanner flags hypotheses that repeat a theorem's conclusion.
The scanners first reported findings for review and later became blocking merge
checks.
Mathematical review continued to inspect definitions and constructions beyond
these patterns (\cref{app:checks}).

Replaying the final proof-debt scanner across historical snapshots showed 63
flagged declarations on 6 May, dropping to zero from 11 May (when it entered
CI) through 5 June (\cref{sec:check-effectiveness}).

Pull-request and issue review also generated repair tasks beyond these static
checks.
The full review corpus contains 21,651 automated and human comments and reports
retrieved on 27 July 2026 across 1,904 closed pull requests and repository
issues (\cref{tab:review-semantic-tree}).
Mathematics and agreement with the paper formed the largest primary category
(55.2\%), which remained the largest under alternative tie-breaking orders
(\cref{app:review-census}).

\begin{table}
  \centering
  \caption{Classification of the 21,651 review comments and reports, one
    subcategory each.
    Shares are fractions of all entries; classification and follow-up analysis
    appear in \cref{app:review-census}.}
  \label{tab:review-semantic-tree}
  \begin{tblr}{
  colspec = {Q[l,wd=0.02\linewidth] X[5.8,l] X[1,r] X[1.1,r]},
  width = 0.72\linewidth,
  row{odd} = {bg=gray!5},
  row{1} = {bg=gray!25, font=\bfseries},
  rowsep = 3pt,
}
  \toprule
  \SetCell[c=2]{l} Review category & & Objects & Share \\
  \midrule
  \SetCell[c=2]{l} \textbf{Mathematics and agreement with the paper}
    & & \textbf{11,949} & \textbf{55.19\%} \\
  & {\scriptsize$\triangleright$} Mathematical content
    & 5,520 & 25.50\% \\
  & {\scriptsize$\triangleright$} Source and blueprint correspondence
    & 3,742 & 17.28\% \\
  & {\scriptsize$\triangleright$} Semantic and API invariants
    & 2,687 & 12.41\% \\
  \SetCell[c=2]{l} \textbf{Exposition and library design}
    & & \textbf{6,998} & \textbf{32.32\%} \\
  & {\scriptsize$\triangleright$} Mathematical exposition
    & 3,307 & 15.27\% \\
  & {\scriptsize$\triangleright$} Library architecture and API
    & 2,059 & 9.51\% \\
  & {\scriptsize$\triangleright$} Reuse and maintainability
    & 1,632 & 7.54\% \\
  \SetCell[c=2]{l} \textbf{Audit and execution infrastructure}
    & & \textbf{2,704} & \textbf{12.49\%} \\
  & {\scriptsize$\triangleright$} Build, CI and review automation
    & 1,064 & 4.91\% \\
  & {\scriptsize$\triangleright$} Reproducibility, security and environment
    & 119 & 0.55\% \\
  & {\scriptsize$\triangleright$} Repository process and evidence
    & 1,521 & 7.03\% \\
  \midrule
  \SetCell[c=2]{l} \textbf{Total}
    & & \textbf{21,651} & \textbf{100.00\%} \\
  \bottomrule
\end{tblr}

\end{table}

\paragraph*{Application to a second theorem.}
\phantomsection\label{sec:second-theorem}
After completing the LIDT formalization, we applied \sysname{} with newer
models to the soundness of the quantum Pauli basis test
(QPBT)~\citep{Ji2021Mip}, a second component of the $\MIPstarRE$ proof.
This development imported the LIDT library and reused the \sysname{} workflow
for task coordination and code review.
Using orchestrating GPT-6 Astra sessions that dispatched parallel subagent
sessions in isolated git worktrees, the run formalized the Pauli basis test
soundness theorem together with three companion results from~\citet{Ji2021Mip}
(completeness of the Pauli basis test, soundness of the simultaneous low-degree
test, and the qubit-coordinate soundness corollary) in 298.5 hours of wall-clock
working time.
It merged 261 pull requests and used 13.7 billion tokens
(\cref{sec:token-usage}).
The resulting Lean library of 118{,}732 lines is available in the
\href{https://github.com/Dengnifer/MIPStarRE-QPBT}{MIPStarRE-QPBT repository}.

\subsection{A Complete Formalization for LIDT}\label{sec:formal-LIDT}

\paragraph*{LIDT and its formalization.}
In the quantum soundness analysis of LIDT, a referee tests whether two
non-communicating provers who share quantum entanglement answer questions by
evaluating a multivariate polynomial $g\colon \F_q^m \to \F_q$ of individual
degree at most $d$.
The referee samples questions according to one of three subtests chosen with
equal probability (\cref{fig:lidt}): the axis-parallel lines test, the
self-consistency test, and the diagonal lines test.
Quantum soundness states that if a projective strategy on two
finite-dimensional Hilbert spaces succeeds with probability $1-\eps$, then
there are global polynomial measurements $G^{\mathrm{A}}$ and
$G^{\mathrm{B}}$ with two forms of approximate agreement.
On average over points $u$, evaluating Bob's global polynomial at $u$ agrees
with Alice's point answer, and evaluating Alice's global polynomial agrees with
Bob's point answer.
The two global polynomial measurements also agree with each other.

\definecolor{lidtnavy}{HTML}{082D6B}
\definecolor{lidtteal}{HTML}{15979B}
\definecolor{lidtgold}{HTML}{D99608}
\definecolor{lidtcoral}{HTML}{EE4938}
\definecolor{lidtbluefill}{HTML}{CEE0F0}
\definecolor{lidttealfill}{HTML}{75C4B3}
\definecolor{lidtgoldfill}{HTML}{FFC953}

\pgfplotsset{
  compat=1.18,
  lidt surface axis/.style={
    view={42}{27},
    axis lines=none,
    tick style={draw=none},
    xtick=\empty,
    ytick=\empty,
    ztick=\empty,
    xmin=-1.2,
    xmax=1.2,
    ymin=-1.2,
    ymax=1.2,
    zmin=-.2,
    zmax=1.2,
    scale only axis=true,
    axis on top,
    clip=false,
    colormap={lidt surface colors}{
      color(0cm)=(lidtteal!15)
      color(1cm)=(lidttealfill)
    }
  }
}

\newcommand{\surfacefunction}{%
  .3 + .3*x - .2*y - .3*x^2 + .4*y^2 + .3*x*y%
}
\newcommand{\drawgsurface}{%
  \addplot3[
    surf,
    shader=interp,
    draw=none,
    domain=-1:1,
    y domain=-1:1,
    samples=17,
    samples y=17
  ] {\surfacefunction};
  \addplot3[
    mesh,
    draw=lidtnavy!58,
    line width=.22pt,
    domain=-1:1,
    y domain=-1:1,
    samples=9,
    samples y=9
  ] {\surfacefunction};
}
\newcommand{\drawgaxes}[1]{%
  \draw[
    draw=lidtnavy,
    line width=.55pt,
    -{Latex[length=1.7mm,width=1.15mm]}
  ] (axis cs:-1.1,-1.1,-.2) -- (axis cs:1.2,-1.1,-.2);
  \draw[
    draw=lidtnavy,
    line width=.55pt,
    -{Latex[length=1.7mm,width=1.15mm]}
  ] (axis cs:-1.1,-1.1,-.2) -- (axis cs:-1.1,1.2,-.2);
  \draw[
    draw=lidtnavy,
    line width=.55pt,
    -{Latex[length=1.7mm,width=1.15mm]}
  ] (axis cs:-1.1,-1.1,-.2) -- (axis cs:-1.1,-1.1,1.2);
  \node[font=#1,anchor=north,text=lidtnavy]
    at (axis description cs:.7,.1) {$x_1$};
  \node[font=#1,anchor=north east,text=lidtnavy]
    at (axis description cs:.7,.4) {$x_2$};
}

\tikzset{
  every node/.style={text=lidtnavy},
  protocol/.style={
    -{Latex[length=3.2mm,width=2.3mm]},
    line width=1.25pt,
    rounded corners
  },
  navy protocol/.style={protocol,draw=lidtnavy},
  teal protocol/.style={protocol,draw=lidtteal},
  gold protocol/.style={protocol,draw=lidtgold},
  cloud outline/.style={
    cloud,
    cloud puffs=13,
    cloud puff arc=105,
    aspect=1.55,
    draw=lidtnavy,
    line width=1pt,
    fill=lidtteal!7
  }
}

\newcommand{\roleicon}[2]{%
  \begin{scope}[shift={(#1)},y=1cm]
    \draw[fill=leanbg,draw=lidtnavy,line width=1.1pt]
      (0,0) circle[radius=.7];
    \draw[fill=#2,draw=lidtnavy,line width=1pt] (0,.2) circle[radius=.3];
    \path[fill=#2,draw=lidtnavy,line width=1pt]
      (-.5,-.5)..controls (-.4,-.1) and (-.2,0)..(0,0)..controls
      (.2,0) and (.4,-.1)..(.5,-.5)
      arc[start angle=-45,end angle=-135,radius=.7]
      -- cycle;
  \end{scope}
}

\newsavebox{\lidtpanelbox}
\newenvironment{lidtpanel}{%
  \begin{lrbox}{\lidtpanelbox}%
  \begin{minipage}[c][46mm][c]{\dimexpr\linewidth-2\fboxsep-2\fboxrule\relax}
  \centering
  \begin{adjustbox}{max totalsize={\linewidth}{44mm},center}}%
 {\end{adjustbox}\end{minipage}\end{lrbox}%
  \fcolorbox{leanframe}{leanbg}{\usebox{\lidtpanelbox}}}

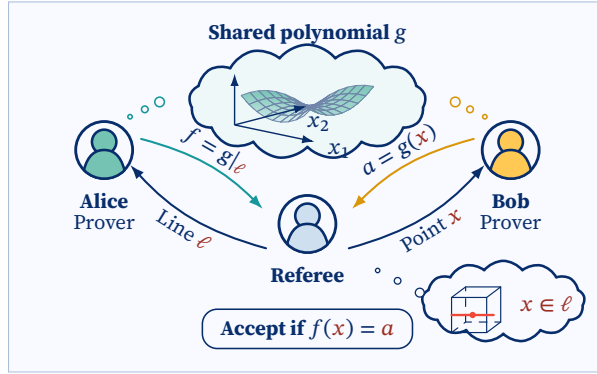
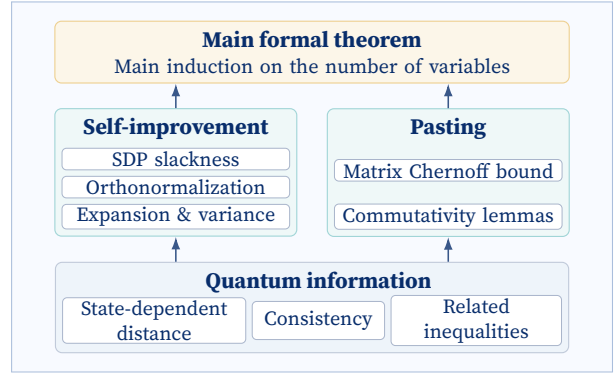
\begin{figure}
  \centering
  \setlength{\fboxsep}{4pt}
  \setlength{\fboxrule}{0.4pt}

  \begin{subfigure}[t]{\textwidth}
    \fcolorbox{leanframe}{leanbg}{%
    \begin{minipage}{\dimexpr\linewidth-2\fboxsep-2\fboxrule\relax}
      \small\raggedright
      \setlength{\parskip}{2pt}
      \textbf{The $(m,q,d)$-low individual-degree test.}
      A referee interacts with Alice and Bob and samples $u\in\F_q^m$
      uniformly. With probability $\nicefrac{1}{3}$ each, it performs:
      \par
      \textbf{\textcolor{lidtnavy}{1.~Axis-parallel line.}}
      Choose a random coordinate direction $i$ and let
      $\ell=\{u+t e_i:t\in\F_q\}$. One prover returns a degree-$d$
      polynomial $f$ on $\ell$; the other returns $a\in\F_q$ at $u$.
      Accept iff $f(u)=a$.
      \par
      \textbf{\textcolor{lidtnavy}{2.~Self-consistency.}}
      Both provers receive $u$ and return $a,b\in\F_q$. Accept iff $a=b$.
      \par
      \textbf{\textcolor{lidtnavy}{3.~Diagonal line.}}
      Sample $i\in\{1,\ldots,m\}$ uniformly, then $v\in\F_q^m$ uniformly with
      $v_{i+1}=\cdots=v_m=0$. Let $\ell=\{u+t v:t\in\F_q\}$. One prover returns
      a degree-$md$ polynomial $f$ on $\ell$; the other returns $a\in\F_q$ at
      $u$. Accept iff $f(u)=a$.
      \par
      In both line tests, the two prover roles are assigned uniformly
      at random.
    \end{minipage}}%
    \caption{The three tests the referee performs.}\label{fig:lidt-checks}
  \end{subfigure}\\[5pt]

  \begin{subfigure}[t]{0.51\textwidth}
    \centering
    \begin{lidtpanel}
\begin{tikzpicture}[
    x=1cm,y=1cm,line cap=round,line join=round,
    every node/.style={text=lidtnavy,font=\footnotesize},
    protocol/.style={-{Latex[length=2mm,width=1.4mm]},
      line width=.8pt,rounded corners}
  ]
    \path[use as bounding box] (-3.35,-.7) rectangle (3.9,3.8);

    \node[
      cloud outline,
      minimum width=3.2cm,
      minimum height=1.55cm,
      aspect=1.9
    ] (commoncloud) at (0,2.7) {};
    \begin{axis}[
      lidt surface axis,
      at={(0cm,2.75cm)},
      anchor=center,
      width=2.15cm,
      height=1.2cm
    ]
      \drawgsurface
      \drawgaxes{\footnotesize}
    \end{axis}
    \node[
      font=\footnotesize\bfseries,
      inner sep=1.2pt
    ] at (0,3.62) {Shared polynomial~$g$};

    \foreach \x/\shade/\name in {-2.75/lidttealfill/Alice, 2.75/lidtgoldfill/Bob}{
      \begin{scope}[shift={(\x,2.05)},scale=.55,transform shape]
        \coordinate (player) at (0,0);
        \roleicon{player}{\shade}
      \end{scope}
      \node[font=\footnotesize\bfseries] at (\x,1.36) {\name};
      \node[font=\footnotesize] at (\x,1.08) {Prover};
    }

    \foreach \side/\shade in {-1/lidtteal,1/lidtgold}{
      \foreach \x/\y/\r in {
        2.4/2.5/.03,
        2.2/2.6/.045,
        2.0/2.7/.065
      }{
        \draw[draw=\shade,fill=leanbg,line width=.6pt]
          (\side*\x,\y) circle[radius=\r];
      }
    }

    \begin{scope}[shift={(0,1.05)},scale=.55,transform shape]
      \coordinate (referee) at (0,0);
      \roleicon{referee}{lidtbluefill}
    \end{scope}
    \node[font=\footnotesize\bfseries] at (0,.36) {Referee};

    \draw[navy protocol] (-.56,.72) to[bend left=16]
      node[midway,sloped,below=1pt,inner sep=1pt]
        {Line $\textcolor{lidtcoral!65!black}{\ell}$} (-2.4,1.85);
    \draw[navy protocol] (.56,.72) to[bend right=16]
      node[midway,sloped,below=1pt,inner sep=1pt]
        {Point $\textcolor{lidtcoral!65!black}{u}$} (2.4,1.85);
    \draw[teal protocol] (-2.25,2.2) to[bend left=15]
      node[midway,sloped,above=1pt,inner sep=1pt]
        {$f=g|_{\textcolor{lidtcoral!65!black}{\ell}}$} (-.6,1.25);
    \draw[gold protocol] (2.25,2.2) to[bend right=15]
      node[midway,sloped,above=1pt,inner sep=1pt]
        {$a=g(\textcolor{lidtcoral!65!black}{u})$} (.6,1.25);

    \node[
      cloud outline,
      fill=leanbg,
      minimum width=2.2cm,
      minimum height=1.1cm,
      aspect=1.95
    ] at (2.65,0) {};
    \foreach \x/\y/\r in {
      .95/.42/.03,
      1.22/.28/.045,
      1.52/.12/.06
    }{
      \draw[draw=lidtnavy,fill=leanbg,line width=.5pt]
        (\x,\y) circle[radius=\r];
    }
    \begin{scope}[
      shift={(1.95,-.42)},
      scale=.95,
      draw=lidtnavy,
      line width=.45pt
    ]
      \draw (0,0) rectangle (.5,.5);
      \draw (0,.5)--(.2,.65)--(.7,.65)--(.5,.5);
      \draw (.5,0)--(.7,.15)--(.7,.65);
      \draw[dashed] (0,0)--(.2,.15)--(.7,.15);
      \draw[dashed] (.2,.15)--(.2,.65);
      \draw[lidtcoral,line width=.9pt] (0,.25)--(.6,.25);
      \fill[lidtcoral] (.3,.25) circle[radius=.045];
    \end{scope}
    \node[font=\footnotesize,anchor=west] at (2.75,-.03)
      {$\textcolor{lidtcoral!65!black}{u\in\ell}$};
    \node[
      draw=lidtnavy,
      line width=.6pt,
      rounded corners=2mm,
      inner xsep=7pt,
      inner ysep=3.5pt,
      font=\footnotesize\bfseries
    ] at (0,-.37)
      {Accept if $f(\textcolor{lidtcoral!65!black}{u})
        =\textcolor{lidtcoral!65!black}{a}$};
  \end{tikzpicture}
    \end{lidtpanel}
    \caption{An honest polynomial strategy.}\label{fig:lidt-protocol}
  \end{subfigure}\hfill
  \begin{subfigure}[t]{0.48\textwidth}
    \centering
    \begin{lidtpanel}
\begin{tikzpicture}[
    >={Latex[length=2mm,width=1.4mm]},
    card/.style={rectangle, rounded corners=3pt, line width=.5pt,
      inner sep=3pt},
    title/.style={font=\normalsize\bfseries, text=lidtnavy, inner sep=1pt},
    plain/.style={font=\small, text=lidtnavy, align=center,
      inner sep=1pt},
    item/.style={rectangle, rounded corners=2pt, line width=.4pt,
      draw=lidtnavy!35, fill=white, font=\small, text=lidtnavy,
      align=center, inner sep=1.2pt, minimum height=0.34cm},
    dep/.style={->, line width=.7pt, draw=lidtnavy!80},
    node distance=2.5pt
  ]
    \node[title] (bt) at (0,0) {Quantum information};
    \node[item, text width=20mm, minimum height=0.6cm, below=of bt] (b2) {Consistency};
    \node[item, text width=28mm, left=of b2, anchor=east] (b1)
      {State-dependent\\ distance};
    \node[item, text width=26mm, right=of b2, anchor=west] (b3)
      {Related\\ inequalities};
    \begin{pgfonlayer}{background}
      \node[card, fill=lidtbluefill!35, draw=lidtnavy!25, fit=(bt)(b1)(b2)(b3)] (cbase) {};
    \end{pgfonlayer}
    \node[item, text width=35mm, anchor=south west]
      (s3) at ([xshift=3pt,yshift=14.5pt]cbase.north west) {Expansion \& variance};
    \node[item, text width=35mm, above=of s3] (s2) {Orthonormalization};
    \node[item, text width=35mm, above=of s2] (s1) {SDP slackness};
    \node[title, above=of s1] (st) {Self-improvement};
    \node[item, text width=35mm, anchor=east]
      (p1) at ([xshift=-3pt]cbase.east |- s1.south) {Matrix Chernoff bound};
    \node[item, text width=35mm] (p2) at (p1 |- s3) {Commutativity lemmas};
    \node[title] (pt) at (st -| p1) {Pasting};
    \node[plain, text width=72mm, anchor=south]
      (mc) at ([yshift=17.5pt]st.north -| cbase) {Main induction on the number of variables};
    \node[title, above=of mc] (mt) {Main formal theorem};
    \begin{pgfonlayer}{background}
      \node[card, fill=lidtteal!7, draw=lidtteal!40, fit=(st)(s1)(s3)] (cself) {};
      \node[card, fill=lidtteal!7, draw=lidtteal!40,
            fit=(pt)(p1)(p2)(s3.south -| p2)] (cpaste) {};
      \node[card, fill=lidtgold!9, draw=lidtgold!45, inner xsep=0pt,
            fit=(mt)(mc)(cbase.west |- mt)(cbase.east |- mt)] (cmain) {};
    \end{pgfonlayer}
    \draw[dep] (cbase.north -| s2) -- (cself.south -| s2);
    \draw[dep] (cbase.north -| p1) -- (cpaste.south -| p1);
    \draw[dep] (cself.north -| s2) -- (cmain.south -| s2);
    \draw[dep] (cpaste.north -| p1) -- (cmain.south -| p1);
  \end{tikzpicture}
    \end{lidtpanel}
    \caption{The proof architecture.}\label{fig:outline}
  \end{subfigure}
  \caption{The low individual-degree test and its formal proof.
    \textbf{\textsf{(a)}}~The referee plays one of three tests with the two
    provers, Alice and Bob.
    \textbf{\textsf{(b)}}~An honest strategy uses a shared polynomial $g$:
    Alice returns $f=g|_\ell$ and Bob returns $a=g(u)$.
    The referee accepts when $f(u)=a$; the roles may be interchanged.
    The surface illustrates $g$ schematically; its domain is $\F_q^m$.
    \textbf{\textsf{(c)}}~The Lean formalization combines a base layer of
    quantum-information inequalities with self-improvement and pasting in
    the main induction.
    For a strategy accepted with high probability, quantum soundness gives
    question-independent polynomial-valued measurements whose evaluations
    approximately agree with the provers' answers.
    The two measurements also give approximately consistent polynomial
    outcomes, even when the provers share entanglement.}
  \label{fig:lidt}
\end{figure}

The proof of the soundness theorem~\citep{jnvwy_ldt} constructs these global
measurements by induction on the number of variables $m$.
It reduces the $m$-variate test to $(m-1)$-variate instances by slicing along
hyperplanes, applying self-improvement, enforcing approximate commutativity,
and pasting local polynomial measurements into a global operator
(\cref{fig:outline}).
Each inductive stage fixes the state space, measurement operators, and error
bounds required by the next stage.

At the pinned snapshot (\PinnedDate), the LIDT library contained \NumLines{}
lines of Lean across \NumFiles{} files.
The library contained no unproven placeholders, and an in-kernel axiom audit
confirmed that the development depends only on Lean's standard foundational
axioms (\cref{app:theorem}).
A human expert, one of the authors of the LIDT paper~\citep{jnvwy_ldt}, audited
the top-level theorem statements and gap resolutions against the published
mathematics, and with the manuscript team reviewed the dependent definitions
down to Lean primitives.

\paragraph*{The discovered proof gaps.}
Formalizing the proof generated \NumGapNotes{} gap notes, which record places
where the formalization and the published proof did not initially agree
(\cref{app:trajectory}; \cref{fig:metrics-evolution}d, \cref{fig:gap-lifecycle},
\cref{tab:gaps-sup-census}).
Some notes led to corrections of the theorem statement or error budget when a
proof obligation could not be closed as published; others led to repairs of the
formalization.
We classified the corrections to the published argument into three principal
categories:
\begin{enumerate}[label=(\roman*)]
  \item \emph{Error-bound corrections}: Much of the LIDT proof bounds
        approximation errors.
        Several bounds used incorrect arithmetic or misapplied an intermediate
        result.

        For example, the printed substitution step leaves the consistency error
        unchanged, but the cited proposition adds an error term.
        The printed proof replaces the polynomial measurement $G^{\mathrm{A}}$
        by $Q^{\mathrm{A}}$, the completion of its orthonormalized
        sub-measurement $P^{\mathrm{A}}$, and carries the consistency error
        $\zeta_1$ across unchanged, whereas the substitution proposition it
        cites adds the square root of the state-dependent distance and gives
        $\zeta_1 + \sqrt{\zeta_2}$.
        The formal proof compares $G^{\mathrm{A}}$ against $P^{\mathrm{A}}$
        instead, where the square root is taken at the smaller scale
        $100\,\zeta_1^{1/4}$, and shows that completion cannot decrease the
        match mass, giving $\zeta_1 + 10\,\zeta_1^{1/8}$ (gap notes
        ``line-169 loss'' and ``sharper fix'' of issue 1099; Supplementary
        Section~\ref{subsec:line169-transport-details}).

        In another example, completing an orthonormalized sub-measurement adds
        an unstated $2\,\zeta_1$ term.
        After this term is absorbed into the cascade parameter, the completion
        error is
        $\zeta_2 = 200\,\zeta_1^{1/4} + 42\,\zeta_1^{1/8}$ rather than the
        printed coefficient $40$ (gap note 904).
  \item \emph{Boundary-case analysis}: Formalization also required explicit
        treatment of boundary cases omitted from the informal argument.
        Gap note 422 shows that the printed conditions permit $k=0$ when $d=0$;
        the formal theorem requires $0<k$.
        Similarly, gap note 930 (main-induction-successor-coefficient) treats
        the $m=1$ boundary of the induction.

  \item \emph{Side-condition handling}: A typographical error changed a
        condition needed later in the proof.
        The printed condition $k \ge md$ on the line-sampling parameter is
        insufficient for the successor pasting stage, whose additive Chernoff
        bound requires $k \ge 400\,md$ (gap note 906).
\end{enumerate}
Together with the total-overlap repair (gap note 1093; \cref{app:trajectory}),
the formalized proof incorporates all such corrections, including the two
side-condition repairs to the published theorem statement ($0 < k$ and
$k \ge 400\,md$) and the repairs to the intermediate error budget; its final
conclusion is unchanged, and the test remains sound with the printed error
expression under the tightened side conditions.

In the subsequent QPBT development (\cref{sec:second-theorem}), we also proved a
strictly stronger explicit bound for the same theorem in Lean, under the same
corrected hypotheses and with the same conclusions.
Its polynomial prefactor improves from $k^{2}m^{4}$ to $k^{1/4}m^{1/2}$, and its
error exponent from $1/40000$ to $1/8192$, compared with the bound in
\cref{eq:nu}.
The gain comes from the final assembly step, where the printed proof applies
the general triangle inequality for approximate equality of measurements
twice, losing a square root each time;
for complete (projective) measurements the inequality is linear, and the
inductive argument is unchanged.

\section{Discussion}\label{sec:discussion}

This work shows that long-horizon autoformalization of a complete research
proof is feasible in an agent-intensive, lightly human-supervised setting.
We formalized the quantum soundness of the classical low individual-degree
test, including corrections to the published statement and intermediate error
bounds~\citep{jnvwy_ldt}.
Under human guidance, models wrote all the Lean code.
Yet scaling from isolated benchmark problems to a full research paper changes
the nature of the task: agents must write not only tactic proofs, but also the
intermediate definitions and lemma statements that connect the argument.
Because the Lean kernel checks only that a proof matches its stated theorem, it
cannot tell whether those statements faithfully capture the mathematics of the
paper.

When an automated proof search stalls, the obstacle is not always a hard proof:
the intermediate claim being attempted may itself be false, either from a
formalization choice or because the published paper left a side condition
implicit or misstated an error bound (\cref{sec:formal-LIDT}).
Unlike one-shot formalization, an iterative workflow with explicit gap
reporting can separate these cases, turning formalization into a mathematical
audit that uncovers and repairs errors in the published proof itself.
When agents can also make a file compile by adding assumptions to intermediate
lemmas, placeholder counts no longer measure true progress: they fall when
obligations are moved into hypotheses and rise when audits restore the paper's
intended statements and isolate partial proofs as conditional helpers
(\cref{sec:local-progress,fig:metrics-evolution}).

Code review in large formalizations must therefore look beyond individual pull
requests to the structure of the full argument.
Approving one change at a time does not guarantee that the resulting pieces fit
together into the intended theorem (\cref{app:composition}), and automated
reviewers can share the same bias toward local compilation as proving agents.
Comparing complete chains of reasoning against the paper (\cref{sec:formalflow})
ensures not only that the final theorem is proved, but also that the Lean
library remains clean, composable, and faithful to the published mathematics.
As language models take on more of the formal coding and routine review,
maintaining this alignment with mathematical intent becomes the central role of
human experts.

Verification procedures in long-running agent systems cannot remain static.
Fixed instructions and checks catch only the failure patterns anticipated in
advance; as agents encountered new obstacles, they found new ways to satisfy
local checks without completing the intended mathematics (\cref{app:checks}).
Maintaining reliability over months of autonomous work required updating the
review criteria and continuous-integration checks as new failure modes
appeared, while protecting the verification rules themselves from agent
modification (\cref{sec:layer-4-axiom-audits}).
Similar co-evolving oversight will likely be needed in other long-horizon agent
applications~\citep{QwenTeam2026Verification, Amodei2016Concrete}.

Most of the LIDT formalization was completed in April and May 2026.
Since then, stronger models and larger developments such as
Erdos90~\citep{Erdos90}, alongside research papers accompanied by Lean
certificates~\citep{OpenAI2026Ten}, have expanded the scope of machine-checked
mathematics.
As models grow more capable at local proof search, the main bottleneck in
formalizing major theorems shifts from proving individual lemmas to maintaining
statement fidelity and composing results across a large development.
Using newer models, the same \sysname{} system has since formalized a second
component of $\MIPstarRE$, the soundness of the quantum Pauli basis test, on
top of the LIDT library while sharpening the LIDT error exponent
(\hyperref[sec:second-theorem]{Results}).

\section{Methods}\label{sec:methods}

We describe how \sysname{} stores the state of the LIDT formalization and
coordinates work across its three nested scales.
Alongside the scripts that define \sysname{}, the repository stores the complete
project record: its history, source paper, gap notes, Lean~4 codebase, and an
interactive blueprint~\citep{Massot2021Leanblueprint}.
We distilled reusable parts of \sysname{} into the \code{oh-my-formalization}
template and supporting tools (\cref{app:starter-tools}) for use in other
mathematical formalization projects.
Because these procedures developed during the project rather than being fixed at
the outset, we distinguish the initial setup from the review and repair
mechanisms introduced later (\cref{app:checks,sec:issue-workflows}).

\paragraph*{Blueprint, repository, and memory.}
At the start of each session, standing instructions direct the agent to read
the relevant paper passage, blueprint nodes, Lean files, gap notes, project
instructions, and a persistent memory store.
Because fresh language-model sessions do not inherit earlier conversations, the
shared GitHub repository records the proof state, open tasks, and review
decisions (\cref{sec:formalflow}), while project documentation and the agent
memory store preserve standing instructions and conventions across sessions.
The repository also archives 147 dated audit and session reports, which agents
retrieve by file name.

Using the LIDT paper's \LaTeX{} source, human supervisors and agents organized
its claims and dependencies in an interactive
blueprint~\citep{Massot2021Leanblueprint}.
Agents then generated the corresponding Lean skeleton of namespaces,
definitions, and unproven theorem signatures.
As the formalization progressed, an orchestrator agent split broad blueprint
statements into intermediate lemmas and added obligations for missing steps and
unstated prerequisites.
Declarations that state results from the paper are linked to blueprint nodes,
which identify the corresponding claims in the paper to support claim-by-claim
comparison.
The blueprint records statements and dependencies in \LaTeX{} and links them to
Lean declarations.
Agents and supervisors update it when they identify missing steps or
definitions that do not meet a later lemma's requirements.

\paragraph*{Task generation and organization.}
We organize proof obligations as GitHub issues.
Tracking issues cover chapters of the paper and theorem families, while
GitHub's native sub-issue links record individual proof tasks
(\cref{fig:workflow-vE2}).
GitHub issues, pull requests, and Actions keep tasks, reviews, and code changes
available to later agent sessions and external readers.
Agents open proof-task issues as work progresses, and issue workflows classify
eligible new issues and attach Mathlib scouting reports.
With post-merge tracking enabled, the issue tracker inspects merged pull
requests and review discussions for newly exposed proof obligations, creates
follow-up issues under the relevant tracker, and recommends a next task whose
prerequisites are satisfied.
A separate weekday workflow summarizes proof activity and open problems.
\Cref{sec:issue-workflows} details these workflows; the repair of the
mathematical pasting step in \hyperref[sec:results]{Results} illustrates how
review findings feed back into new proof tasks (\cref{fig:workflow-vE2}).

\paragraph*{The formalization session.}
We use TeXRA, Claude Code, OpenCode, and Codex to execute proof tasks in the
shared repository.
TeXRA's multi-agent system, described in our earlier
work~\citep{Lu2026Multiagent}, allows independent tasks to proceed concurrently
in separate git worktrees.
\Cref{sec:division-of-labor,sec:token-usage} report the recorded agent
contributions and model token usage.

An agent selects an open issue whose prerequisites are resolved and follows the
repository's standing instructions (\cref{sec:issue-session-instructions}),
which specify reading the paper passage first, then the blueprint node, and
finally the Lean files.
Before writing tactics, the agent documents the mathematical proof strategy
from the paper so that reviewers can compare the proposed proof against the
intended argument, and searches Mathlib and the codebase (assisted by scouting
reports) for reusable lemmas.
The instructions favor modular lemmas that compose cleanly across files.
When a session modifies a paper-linked theorem, the instructions require a
statement audit comparing its hypotheses and conclusion directly against the
paper, synchronizing the blueprint markers in the same pull request, and
reporting the verification steps performed.

\paragraph*{Tool loops, autonomous goals, and stopping criteria.}
\phantomsection\label{sec:stopping-criteria}
Proving sessions operate autonomously toward an assigned mathematical
obligation under a strict stopping criterion.
The agent iteratively explores proof tactics, queries Mathlib, and inspects
compiler diagnostics; if the session becomes idle, a continuation prompt
returns it to the target objective.
Execution terminates only when the target declaration compiles cleanly under
\code{lake env lean} without \code{sorry} placeholders or unapproved axioms,
or when the iteration budget is exhausted and the task is escalated for human
review (\cref{app:goal-loop}).

Repository instructions require the agent to type-check every edited Lean file
with \code{lake env lean}, scan it for \code{sorry} and \code{axiom}, and run
\code{lake build} whenever imports or shared declarations change.
Declarations that retain \code{sorry} tokens cannot mark a blueprint node
complete and remain tracked as open proof debt.
The agent records its verification commands in its task plan, and statement
review separately checks whether the proved declaration discharges the intended
mathematical obligation.

\begin{figure}
  \centering
  \begin{tikzpicture}[
      >={Stealth[length=3.2pt]},
      every node/.style={font=\footnotesize},
      box/.style={draw, rounded corners=1.5pt, inner sep=4pt,
                  align=center, minimum height=0.6cm},
      stage/.style={box, draw=black!70, fill=white},
      kchk/.style={draw=green!60!black, fill=green!14},
      revcore/.style={draw=coreaccent, line width=1pt, densely dashed,
                     fill=coreaccent!10},
      actor/.style={text=agentaccent!80!black, align=center},
      flow/.style={->, semithick, draw=black!80},
      aud/.style={->, semithick, draw=black!70, densely dashed},
      elbl/.style={fill=white, inner sep=2pt, align=center},
      legsw/.style={draw, minimum width=0.38cm, minimum height=0.25cm,
                    inner sep=0pt, rounded corners=0.5pt},
      x=1cm, y=1cm
    ]

    \node[box, corebox, minimum width=1.7cm] (paper) at (-3.15, 0.65)
      {Paper\\\scriptsize Frozen \LaTeX};
    \node[stage, minimum width=1.6cm] (bp) at (0, 0.65)
      {Blueprint\\\scriptsize \NumBlueprintNodes nodes};
    \node[box, kchk, minimum width=1.65cm] (lean) at (3.15, 0.65)
      {Lean code\\\scriptsize \NumLines lines};
    \node[box, revcore, minimum width=2.4cm] (gaplog) at (-1.6, -0.5)
      {Paper-gap notes\\\scriptsize \NumGapNotes notes};
    \node[box, evolvebox, minimum width=2.55cm] (scripts) at (1.5, -0.5)
      {Prompts, scripts,\\automated checks};
    \begin{pgfonlayer}{background}
      \node[resourcebox, rounded corners=3pt, inner sep=7pt,
            fit=(paper)(bp)(lean)(gaplog)(scripts)] (repo) {};
    \end{pgfonlayer}
    \node[anchor=south west, font=\footnotesize\itshape]
      at (repo.north west) {Shared repository};
    \draw[flow] (paper.east) --
      node[elbl, fill=none, above, font=\scriptsize]{Transcribe} (bp.west);
    \draw[flow] (bp.east) --
      node[elbl, fill=none, above, font=\scriptsize]{Formalize}
      node[elbl, fill=none, below, font=\scriptsize]{and prove} (lean.west);

    \node[stage, minimum width=3cm, minimum height=0.85cm] (tasksB)
      at (0.16, 3.26) {};
    \node[stage, minimum width=3cm, minimum height=0.85cm] (tasks)
      at (0, 3.15) {\textbf{Open proof tasks}\\\scriptsize GitHub issues};
    \node[stage, minimum width=2.6cm, minimum height=0.85cm] (prB)
      at (5.96, 0.76) {};
    \node[stage, minimum width=2.6cm, minimum height=0.85cm] (pr)
      at (5.8, 0.65) {\textbf{Pull requests}\\\scriptsize Proposed edits};
    \node[stage, minimum width=2.6cm] (merge) at (-6.2, 0.1)
      {\textbf{Merge}\\Reviewed changes};
    \node[box, humanbox] at (-6.2, 1.55)
      {\textbf{Authors}\\Steer and assess gaps};

    \draw[flow] (bp.north) --
      node[elbl, actor, pos=0.65, anchor=west]
      {Issue-tracking agents\\organize follow-up tasks} (tasks.south);
    \draw[flow] (tasks.east) -|
      node[elbl, actor, pos=0.31]
      {Task agents\\prove and submit edits} (pr.north);
    \draw[flow] (merge.east) -- (repo.west |- merge.east);

    \node[box, revcore, minimum width=10.6cm, inner sep=7pt]
      (checks) at (0, -3.4)
      {\textbf{Independent checks}\\[5pt]
       \begin{tabular}{@{}c@{\hspace{0.65cm}}c@{}}
       \textbf{Meaning: review agents} & \textbf{Validity: Lean kernel}\\[2pt]
       Compare statements with & Check proofs and audit axioms\\
       the blueprint and paper & \NumAxAssert axiom assertions
       \end{tabular}};
    \draw[flow] (pr.south) |- (checks.east);
    \draw[flow] (checks.west) -| (merge.south);
    \node[elbl, anchor=east] at (-5.8, -2)
      {Check\\results};

    \draw[aud] ([xshift=-1.6cm]checks.north) --
      node[elbl, pos=0.5]
      {Agents document\\deviations from the paper} (gaplog.south);
    \draw[aud, draw=agentaccent] (pr.south west) -- (scripts.east);
    \node[elbl, actor] at (3.25, -1.65)
      {Agents may propose changes\\to prompts and checks};

    \node[legsw, corebox] (lg1) at (-6.9, -4.7) {};
    \node[anchor=west, font=\scriptsize] at (lg1.east)
      {Frozen paper};
    \node[legsw, revcore] (lg2) at (-3.9, -4.7) {};
    \node[anchor=west, font=\scriptsize] at (lg2.east)
      {Changes require review};
    \node[legsw, kchk] (lg3) at (0.45, -4.7) {};
    \node[anchor=west, font=\scriptsize] at (lg3.east)
      {Kernel-checked};
    \node[legsw, evolvebox] (lg4) at (3.8, -4.7) {};
    \node[anchor=west, font=\scriptsize] at (lg4.east)
      {Agents may modify};
    \node[legsw, humanbox] (lg5) at (-6.9, -5.2) {};
    \node[anchor=west, font=\scriptsize] at (lg5.east) {Authors};
    \node[legsw, resourcebox] (lg6) at (-3.9, -5.2) {};
    \node[anchor=west, font=\scriptsize] at (lg6.east)
      {Shared repository};
    \node[stage, legsw] (lg7b) at (0.53, -5.14) {};
    \node[stage, legsw] (lg7) at (0.45, -5.2) {};
    \node[anchor=west, font=\scriptsize] at (lg7b.east)
      {Stacked cards: concurrent tasks or pull requests};
  \end{tikzpicture}
  \caption{The \sysname workflow.
    Task agents work on open proof tasks and submit their changes as pull
    requests.
    Issue-tracking agents organize follow-up proof obligations.
    Review agents compare statements with the blueprint and paper; Lean
    checks proofs and the axiom audit checks their dependencies.
    These checks run independently, according to the files changed: Lean and
    build changes trigger compilation and the axiom audit, while changes to
    Lean, blueprint, gap notes, or project instructions trigger review.
    Agents merge reviewed changes under the project instructions; authors
    steer the work and assess mathematical gaps.
    Agents document deviations from the paper in separate paper-gap notes.
    They may also propose changes to prompts, scripts, and automated checks
    through pull requests.
    The review prompt is read from the base branch, so a proposed prompt edit
    does not govern its own review.}
  \label{fig:workflow}
\end{figure}
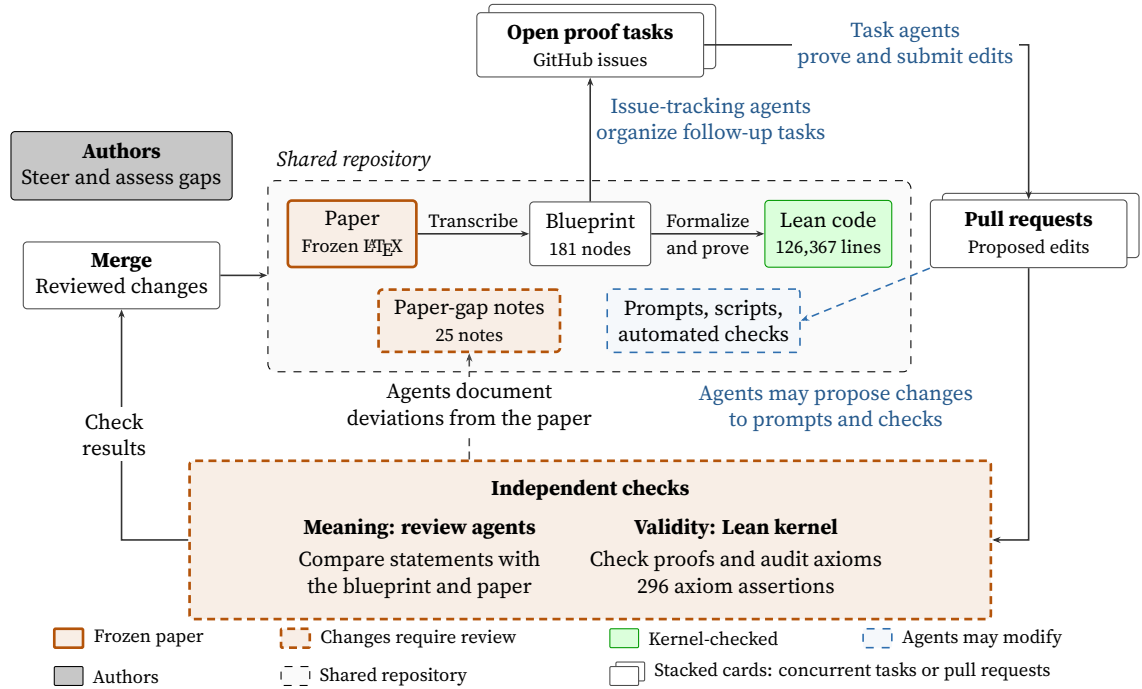

\paragraph*{Review and integration.}
Each worktree uses the Lean toolchain~\citep{DeMoura2015Lean} and
Mathlib~\citep{TheMathlibCommunity2020Lean}.
Build and review workflows trigger on pull requests according to the files
modified (\cref{fig:workflow}).
For Lean and build-configuration changes, continuous integration runs full
compilation (\code{lake build}) and an in-kernel axiom audit
(\code{Lean.collectAxioms}) checking transitive dependencies against the
approved axioms, while separate workflows check blueprint and source integrity
(\cref{app:theorem,app:checks}).
Automated review agents evaluate proposed Lean changes for mathematical
fidelity against the paper, the blueprint, and the library architecture.
When a build or blueprint check fails, GitHub Actions starts a repair agent on
the pull-request branch and supplies the compiler diagnostics.
When review comments identify discrepancies requiring repair, the agent
addresses unresolved threads under maintainer authorization.
Repair prompts (\cref{app:repair-prompt}) require local verification before
pushing and instruct agents to isolate mathematical obstructions rather than
weaken theorem statements.

The automatic repair workflow halts after five consecutive tagged repair
commits, escalating further intervention to maintainers
(\cref{sec:repository-repair}).
When concurrent branches conflict, an automated integration agent rebases the
pull request in a dedicated worktree, resolves syntactic and semantic
conflicts, and verifies full compilation before merging.
Edits to standing instructions, review prompts, and continuous-integration
scripts follow the same pull-request review process; whenever a review workflow
runs, it loads its prompt from the protected base branch so an agent cannot
weaken the criteria used to evaluate its own submission.

\paragraph*{The proof-gap protocol.}
A failed proof attempt may stem from a missing intermediate lemma, a defective
Lean definition, or an error in the published paper.
For a suspected discrepancy with the paper, the agent files a \LaTeX{} note
describing the claim, discrepancy, mathematical cause, proposed repair, and
affected results.
Human supervisors decide whether to correct the paper, repair the
formalization, or close the note without changing either statement.
An approved paper correction records the revised bound or hypothesis, such as
the $400\,md$ side condition, together with the required updates to dependent
theorems; a formalization repair instead restores the intended mathematical
object, for example by correcting state normalization.
Until the note is resolved, affected declarations and their dependents retain
metadata markers linking to the note (\cref{app:trajectory}).

\paragraph*{Auditing and discrepancy resolution.}
During the repair phase, maintainers launched a repository-wide audit-and-repair
prompt (\cref{app:repair-prompt}) instructing agents to compare formal
declarations against the blueprint and the paper, flag assumed intermediate
results or altered hypotheses, and log unresolved proof obligations as issues.
We executed this prompt in an autonomous run with repeated progress checks:
whenever a working session ended, a progress-check subagent evaluated the
repository state to recommend whether the orchestrator should re-invoke the
working agent, continuing until a subsequent audit pass identified no new
discrepancies and driving the repository toward the fixed point of faithful
statements and closed obligations.
Completing the proof then still required the in-kernel axiom audit and human
mathematical review described below.

\paragraph*{Automated checks.}
\phantomsection\label{sec:automated-checks}
We cataloged recurring failure modes and used them to refine review instructions
and add static checks for subsequent pull requests (\cref{app:checks}).
Automated review agents continued to inspect changes for semantic defects beyond
static rules: for example, review agents flagged an intermediate theorem whose
auxiliary hypotheses assumed variance bounds not present in the paper,
prompting an automated repair pull request (PR~\#1466) that preserved the
partial deduction as a conditional helper, restored the paper's statement, and
logged the missing step as a tracked obligation proved in a subsequent session
(PR~\#1496).
Similarly, in the induction section, PR~\#1664 restated the internal
self-improvement adapter to consume the Chapter~7 polynomial-measurement
interface and removed its completion marker while an underlying dependency
remained open.
A transitive proof-status script checks whether any blueprint node marked
complete depends on unfinished proofs or unapproved axioms (invoked in full
mode by the local pre-push hook and requested by the blueprint CI job at the
pinned snapshot), while companion scripts verify blueprint-to-Lean links and
flag conclusion-shaped hypotheses or forbidden proof-debt stems for reviewer
evaluation (\cref{app:checks}).

\paragraph*{Checking the completed proof.}
\phantomsection\label{sec:checking-proof}
The Lean compiler checks each proof against its stated theorem, and an
in-kernel axiom audit checks transitive axiom dependencies across selected
declarations using \NumAxAssertA{} standard-axiom assertions and
\NumAxAssertB{} no-unproven-step assertions.
The completed build contains no unfinished proofs, and the main theorem depends
only on the three standard axioms of classical mathematics (\code{propext},
\code{Classical.choice}, and \code{Quot.sound}; \cref{app:theorem}).
A standalone comparator repository independently verifies that the formalized
theorem statement matches a separately declared target signature.
Because the Lean kernel cannot verify whether that target signature faithfully
captures the informal theorem in the paper, one of the authors of the LIDT
paper~\citep{jnvwy_ldt} and our team audited the top-level theorem and its
transitive definitions down to Lean primitives to rule out weakened conclusions
or unintended hypotheses.
The repository metrics in Table~\ref{tab:artifact-process-metrics} report the
library at the pinned snapshot (\PinnedDate) under Lean toolchain
\NumToolchain.

\paragraph*{Use of language models.}
All Lean code and all repository commits, including the blueprint files, were
written by language-model agents; the human supervisors organized the blueprint
with the agents, set objectives, resolved discrepancies with the paper, and
audited the result (\cref{sec:division-of-labor}).
The models used are listed in \cref{sec:token-usage}.
The review census in \cref{tab:review-semantic-tree} used a GPT-5.4 reading
pass followed by a deterministic term-matching classifier
(\cref{app:review-census}).
We used language-model tools to draft and edit parts of the manuscript text and
the figure code; the authors checked all content and take responsibility for
it.

\section*{Data availability}
The data supporting the findings of this study are available in the public
\href{https://github.com/LionSR/MIPStarRE}{MIPStarRE} and
\href{https://github.com/Dengnifer/MIPStarRE-QPBT}{MIPStarRE-QPBT} GitHub
repositories.
The main-theorem proof was completed in May 2026.
Library integration and statement alignment continued in June, including
removal of the legacy same-space route and proof of a heterogeneous
sub-measurement consistency lemma (\cref{app:trajectory}).
The \code{MIPStarRE} repository also contains the
\href{https://sirui-lu.com/MIPStarRE/blueprint/}{interactive blueprint} and
the gap notes.

\section*{Code availability}
The Lean proof and supporting library are available in the
\href{https://github.com/LionSR/MIPStarRE}{MIPStarRE repository} under the
Apache License 2.0.
The reported library measurements use snapshot \PinnedHead{} and Lean~4
\NumToolchain{}; \cref{app:theorem} gives the build and
verification instructions.
The formalization of the quantum Pauli basis test, which also includes the
LIDT main theorem with the improved $k^{1/4}m^{1/2}$ prefactor and $1/8192$
error exponent (\cref{sec:formal-LIDT}), is available in the
\href{https://github.com/Dengnifer/MIPStarRE-QPBT}{MIPStarRE-QPBT repository}.
The formalization is registered in the
\href{https://palomar-registry.org/entry.html?%
id=PALOMAR-2026-08-18-000001}{Palomar Registry}
as PALOMAR-2026-08-18-000001: the registry re-checks the proofs with Lean
from its registered source and publishes the exact statement
(\code{MIPStarRE.LDT.Test.mainFormal}), the libraries it uses, and the
review's comments.
The registered source is the companion
\href{https://github.com/LionSR/LDT-comparator}{comparator repository}, which
certifies that
the library's main theorem matches the registered target statement.

\section*{Author contributions}
S.L., R.D., and Z.J. developed the methodology and software, conducted the
investigation, and validated the results.
Y.Z. supported the computational infrastructure for the large-scale model
runs.
All authors wrote and revised the manuscript.

\section*{Competing interests}
The authors declare no competing interests.

\section*{Acknowledgements}
We thank Thomas Vidick for suggesting to organize a comparator and
Challenge.lean for the Lean formalization.
We thank the Palomar team for building the registry.

This work is partially supported by National Key Research and Development
Program of China (Grant No.\ 2023YFA1009403), National Natural Science
Foundation of China (Grant No.\ 12347104), and Beijing Science and Technology
Planning Project (Grant No.\ Z25110100810000).
The work is partially supported by the Deutsche Forschungsgemeinschaft (DFG,
German Research Foundation) under Germany's Excellence Strategy -- EXC-2111 --
390814868.
This research is part of the Munich Quantum Valley, which is supported by the
Bavarian state government with funds from the Hightech Agenda Bayern Plus.

Yanqiao Zhu's contribution to this publication was made in an individual
capacity, not in connection with duties or responsibilities to UCLA.
Any opinions expressed in this work are those of Yanqiao Zhu; they are not
official positions or endorsements of UCLA and do not represent Yanqiao Zhu's
research at UCLA.

\beginsupplement

\startcontents[si]
\printcontents[si]{}{1}{\setcounter{tocdepth}{2}}
\vspace{10pt}{\color{sciaccent!45}\rule{\linewidth}{0.6pt}}\par\vspace{10pt}

We organize the supplementary material around the formal soundness theorem, the
statement repairs required to complete its inductive proof, and the verification
and task-tracking procedures that governed the formalization.
\Cref{app:theorem} states the formal theorem \code{Test.mainFormal}, maps its
hypotheses and conclusions to the objects of the low individual-degree test, and
explains how to verify the completed proof in the Lean kernel.
\Cref{app:trajectory} traces the sequence of statement repairs from the initial
defective declarations to their final kernel-checked forms.
\Cref{app:composition} examines how these repaired lemmas compose across the
inductive pasting proof, tracking their hypotheses, operator representations,
and error bounds.
\Cref{app:checks} specifies the automated checks and mathematical review
procedures used to detect unproved obligations and statement discrepancies.
\Cref{app:task-tracking} documents the GitHub task-tracking workflow, agent
contributions, and model token usage.
\Cref{app:review-census} details the construction and classification of the
review dataset analyzed in the main text.
\Cref{app:repair-prompt} reproduces the prompt used to audit formal declarations
against the paper and repair statement discrepancies across the repository.
Finally, \cref{app:starter-tools} describes the project template, blueprint
tools, and agent instructions distilled from this formalization for reuse in new
projects.

\IfStandalone{%
  \tableofcontents
  \clearpage
}{}

\suppappendix{The formal theorem and its verification}
\label{app:theorem}

\subsection{The formal statement}
\label{sec:formal-statement}

\cref{fig:statement} presents the verified Lean declaration
\code{Test.mainFormal} alongside Theorem~3.10 of~\citet{jnvwy_ldt}.
For an $(m,q,d)$-low individual-degree test instance with field size $q$,
variable count $m$, degree bound $d$, soundness error $\varepsilon$, and
line-sampling parameter $k$, any finite-dimensional two-prover projective
strategy succeeding with probability at least $1-\varepsilon$ admits global
polynomial measurements $G^{\mathrm{A}}, G^{\mathrm{B}} \in \polymeas{m}{q}{d}$
that approximate the provers' point measurements and agree with each other.
The formal declaration establishes the published conclusion and error bound
under two repaired hypotheses on $k$: the sampling threshold $k \ge 400\,md$
(\code{hk}, replacing $k \ge md$) and strict positivity $0 < k$ (\code{hk0},
excluding $k = 0$ when $d = 0$).
Under these conditions, point consistency and self-consistency hold within the
final soundness error
\begin{equation}
  \nu = 100000\,k^{2}m^{4}
  \left(\varepsilon^{\frac{1}{40000}} + {(d/q)}^{\frac{1}{40000}}
  + \exp(-k/(2560000\,m^{2}))\right).
  \label{eq:nu}
\end{equation}

\FloatBarrier
\begin{figure}
  \newlength{\statementpanelheight}
  \newlength{\statementframesep}
  \newlength{\statementframerule}
  \setlength{\statementpanelheight}{7.5cm}
  \newsavebox{\statementformalbox}
  \setlength{\statementframesep}{4pt}
  \setlength{\statementframerule}{0.4pt}
  \begingroup
  \setlength{\fboxsep}{\statementframesep}
  \setlength{\fboxrule}{\statementframerule}
  \centering
  \begin{subfigure}[b]{0.39\textwidth}
    \raggedright
    \fcolorbox{leanframe}{leanbg}{\begin{minipage}[t][\statementpanelheight][t]
    {\dimexpr\linewidth-2\fboxsep-2\fboxrule\relax}
    \small
    \textbf{Informal soundness statement.}
    Consider a projective strategy
    $(\psi,A^{\mathrm{A}},B^{\mathrm{A}},L^{\mathrm{A}},
    A^{\mathrm{B}},B^{\mathrm{B}},L^{\mathrm{B}})$ that passes the
    $(m,q,d)$-low individual-degree test with probability at least
    $1-\eps$.
    Let $k \geq md$ be an integer and $\nu$ be defined as in \cref{eq:nu}.
    Then there are projective measurements
    $G^{\mathrm{A}},G^{\mathrm{B}}\in\polymeas{m}{q}{d}$ such that:
    \begin{enumerate}[leftmargin=1.25em,itemsep=1pt,topsep=2pt]
      \item \textit{Consistency with $A$.}
        On average over $\mathbf{u}\sim\F_q^m$,
        \[
          \begin{aligned}
            A^{\mathrm{A},u}_a\otimes I&\simeq_\nu
            I\otimes G^{\mathrm{B}}_{[g(u)=a]},\\[-2pt]
            I\otimes A^{\mathrm{B},u}_a&\simeq_\nu
            G^{\mathrm{A}}_{[g(u)=a]}\otimes I.
          \end{aligned}
        \]
      \item \textit{Self-consistency.}
        \[
          \begin{aligned}
            G^{\mathrm{A}}_g\otimes I&\simeq_\nu
            I\otimes G^{\mathrm{B}}_g.
          \end{aligned}
        \]
    \end{enumerate}
    \end{minipage}}
    \caption{Informal statement.}\label{fig:informal-statement}
  \end{subfigure}\hfill
  \begin{subfigure}[b]{0.60\textwidth}
    \raggedright
    \begin{lrbox}{\statementformalbox}
    \begin{minipage}{\dimexpr\linewidth-2\fboxsep-2\fboxrule\relax}
    \begin{lstlisting}[basicstyle=\ttfamily\scriptsize,breaklines=false,
      frame=none,xleftmargin=0pt,
      backgroundcolor=\color{leanbg},aboveskip=0pt,belowskip=0pt]
theorem mainFormal
    (params : Parameters) [FieldModel params.q] {ιA ιB : Type*}
    [Fintype ιA] [DecidableEq ιA] [Fintype ιB] [DecidableEq ιB]
    (strategy : ProjStrat params ιA ιB) (eps : Error)
    (hpass : strategy.PassesLowIndividualDegreeTest eps) (k : ℕ)
    -- paper prints k ≥ md
    (hk : 400 * params.m * params.d ≤ k)
    -- absent from the paper
    (hk0 : 0 < k) :
    ∃ G_A : ProjMeas (Polynomial params) ιA,
    ∃ G_B : ProjMeas (Polynomial params) ιB,
      ConsRel strategy.state
          (uniformDistribution (Point params))
          (IdxProjMeas.toIdxSubMeas strategy.pointMeasurementA)
          (polynomialEvaluationFamily params G_B.toSubMeas)
          (mainFormalError params k eps) ∧
      ConsRel strategy.state
          (uniformDistribution (Point params))
          (polynomialEvaluationFamily params G_A.toSubMeas)
          (IdxProjMeas.toIdxSubMeas strategy.pointMeasurementB)
          (mainFormalError params k eps) ∧
      ConsRel strategy.state
          (uniformDistribution Unit)
          (constSubMeasFamily G_A.toSubMeas)
          (constSubMeasFamily G_B.toSubMeas)
          (mainFormalError params k eps)
    \end{lstlisting}
    \end{minipage}
    \end{lrbox}
    \fcolorbox{leanframe}{leanbg}{\begin{minipage}[t][\statementpanelheight][t]
    {\dimexpr\linewidth-2\fboxsep-2\fboxrule\relax}
    \usebox{\statementformalbox}
    \end{minipage}}
    \caption{Formal statement.}\label{fig:formal-statement}
  \end{subfigure}
  \caption{The soundness theorem in two forms.
    \textbf{\textsf{(a)}}~The statement printed in the paper~\citep{jnvwy_ldt}.
    \textbf{\textsf{(b)}}~The statement that the Lean kernel has checked.
    Each hypothesis and each conclusion on the left has a formal
    counterpart on the right, including the error bound.
    The comments on the right mark two corrected side conditions: the formal theorem requires
    $k \ge 400\,md$ where the paper prints $k \ge md$, and it requires
    $0 < k$, which the paper omits.}
  \label{fig:statement}
  \endgroup
\end{figure}

\subsection{The error parameters}
\label{sec:error-parameters}

\cref{tab:error-parameters} traces how the intermediate error bounds from the
inductive pasting step and the final assembly stage accumulate into the
soundness error $\nu$ of \cref{eq:nu}.
In the pasting step, the base-case line and plane consistency errors
$\nu_1$--$\nu_3$ are absorbed into the $\widehat{H}$--$B$ consistency bound
$\nu_5$, which we write in terms of the combined error factor
$E \coloneqq \eps^{1/32} + \delta^{1/32} + \gamma^{1/32} + \zeta^{1/32}
+ (d/q)^{1/32}$.
The table highlights the three intermediate bounds that required correction
during formalization and lists the printed expressions alongside them.
At the pasting stage, replacing the linear factor $km$ in $\nu_8$ by the
quadratic binomial-tail factor $k^2 m$ preserves the completeness bound
$\nu_7 + \nu_8 \le \nu$.
At the final assembly stage in Section~5 of the LIDT paper~\citep{jnvwy_ldt},
the self-consistency error $\zeta_3 = 6\zeta_1 + 6\zeta_2$ (which incorporates
the repaired completion error $\zeta_2$) and the repaired point-consistency
substitution term $\zeta_1 + 10\zeta_1^{1/8}$ combine to yield the repaired
point-consistency bound $\zeta_4^{\mathrm{repaired}}$, which remains bounded by
the final error $\nu$ in \cref{eq:nu}.

\begin{table}
  \centering
  \caption{The error parameters of the proof.
    The final error in \cref{eq:nu} bounds the accumulated
    errors.
    Here $E = \eps^{1/32}+\delta^{1/32}+\gamma^{1/32}+\zeta^{1/32}
    +(d/q)^{1/32}$.
    In the final-assembly rows, $\sigma$ is the induction error for the
    symmetrized $(3\eps,3\eps,3\eps)$-good strategy; in the last row it
    denotes the output of one pasting step.
    The paper reuses the letter $\nu$ at three scales: the pasting-stage
    $\nu$ of the last row, the induction error
    $\nu = 1000k^2m^2(\eps^{1/1024}+\delta^{1/1024}+\gamma^{1/1024}
    +(d/q)^{1/1024})$ that absorbs it, and the final error of \cref{eq:nu}.
    \cref{subsec:zeta2-cascade-residual,subsec:bernoulli-telescoping} derive
    the corrected $\zeta_2$ and $\nu_8$.
    \cref{subsec:line169-transport-details} gives
    the repaired point-consistency bound and shows how the final error
    bound includes it.}
  \label{tab:error-parameters}
  \begin{tblr}{
      colspec = {Q[l] X[5,l] X[8,l] X[8,l]},
      row{odd} = {bg=gray!5},
      row{1}   = {bg=gray!25, font=\bfseries},
      rowsep = 3pt,
      colsep = 4pt,
    }
    \toprule
    Parameter & Defined in & Expression or bound & Absorbed into \\
    \midrule
    $\eps$ & Test definition (good strategy) &
      Axis-parallel lines test fails w.p.\ $\le \eps$ &
      $\zeta$; $\nu_5$--$\nu_7$; $\nu$ \\
    $\delta$ & Test definition (good strategy) &
      Self-consistency test fails w.p.\ $\le \delta$ &
      $\zeta$; $\nu_5$--$\nu_7$; $\nu$ \\
    $\gamma$ & Test definition (good strategy) &
      Diagonal lines test fails w.p.\ $\le \gamma$ &
      $\nu_4$--$\nu_8$; $\nu$ \\
    $\zeta$ & Self-improvement (helper lemma) &
      $100m\,(\eps^{1/2}+\delta^{1/2}+(d/q)^{1/2})$ &
      The pasting hypotheses; $\nu_4$--$\nu_8$ \\
    $\zeta_1$ & Final assembly (symmetrized pair) &
      $2\sigma + 2\sqrt{3\eps+2\sigma} + md/q$ &
      $\zeta_2$; $\zeta_3$; $\zeta_4^{\mathrm{repaired}}$ \\
    $\zeta_2$ & Final assembly (completion) &
      $\mathrepair{200\zeta_1^{1/4}+42\zeta_1^{1/8}}$\newline
      (printed $200\zeta_1^{1/4}+40\zeta_1^{1/8}$) &
      $\zeta_3$; $\zeta_4^{\mathrm{repaired}}$ \\
    $\zeta_3$ & Final assembly (self-consistency) &
      $6\zeta_1+6\zeta_2$ &
      $\zeta_4^{\mathrm{repaired}}$; final self-consistency \\
    $\zeta_4^{\mathrm{repaired}}$ & Final assembly (point consistency) &
      $2\sigma+2\sqrt{\mathrepair{\zeta_1+10\zeta_1^{1/8}}+\zeta_3/2}$\newline
      (printed $2\sigma+2\sqrt{\zeta_1+\zeta_3/2}$) &
      Final error in \cref{eq:nu} \\
    $\nu_4$ & Pasting (commuting past $\widehat{G}$'s) &
      $426\,k^2m\,(\gamma^{1/16}+\zeta^{1/16}+(d/q)^{1/16})$ &
      $\nu_5$ via $\nu_1 + 2\sqrt{\nu_4}$; $\nu_8$ \\
    $\nu_5$ & Pasting ($\widehat{H}$--$B$ consistency) &
      $43\,km\,E$ &
      $\nu_6$ via $k^2/q + k\nu_5$; $\nu_7$ \\
    $\nu_6$ & Pasting ($H$--$B$ consistency) &
      $44\,k^2m\,E$ &
      $H$--$A$ consistency: $\nu_6+\sqrt{8m\eps+4\delta} \le \nu$ \\
    $\nu_7$ & Pasting (over all outcomes) &
      $46\,k^2m\,E$ &
      Completeness of $H$: $\nu_7+\nu_8 \le \nu$ \\
    $\nu_8$ & Pasting (from $\widehat{H}$ to $G$) &
      $\mathrepair{46\,k^2m\,(\gamma^{1/32}+\zeta^{1/32}+(d/q)^{1/32})}$\newline
      (printed $46\,km\,(\cdots)$) &
      Completeness of $H$: $\nu_7+\nu_8 \le \nu$ \\
    $\nu$ & Pasting theorem &
      $100\,k^2m\,E$ &
      $\sigma = \kappa\,(1+\tfrac{1}{100m}) + 2\nu + e^{-k/(80000m^2)}$;
      the induction; \cref{eq:nu} \\
    \bottomrule
  \end{tblr}
\end{table}

\subsection{Scope}
\label{sec:theorem-scope}

The formal proof of \code{Test.mainFormal} depends only on the three standard
axioms of classical mathematics in Lean: propositional extensionality
(\code{propext}), classical choice (\code{Classical.choice}), and quotient
soundness (\code{Quot.sound}).
The quantum soundness proof does not invoke the two classical precursor
theorems, Raz--Safra~\citep{Raz1997Subconstant} and
Polishchuk--Spielman~\citep{Polishchuk1994Nearlylinear}.

\subsection{Checking the completed proof}
\label{sec:verification-artifacts}

The theorem and proof statistics in this paper refer to the \code{MIPStarRE}
pinned library snapshot \PinnedHead{} (\PinnedDate), at which the library
contains no unfinished proofs; the standalone comparator checks use a subsequent
July revision.
At this June library snapshot, an external reader verifies the theorem
through Lean kernel typechecking (\code{lake build}), in-kernel axiom
inspection (\code{\#print axioms Test.mainFormal}), and a source-level
declaration audit.
Within the build, the in-kernel axiom audit module automates this axiom check
across \code{Test.mainFormal} and other target declarations using
\code{Lean.collectAxioms}.
Its \code{assert\_standard\_axioms} command verifies that \code{Test.mainFormal}
depends only on \code{propext}, \code{Classical.choice}, and \code{Quot.sound},
while its \code{assert\_no\_sorry\_axiom} command rejects any dependency on
\code{sorryAx}, Lean's axiom for an unfinished proof.
Complementing the kernel check, the declaration audit script scans the library
tree and rejects explicit \code{axiom} and \code{constant} commands.
For repository status display, a text scanner also checks tracked files for
\code{sorry} tokens and finds zero at this snapshot, while formal verification
relies on the in-kernel axiom collection above.

To allow external auditing of the theorem statement without traversing the full
library, we exported \code{Test.mainFormal} and its transitive definitional
closure into a single file, \path{Challenge.lean}, that imports only Mathlib
and replaces the proof body with an intentional \code{sorry}
(\cref{fig:statement}).
In \href{https://github.com/LionSR/LDT-comparator}{\code{LDT-comparator}}, the
official Lean comparator checks at the kernel level that the proof term in
\code{MIPStarRE} solves this standalone challenge declaration without
additional axioms.
We added the closure-export script to \code{MIPStarRE} in
\mipcommit{9e882d49c} on 16 July 2026 and added an automated drift check for the
exported statement in \mipcommit{5e8e64ea6} on 17 July, both after the
June library snapshot (\PinnedDate).

\subsection{Software configuration}
\label{sec:software-configuration}

The June library snapshot (\PinnedDate) pins both the Lean toolchain and
Mathlib to \NumToolchain, whereas the July comparator revision pins both to
\code{v4.32.0}.
In the June repository state, the continuous-integration build workflow compiles
the library with \code{lake build} and executes the in-kernel axiom audit on
every pull request modifying Lean sources, the Lake configuration, the
toolchain, or the workflow file.
Separate workflows execute the blueprint and proof-integrity checks detailed in
\cref{app:checks}.

\subsection{Reproducing the checks}
\label{sec:reproducing-the-checks}

An external reader can reproduce the kernel verification in two stages.
First, checking out the June library snapshot (\PinnedDate) with its pinned
toolchain and running \code{lake build} together with the declaration and
in-kernel axiom audits verifies that \code{Test.mainFormal} typechecks in the
Lean kernel and depends only on the three standard classical axioms.
Second, checking out the July comparator revision regenerates the standalone
target statement, confirms that it matches the certified target byte for byte,
and runs the Lean comparator to check the library proof against that target
under the same revision's toolchain and Mathlib dependency.

\suppappendix{Trajectory of the formal statements}
\label{app:trajectory}
\subsection*{Structure, notation, and color conventions}
\addcontentsline{toc}{subsection}{Structure, notation, and color conventions}

We trace how the formal statements evolved during the formalization of the
quantum low individual-degree test (LIDT)~\citep{jnvwy_ldt} across the
\NumCommits-commit pinned history of the \code{MIPStarRE} repository.
Each transition cites the pull request, commit, or issue that introduced or
repaired the statement.
We first examine three representative statement trajectories and the repository
chronology, then summarize the statement milestones across the ten blueprint
chapters, and finally catalog the formal gap notes that adjudicated
discrepancies with the paper.

\begin{itemize}
  \item \textbf{Theorem and lemma numbering:} When an environment heading
    names a statement in the paper, its base number matches the paper's
    numbering~\citep{jnvwy_ldt}.
    Other labels are local to this appendix.
  \item \textbf{Defect highlighting:} \mathdefect{\text{Light rose
    highlighting}} marks uncoupled existential witnesses, bundled auxiliary
    hypotheses, omitted side conditions, or invalid algebraic steps.
  \item \textbf{Repair highlighting:} \mathrepair{\text{Light teal
    highlighting}} marks proved constructions, corrected domain bounds, and
    completed proofs.
  \item \textbf{Commit and pull request identifiers:} Pull requests
    (\mippr{123}), issues (\mipissue{456}), and commits (\commit{8ae0cc7e})
    refer to historical records in the public repository.
\end{itemize}

Throughout this appendix, \emph{measurement}, \emph{sub-measurement}, and
\emph{projective sub-measurement} denote the finite-matrix structures defined
for the LIDT paper.
A sub-measurement consists of positive semidefinite outcomes whose sum is at
most the identity; a measurement has sum equal to the identity; and a
projective sub-measurement additionally requires idempotent outcomes.
These conditions distinguish an arbitrary family of projective matrices from a
projective sub-measurement, and both from a complete projective measurement.
In all formal declarations, operator families act on finite-dimensional
Hilbert spaces and are represented as complex matrices
$\operatorname{Matrix}(\iota,\iota;\mathbb C)$ indexed by finite decidable
carrier types $\iota$.

\subsection{Three worked trajectories}
\label{sec:exemplar-trajectories}

A \emph{formally valid shortcut} is a compiler-accepted construction that
avoids part of the intended proof or changes the statement without proving
the paper's claim: Lean verifies that a proof term establishes its stated
proposition, but whether that proposition expresses the original mathematical
claim requires a separate check~\citep{Lu2026Multiagent}.
During early formalization passes, compiler-accepted declarations departed from
the paper's statements through uncoupled existential witnesses, tautological
definitional equalities, bundled bridge hypotheses, and weakened conclusions or
side conditions.
We trace three representative statements from these initial declarations to
their final verified forms, showing how downstream proof steps and statement
audits required each repair.

\subsubsection{Trajectory 1: rounding to projectors}
\label{subsec:rounding-trajectory}
Lemma~5.6 in the paper takes an almost-projective measurement $\{A_a\}$
on a state $\psi$ with rounding defect at most $2\zeta$ and produces a family of
projective matrices $\{R_a\}$ close to $A_a$ on $\psi$, satisfying
$A_a \otimes I \approx_{2\sqrt{\zeta}} R_a \otimes I$ and
$\sum_a R_a \le (1+2\sqrt{\zeta})I$.
The initial formalization at \mipcommit{8ae0cc7e} supplied an existential
witness uncoupled from both $A$ and $\psi$, defining a constant $1 \times 1$
projection matrix $[1]$ on a one-dimensional carrier
(\mathdefect{\code{carrier~:= PUnit}}).
That witness satisfied the idempotence and operator-sum bounds on \code{PUnit}
while omitting the state-dependent closeness relation to the input measurement
$A$.
Downstream proof steps in Chapter~7 could not apply this declaration because
they required state-dependent closeness on the input carrier $\mathcal{H}_A$
(\mippr{287}).
The final verified theorem (\mippr{1126}, \mippr{1632}) constructs the
projective matrices on $\mathcal{H}_A$ by continuous functional calculus in
three branches: an exact eigenspace projection at $\zeta=0$, a spectral cutoff
for $0<\zeta\le 1/4$, and zero projectors for $\zeta > 1/4$, establishing the
sharp $2\sqrt{\zeta}$ closeness bound
(\code{projectiveNonMeasurement\_of\_sourceAlmostProjective\_two\_mul\_full}).
Comparing the initial declaration with the repaired theorem gives:
\begin{equation}
  \begin{gathered}
    \sum_a\langle\psi|(A_a-A_a^2)\otimes I|\psi\rangle\le2\zeta
    \quad\Longrightarrow\\
    \mathdefect{\begin{gathered}
      \exists\{R_a\}\subseteq\operatorname{End}(\mathbb{C}^1),\quad
      R_a^2=R_a=R_a^\dagger,\quad
      \sum_{a}R_{a}\le(1+2\sqrt{\zeta})I
    \end{gathered}}
    \quad\text{(defective witness on $\code{PUnit}$)} \\
    \mathrepair{\begin{gathered}
      \exists\{R_a\}\subseteq\operatorname{End}(\mathcal H_A),\quad
      R_a^2=R_a=R_a^\dagger,\\
      A_a\otimes I\approx_{2\sqrt\zeta}R_a\otimes I,\quad
      \sum_{a}R_{a}\le(1+2\sqrt{\zeta})I
    \end{gathered}}
    \quad\text{(repaired theorem on $\mathcal{H}_A$)}.
  \end{gathered}
\end{equation}
Here $A$ is a measurement on $\mathcal{H}_A$ and $\psi$ is a normalized
bipartite state.
Retaining the input carrier $\mathcal{H}_A$ and the closeness bound
$A_a\otimes I\approx_{2\sqrt\zeta}R_a\otimes I$ allows the subsequent
orthonormalization step to bound the operator-sum defect on the original
Hilbert space.

\subsubsection{Trajectory 2: Naimark sub-measurement dilation}
\label{subsec:naimark-alphabet-repair}
Theorem~5.1 in the paper takes sub-measurements $\{A_a^x\}$ and
$\{B_b^y\}$ on a bipartite state $\psi$ and describes the dilated outputs as
projective measurements on the original outcome alphabets that preserve all
joint outcome correlations.
The first formal declaration (\commit{82d70fef}) encoded this prose literally
by asserting completeness $\sum_a \widehat A^x_a = I$ on the unextended outcome
alphabet.
When an input sub-measurement has strictly deficient total mass, however,
complete projective outcomes on the unextended alphabet cannot preserve all
joint probabilities.
For example, on a one-dimensional space with a single outcome $*$,
\begin{equation}
  A_*=0,\quad B_*=I
  \quad\Longrightarrow\quad
  \langle\psi|A_*\otimes B_*|\psi\rangle=0,
  \qquad
  \mathdefect{\widehat A_*=\widehat B_*=I}
  \quad\Longrightarrow\quad
  \langle\widehat\psi|\widehat A_*\otimes\widehat B_*|\widehat\psi\rangle=1.
\end{equation}
To preserve outcome probabilities for deficient inputs, the repaired
construction first adjoins a complementary outcome $\bot$ to form a complete
measurement on the extended alphabet:
\begin{equation}
  A^{x,+}_a=A^x_a,\qquad
  A^{x,+}_{\bot}=I-\sum_a A^x_a.
\end{equation}
Applying Naimark dilation to $A^{x,+}$ and restricting back to the original
outcomes yields a projective sub-measurement on the original alphabet.
In the bipartite setting, an intermediate questionwise construction at
\commit{6311ba3f} preserved single-prover marginals but did not establish joint
two-prover correlation preservation.
The final bipartite construction (\mippr{1782}, \commit{440cb0f8}) proves the
joint correlation identity on the four-register space
$(\mathcal H_A\otimes\mathcal H_{\mathrm{aux},A})\otimes
(\mathcal H_B\otimes\mathcal H_{\mathrm{aux},B})$:
\begin{equation}
  \mathrepair{\langle\psi|A^x_a\otimes B^y_b|\psi\rangle
  =\langle\widehat\psi|\widehat A^x_a\otimes
    \widehat B^y_b|\widehat\psi\rangle},
  \qquad
  \sum_a\widehat A^x_a\le I,\quad
  \sum_b\widehat B^y_b\le I.
\end{equation}
Here $\widehat\psi$ is the input state tensored with a product auxiliary
state, with registers permuted so that Alice's and Bob's dilated operators act
on their respective local tensor factors.

\subsubsection{Trajectory 3: the soundness statement}
\label{subsec:soundness-trajectory}
Theorem~3.10 in the paper states soundness of the low individual-degree test
under the sampling condition $k\ge md$.
In the repository history, the formal soundness statement evolved across five
stages as carrier assumptions, bundled hypotheses, and side conditions were
resolved.
First, the initial declaration at \commit{f659daae} restricted Alice and Bob to
a single shared carrier and accepted a bundled bridge structure
(\code{MainFormalBridgePackage}) containing the target conclusion as an input
field.
Second, after \commit{5fade34e} removed that bridge structure and exposed an
explicit \code{sorry} obligation, \mippr{912} (\commit{805e890d9}) tightened
the shared-carrier declaration \code{mainFormal} to the side conditions
$400md\le k$ and $0<k$.
Third, at \commit{440cb0f8}, a parallel declaration
\code{mainFormal\_sourceStatement} introduced independent prover carriers
$\iota_A$ and $\iota_B$, but retained the paper's printed condition $md\le k$
without requiring $0<k$:
\begin{equation}
  \begin{aligned}
    \code{mainFormal}:\quad&\text{one shared carrier},\quad
      400md\le k,\quad 0<k,\\
    \code{mainFormal\_sourceStatement}:\quad&\text{independent carriers},
      \quad\mathdefect{md\le k}\quad\text{with no }0<k.
  \end{aligned}
\end{equation}
Fourth, completing the inductive proof on independent carriers required two
corrections to \code{mainFormal\_sourceStatement}: Bernoulli pasting requires
$k\ge400md$ (\mipissue{1507}, added in \commit{78d56aa}), and the endpoint
$d=k=0$ causes the printed soundness error to vanish (\mipissue{422}, excluded
by adding $0<k$ in \commit{27df275}).
Fifth, \commit{28d5417d} removed the shared-carrier declaration and renamed the
independently quantified theorem to \code{Test.mainFormal}.
Its final verified declaration appears in \cref{sec:formal-statement}, and the
zero-sampling counterexample and pasting threshold are derived in
\cref{subsec:pasting-degree-zero,subsec:bernoulli-telescoping}.

\subsection{Timeline of formal transitions}
\label{sec:timeline}

\cref{tab:timeline-shortcuts} chronicles the statement transitions and proof
repairs from March to June 2026, and \cref{fig:metrics-evolution} places these
phases on the same calendar alongside the daily blueprint and \code{sorry}
counts.
Replacing early stand-in witnesses and bundled bridge hypotheses with faithful
statements repeatedly reopened proof obligations in modules that had previously
compiled without \code{sorry} markers, producing the non-monotone declaration
and placeholder trajectories analyzed in \cref{sec:blueprint-progress}.

\FloatBarrier

\begin{longtblr}[
  caption = {\textbf{Chronology of formal transitions and repairs
    (March--June 2026).}
  In March and April most entries replace an intermediate stand-in with a
  proved construction.
  The May and June entries document proofs of the bridge hypotheses, corrected
  side conditions, and checks required before merging.},
  label = {tab:timeline-shortcuts},
]{colspec = {Q[l] X[6.6,l] X[5.35,l] X[6.3,l] X[21.75,l]},
  row{odd} = {bg=gray!5},
  row{1} = {bg=gray!25, font=\bfseries},
  rowsep = 3pt, colsep = 4pt, rowhead = 1}
\toprule
Date & Milestone & Module & Phenomenon & Change \\
\midrule
\textbf{24 Mar} & \mippr{19}, \mippr{36} & Ch.~5 (Expansion) & Tautological \code{rfl} & Defined Fourier inner product directly as Kronecker delta; proved by reflexivity. \\
\textbf{27 Mar} & \mipcommit{78a14cd4} & Ch.~5 (Expansion) & Definitional alias & Defined Laplacian difference form as $L_{\mathrm{diff}} \coloneqq L$; proved by \code{rfl}. \\
\textbf{03 Apr} & \mipcommit{826b5327} & Ch.~2 (Registers) & Tensor registers & Added explicit left/right tensor placement, while some of Bob's operators were still placed with \code{liftLeft}. \\
\textbf{04 Apr} & \mippr{148} & Ch.~2 (Registers) & Register placement & Corrected \code{liftLeft} $\to$ \code{liftRight} mislabeling for Bob ($A \otimes B$). \\
\textbf{05 Apr} & \mippr{210} & Ch.~9 (Pasting) & Empty support & Case-split collision bound ($k \le q$) and empty support ($k > q$) in \code{ldDnoteq}. \\
\textbf{09 Apr} & \mippr{251} & Ch.~7 (SDP) & Formulation fix & Refactored SDP primal from complete measurement to sub-measurement ($\sum T_g \le I$). \\
\textbf{09 Apr} & \mipcommit{bfd9ed09} & Ch.~7 (SDP) & Trivial delta & Solved the reduced SDP using single-outcome delta $T_g = \delta_{g,g_0}I$ and dual $Z=I$. \\
\textbf{09 Apr} & \mipcommit{8ae0cc7e} & Ch.~4 (Rounding) & Unrelated 1D witness & Proved rounding using trivial witness $R=[1]$ on an unrelated 1D space $\mathbb{C}^1$ (\code{carrier~:= PUnit}). \\
\textbf{09 Apr} & \mipcommit{48d1475c} & Ch.~7 (Self-Imp.) & Supplied hypotheses & Supplied four bridge fields: state permutation invariance, helper self-consistency, evaluation data processing, and final fields. \\
\textbf{09 Apr} & \mippr{287} & Ch.~4 (Rounding) & Input-dependent statement & Required the rounding statement to depend on input measurement $A$ and state $\psi$. \\
\textbf{11 Apr} & \mippr{303} & Ch.~4 (SVD) & Assumption bundling & Stored the $Q \approx P$ approximation bound as an unproved field in \code{QXPLayerData}. \\
\textbf{12 Apr} & \mipcommit{f659daae} & Ch.~10 (Assembly) & Assumed conclusion & Required the target conclusion as input field \code{witness} in \code{MainFormalBridgePackage}. \\
\textbf{15 Apr} & \mippr{375}, \mippr{385} & Ch.~2 (Test) & Branch correction & Removed false point-agreement and per-prover goodness claims (\mipissue{360}); restored cross-prover agreement branch and proved symmetrized strategy's self-consistency equal to it. \\
\textbf{15 Apr} & \mippr{407} & Ch.~1 (Precursors) & Conditional statement &
  Merged a classical soundness statement containing \code{sorry};
  \mipissue{408} flagged the defect and \commit{c530eb2} reverted it the
  same day. \\
\textbf{16 Apr} & \mippr{416} & Ch.~1 (Precursors) & Conditional statement & Replaced \code{sorryAx} classical statements with conditional versions and added the axiom audit. \\
\textbf{18 Apr} & \mippr{473}, \mippr{526} & Ch.~4 (Rounding) & Caller-supplied data & Converted \code{RankReductionBridgePackage} into caller-supplied parameters. \\
\textbf{19 Apr} & \mipcommit{de9a974e} & Ch.~6 (Variance) & Fourier coefficient & Replaced \code{default} with a witness storing the constant Fourier coefficient; orthogonal slots remained zero. \\
\textbf{20 Apr} & \mippr{542} & Ch.~6 (Variance) & Trace assembly & Used the decomposition to prove the global-variance trace identity. \\
\textbf{21 Apr} & \mippr{495}, \mippr{561} & Ch.~9 (Pasting) & Sentinel removal & Deleted \code{extractSliceOr0}; introduced \code{InterpolationSupportWitness}. \\
\textbf{22 Apr} & \mippr{576} & Ch.~5 (Expansion) & Character sums & Proved Fourier orthonormality from additive character sums via Mathlib. \\
\textbf{22 Apr} & \mippr{552} & Ch.~10 (Induction) & Coefficient repair & Proved pasting error factorization $\nu_{\mathrm{paste}} \le \frac{1}{5}\nu_{\mathrm{ind}}$, resolving $m=1$ arithmetic gap. \\
\textbf{25 Apr} & \mipcommit{f553d9a0}, \mipcommit{1d353279} & Ch.~4 (SVD) & SVD field removal & Removed explicit SVD fields from \code{QXPLayerData} and proved $P \approx Q$ from primitive fields. \\
\textbf{25 Apr} & \mippr{726} & Ch.~4 (Rounding) & Projector range ONB & Constructed the auxiliary space from projector ranges \code{IsProj.rangeONB}. \\
\textbf{27 Apr} & \mipcommit{32377678} & Ch.~2 (Registers) & Direct-sum helpers & Added \code{BiProjStrat} direct-sum symmetrization helpers on the role-tagged carrier. \\
\textbf{27 Apr} & \mipcommit{15df91dd} & Ch.~9 (Pasting) & Quadratic error & Corrected \code{lem:from-H-to-G} telescoping sum error from linear $46km$ to quadratic $46k^2m$. \\
\textbf{29 Apr} & \mipissue{904}, \mippr{909} & Ch.~10 (Induction) & Cascade absorption & Absorbed $+2\zeta_1$ completion term into cascade parameter $\zeta_2$ (coeff $42$). \\
\textbf{01 May} & \mippr{958}, \mipcommit{c3141973} & Ch.~2 (Registers) & Independent carriers & Generalized \code{ProjStrat} to separate Alice and Bob carrier types. \\
\textbf{01 May} & \mipcommit{a3fcfcb} & Ch.~5 (Expansion) & Combinatorial sum & Proved Laplacian edge identity from entrywise indicator symmetries and bijections. \\
\textbf{01 May} & \mipcommit{9343ff27}, \mipissue{933} & Ch.~3 (Prelim) & State normalization & Added explicit \code{QuantumState.IsNormalized} hypothesis required by the paper's expectation bounds. \\
\textbf{03 May} & \mippr{1126} & Ch.~4 (Rounding) & Zero-error case & Formulated 3-branch spectral cutoff with exact $1$-eigenspace projection at $\zeta=0$. \\
\textbf{04 May} & \mippr{1190}, \mipissue{1099}& Ch.~4 (Line-169) & Pre-completion & Proved pre-completion transport $\zeta_1 + 10\zeta_1^{1/8}$ via match-mass monotonicity. \\
\textbf{04 May} & \mippr{1210}, \mipcommit{da3e3fe} & Ch.~4 (SVD) & Rectangular identities & Proved rectangular-SVD identities under supplied factors and their algebraic hypotheses. \\
\textbf{05 May} & \mipcommit{773228dc} & Ch.~4 (SVD) & Positive-Gram extension & Constructed coisometric factor from positive Gram matrix under required cardinality bound. \\
\textbf{05 May} & \mipissue{1093}, \mippr{1239}, \mipcommit{dc2faed5}& Ch.~7 (SDP) & Total overlap & Proved total-overlap displacement $\eta \le \sqrt{|\mathbb{F}_q|\varepsilon_{\mathrm{DP}}}$ after setup in \mippr{1222}. \\
\textbf{06 May} & \mipcommit{3685be1c} & Ch.~7 (SDP) & False dominance hypothesis & Introduced block SDP, but assumed saturation via the unproved hypothesis $I \preceq Z$. \\
\textbf{11 May} & \mippr{1462} & Ch.~7 (Self-Imp.) & Conditional theorem & Isolated \code{selfImprovement\_assumingFinalFields}; restored the theorem statement. \\
\textbf{11 May} & \mipissue{1456}, \mippr{1466}, \mippr{1496} &
  Ch.~6 (Variance) & Statement restoration &
  Removed three supplied variance-bound hypotheses from
  \code{globalVarianceOfPoints}; restored paper statement and proved it via
  local transport. \\
\textbf{13 May} & \mippr{1539} & Ch.~10 (Induction) & Conditional API removal &
  Removed \code{mainInductionPublicWrapper} and conditional induction APIs. \\
\textbf{15 May} & \mippr{1632} & Ch.~4 (Ortho) & Sharp constant &
  Proved sharp paper constant $100\zeta^{1/4}$ (via $2\zeta$ scale), avoiding
  the doubled $4\zeta$ error. \\
\textbf{16 May} & \mippr{1638} & Ch.~7 (Self-Imp.) & Theorem assembly &
  Assembled self-improvement from strategy and consistency inputs using the
  tracked SDP theorem. \\
\textbf{17 May} & \mippr{1643} & Ch.~9 (Pasting) & Degree-zero branch &
  Implemented dedicated $d=0$ constant-polynomial pasting construction. \\
\textbf{18 May} & \mippr{1664}, \mipissue{1645} &
  Ch.~10 (Induction) & Corrected input &
  Restated \code{selfImprovementInInductionSection} to take a polynomial
  measurement $G$ and proved it from the Chapter~7 theorem. \\
\textbf{19 May} & \mippr{1708} & Ch.~7 (SDP) & Duality proof &
  Proved cone separation and slack mass saturation lemma without
  $I \preceq Z$. \\
\textbf{22 May} & \mipcommit{440cb0f8}, \mippr{1782} &
  Ch.~3 (Naimark) & Tensor assembly &
  Proved 4-register two-sided correlation preservation
  \code{naimarkTensorProductCorrelation}. \\
\textbf{23 May} & \mipcommit{78d56aa} & Ch.~10 (Induction) & Factor-400 range &
  Confirmed $400md \le k$ for the revised induction statement; the restricted
  same-space theorem already required it (\mippr{912}). \\
\textbf{23 May} & \mipcommit{27df275} &
  Ch.~10 (Soundness) & Positive sample count &
  Added $0 < k$ to the reopened two-space statement to exclude the zero error
  at $d=k=0$. \\
\textbf{10 Jun} & \mipcommit{28d5417d} &
  Ch.~2 (Registers) & Same-space removal &
  Removed legacy same-space strategy definitions; renamed two-space theorem to
  \code{Test.mainFormal}. \\
\textbf{19 Jun} & \mippr{2348} & Ch.~3 (Prelim) & Heterogeneous lemma &
  Proved heterogeneous consistency lemma \code{consSubMeas\_heterogeneous} on
  distinct carriers. \\
\textbf{20 Jun} & \mipcommit{3e09f41} & Ch.~7 (SDP) & Dominance removal &
  Removed all \code{*WithDominance} declarations; used canonical slack
  saturation throughout. \\
\bottomrule
\end{longtblr}

\FloatBarrier

\subsection{Ten-chapter formalization milestones}
\label{sec:chapter-ledger}

The formalization is organized into ten blueprint chapters that divide into the
protocol setup (Chapters~1--2), the single-step analytical core
(Chapters~3--7), and the dimension induction (Chapters~8--10).
Composing declarations across chapter boundaries exposed register placement
mismatches, uncoupled witnesses, bundled bridge hypotheses, and omitted side
conditions that were invisible when individual modules were compiled in
isolation.

Chapters~1 and~2 specify the conditional classical Polishchuk--Spielman
soundness interface (\mippr{416}, \mipissue{408}) and formalize the bipartite
tensor-product register model with independent prover carriers $\iota_A$ and
$\iota_B$ (\mippr{148}, \mippr{958}, \mipissue{560}).
Chapters~3 through~7 then execute the five analytical stages of a single
inductive step, where each stage's output interface must match the next stage's
hypotheses:
\begin{itemize}
  \item \textbf{Chapter~3 (Naimark dilation and preliminaries):} To apply
    projective matrix inequalities to sub-measurement strategies, Chapter~3
    dilates sub-measurements to projective sub-measurements while preserving
    bipartite correlations.
    Early declarations omitted state normalization (\mipissue{933}) and asserted
    complete output measurements on the unextended outcome alphabet
    (\commit{82d70fef}); adjoining a complementary outcome $\bot$ on
    $\operatorname{Option}(\mathcal{A})$ and constructing a four-register
    bipartite dilation resolved both mismatches (\mippr{332}, \mippr{1782}).
  \item \textbf{Chapter~4 (Spectral rounding and orthonormalization):} To
    convert almost-projective consistency into orthogonal projectors, Chapter~4
    rounds measurements by spectral thresholding and orthonormalizes them via
    polar decomposition.
    Replacing the uncoupled \code{PUnit} rounding witness (\commit{8ae0cc7e})
    and bundled \code{QXPLayerData} approximation fields required a three-branch
    continuous functional calculus construction (\mippr{287}, \mippr{1126}), a
    six-commit positive-Gram polar construction chain ending at
    \commit{a00c4474} alongside a parallel unitary-extension commit
    (\commit{773228dc}), and pre-completion match-mass transport (\mippr{1190},
    \mipissue{1099}).
  \item \textbf{Chapter~5 (Hypercube spectral expansion):} To bound how
    disagreement propagates across axis-aligned lines, Chapter~5 establishes the
    spectral gap of the hypercube random-walk Laplacian.
    Initial declarations defined the Fourier inner product as a Kronecker delta
    by \code{rfl} (\commit{de9a974e}) and aliased the difference form as
    $L_{\mathrm{diff}} \coloneqq L$ (\commit{78a14cd4}); integration with
    Chapter~6 replaced both aliases with character-sum orthonormality
    (\mippr{576}) and the combinatorial edge-difference Dirichlet form
    (\mippr{721}, \commit{a3fcfcb}).
  \item \textbf{Chapter~6 (Global variance):} To convert the spectral gap into
    an average-case closeness bound across affine subspaces, Chapter~6 bounds
    the global variance of point measurements.
    Early proofs supplied a zero-operator \code{default} witness for orthogonal
    Fourier components and took three variance bounds as hypotheses
    (\mipissue{1456}); proving the centered residual decomposition $A_\perp^u =
    A^u - A_{\mathrm{avg}}$ and its lifted trace identity eliminated those
    hypotheses (\mippr{542}, \mippr{1466}, \mippr{1496}).
  \item \textbf{Chapter~7 (Self-improvement and semidefinite programming):} To
    consolidate approximately consistent point measurements into a single
    low-degree polynomial sub-measurement, Chapter~7 applies finite-dimensional
    semidefinite programming duality.
    Initial declarations bundled the target conclusions in
    \code{SelfImprovementBridgePackage} (\mipissue{1453}), solved a collapsed
    single-outcome SDP, and assumed the false operator inequality $I \preceq Z$;
    proving cone separation, slack mass saturation, and the total-overlap bound
    for $\eta$ discharged these assumptions (\mippr{1239}, \mippr{1462},
    \mippr{1708}).
\end{itemize}
Finally, Chapters~8 through~10 close the induction across the ambient dimension
$m$: Chapter~8 proves approximate commutativity on successor slices
(\mippr{730}, \mipissue{930}, \mipissue{713}, \commit{6824be86}), Chapter~9
pastes axis-aligned polynomial measurements across $k$ random evaluation points
via support-certified Lagrange interpolation (\mippr{210}, \mippr{495},
\mippr{1643}), and Chapter~10 combines the successor error recurrence with the
classical soundness premise under the corrected side conditions $400md \le k$
and $0 < k$ (\mippr{552}, \mippr{909}, \mippr{1664}, \mipissue{1507},
\mipissue{422}, \mipissue{1645}).

\cref{tab:chapter-milestone-ledger} summarizes the ten chapter milestones.
For each chapter, the table records the initial declaration, the integration or
audit defect identified by downstream compilation failures, automated CI
scanners, or maintainer review, the repaired declaration, the corresponding
repository pull requests or issues, and the mathematical discussion in
\crefrange{app:theorem}{app:composition}.

\FloatBarrier

\begin{longtblr}[
  caption = {\textbf{Ten-chapter formalization milestones.}
    Initial declarations, integration defects, repaired declarations,
    and mathematical case studies in
    \crefrange{app:theorem}{app:composition}.},
  label = {tab:chapter-milestone-ledger}
]{colspec = {X[13,l] X[21,l] X[20.5,l] X[24,l] X[11.5,l] X[10,l]},
  row{odd} = {bg=gray!5},
  row{1}   = {bg=gray!25, font=\bfseries},
  rowsep = 3pt,
  colsep = 2pt,
  rowhead = 1}
    \toprule
    Chapter \& topic & Initial statement
      & Problem found & Final statement
      & PR / issue & Math ref \\
    \midrule
    \textbf{Ch.~1: Classical soundness}\newline
      (15 Apr--23 May)
      & Quoted classical soundness directly as a Lean theorem using placeholder \code{sorry} (\commit{187c2c4a}).
      & \mipissue{408} identified \code{sorryAx} among the theorem's transitive axioms; PR reverted same day (\commit{c530eb23}).
      & Stated as conditional premise
        \texttt{PolishchukSpielman\allowbreak ClassicalSoundness\allowbreak
        Statement}; quantum
        soundness verified axiom-clean.
      & \mippr{416}, \mipissue{408}
      & App.~A \S\ref{sec:theorem-scope} \\
    \textbf{Ch.~2: Quantum registers}\newline
      (3 Apr--1 May)
      & Single local carrier $\iota$ with both operators multiplied on the left register $(A \cdot B) \otimes I$ (\commit{a0bec1d4}).
      & Audit revealed both provers acting on the left register; Bob's operator lifts set to \code{liftLeft} (\mippr{148}).
      & Independent carriers $\iota_A$, $\iota_B$ with explicit $A \otimes I$ vs $I \otimes B$; direct-sum role carrier $\operatorname{Role} \times (\iota_A \oplus \iota_B)$.
      & \mippr{148}, \mippr{958}, \mipissue{560}
      & App.~C Case~I (\S\ref{pf:case-mainformal}) \\
    \textbf{Ch.~3: Naimark dilation}\newline
      (30 Mar--22 May)
      & Complete output asserted for sub-measurement input on original alphabet $\mathcal{A}$ (\commit{82d70fef}); single-measurement local dilation.
      & \mipissue{933} (unnormalized state scaling); \commit{6311ba3f} (questionwise interface lacked joint correlation preservation).
      & Projective sub-measurements on $\mathcal{A}$ via $\operatorname{Option}(\mathcal{A})$ completion; 4-register bipartite dilation space.
      & \mippr{332}, \mippr{1782}, \mipissue{933}
      & App.~B \S\ref{subsec:naimark-alphabet-repair} \\
    \textbf{Ch.~4: Spectral rounding}\newline
      (9 Apr--3 May)
      & Rounding proved on an unrelated $1$D space \code{carrier~:= PUnit} (\commit{8ae0cc7e}); $P \approx Q$ bound bundled in \code{QXPLayerData}.
      & \mippr{287} required rounding to depend on input $A$; missing rectangular SVD in Mathlib; Line-169 transport incurred $\sqrt{\cdot}$ loss (\mipissue{1099}).
      & Rounding by continuous functional calculus in 3 cases (\mippr{1126});
        positive-Gram polar extension (\commit{a00c4474}, \commit{773228dc});
        match-mass Line-169 transport.
      & \mippr{287}, \mippr{1126}, \mippr{1190}
      & App.~B \S\ref{subsec:rounding-trajectory}; App.~C Case~IV.a (\S\ref{subsec:line169-transport-details})  \\
    \textbf{Ch.~5: Hypercube graph}\newline
      (24 Mar--22 Apr)
      & Fourier inner product defined as Kronecker delta via \code{rfl}
        (\commit{de9a974e}); Laplacian difference form defined as
        $L_{\mathrm{diff}} \coloneqq L$ (\commit{78a14cd4}).
      & \mippr{721} review found that the definition did not express the random-edge Dirichlet form.
      & Fourier orthonormality proved from character sums (\mippr{576}); combinatorial Laplacian proved entrywise from edge symmetries.
      & \mippr{576}, \mippr{721}, \commit{a3fcfcb}
      & App.~C Case~V (\S\ref{pf:case-expansion-variance}) \\
    \textbf{Ch.~6: Global variance}\newline
      (24 Mar--22 Apr)
      & Variance decomposition used \code{default} zero-operator witness;
        \path{globalVarianceOfPoints} took 3 supplied hypotheses
        (\mipissue{1456}).
      & Fields for orthogonal Fourier data set to zero; \mipissue{1456} found that hypotheses assumed the required conclusions.
      & Centered residual family $A_\perp^u = A^u - A_{\mathrm{avg}}$ satisfying $\sum_u A_\perp^u = 0$; lifted trace identity with prefactor $1/M$.
      & \mippr{542}, \mippr{1466}, \mippr{1496}
      & App.~C Case~V (\S\ref{pf:case-expansion-variance}) \\
    \textbf{Ch.~7: Self-imp. \& SDP}\newline
      (9 Apr--19 May)
      & \texttt{SelfImprovement\allowbreak BridgePackage} assumed
        the conclusions as fields; SDP
        solved with 1-outcome delta witness; saturation assumed $I \preceq Z$.
      & \mipissue{1453} found that the proof returned a supplied conclusion; $I \preceq Z$ proved false in general; substitution of a sub-measurement required control of total overlap $\eta$.
      & Finite-dimensional canonical strong duality; slack saturation from weak-duality inequalities; total-overlap bound $\eta \le \widehat\zeta+2\sqrt{\widehat\zeta_{\mathrm{ortho}}}$ from completeness transfer.
      & \mippr{1239}, \mippr{1462}, \mippr{1708}
      & App.~C Cases~II, III, IV.b (\S\ref{pf:case-self-improvement}--\ref{subsec:total-overlap-eta}) \\
    \textbf{Ch.~8: Commutativity}\newline
      (3 Apr--1 May)
      & Theorem header typed at $m$ while point consistency evaluated on $\mathbb{F}_q^{m+1}$ slices; scalar-to-tensor transport assumed no loss.
      & \mipissue{930} identified dimension mismatch on transverse slices; \mipissue{713} exposed $2\sqrt{\zeta}$ tensor transport loss.
      & Strategy stated at successor dimension $m+1$ via \code{params.next}; full-slice scalar-to-tensor bridges proved with $\sqrt{\zeta}$ loss.
      & \mippr{730}, \mipissue{930}, \commit{6824be86}
      & Gap notes (\S\ref{sec:gap-audit}) \\
    \textbf{Ch.~9: Pasting}\newline
      (5 Apr--29 Apr)
      & The function \code{extractSliceOr0} returned $0$ on empty slices; linear error $46km$ in telescoping sum; missing $d=0$ constant branch.
      & The default zero value invalidated point consistency on unmeasured slices; summing the recurrence with $\sqrt{\nu_4} \propto k$ gave quadratic $k^2$ loss.
      & Support-certified interpolation via
        \texttt{Interpolation\allowbreak SupportWitness}; quadratic error
        $46k^2m$;
        dedicated $d=0$ pasting construction.
      & \mippr{210}, \mippr{495}, \mippr{1643}
      & App.~C Cases~VI, VII (\S\ref{pf:case-pasting}, \S\ref{subsec:bernoulli-telescoping}) \\
    \textbf{Ch.~10: Soundness induction}\newline
      (12 Apr--23 May)
      & Printed condition $k \ge md$ admitted $md \le k < 400md$ and $k=0$; successor error estimate $(1.01)(4)=4.04 > 4$ failed at $m=1$.
      & \mipissue{1507} exposed factor-400 pasting gap; \mipissue{422} found that the error vanishes at $k=0$; \mipissue{1645} identified a mismatched input.
      & Tightened side conditions $400md \le k$ and $0 < k$; successor error bound $\nu_{\mathrm{paste}} \le \frac{1}{5}\nu_{\mathrm{ind}}$ repairing $m=1$ to $2.42 \le 4$.
      & \mippr{552}, \mippr{909}, \mippr{1664}
      & App.~C \S\ref{subsec:successor-m1-gap},
        \S\ref{subsec:bernoulli-telescoping},
        \S\ref{subsec:pasting-degree-zero} \\
    \bottomrule
\end{longtblr}

\FloatBarrier

\subsection{The blueprint as a measure of progress}
\label{sec:blueprint-progress}

The project blueprint, built with Massot's \texttt{leanblueprint}
tool~\citep{Massot2021Leanblueprint}, tracks the dependency graph between target
statements.
Each target declaration is marked \emph{not ready} while its formulation or
dependencies remain unsettled, \emph{stated} once a Lean declaration exists,
and \emph{fully formalized} once both its statement and proof are verified (the
\code{leanok} marker).
Panels~b and~c of \cref{fig:metrics-evolution} plot these daily status counts
alongside the \code{sorry} placeholder count in the Lean source files.
All counts are evaluated across the 1,793 first-parent commits reachable at the
pinned snapshot, which form the main-branch integration sequence within the
repository's \NumCommits{} total commits.

The placeholder count and the blueprint status counts measure distinct
verification criteria: eliminating \code{sorry} tokens from a Lean module
ensures that the kernel typechecks its declared propositions, whereas blueprint
readiness tracks whether those declarations faithfully express the paper's
lemmas and compose across the dependency graph.
The declaration counts in Panels~b and~c track unique Lean declarations linked
via the blueprint's \code{\textbackslash lean} tags (often several declarations
per statement environment, with remarks excluded); at the pinned snapshot, the
main text's 181 blueprint statement environments link to 598 Lean declarations.
On 29~April 2026, the repository \code{sorry} count dropped to one (the main
theorem itself) and remained at one for 335 first-parent commits; on that date,
the blueprint listed 283 target declarations, of which 114 were not yet fully
formalized (74 \emph{not ready} and 40 \emph{stated}).
While that single \code{sorry} persisted, the blueprint continued to expand as
missing intermediate lemmas were added, reaching 672 target declarations with
292 not yet fully formalized on 8~May.
The blueprint trajectories were also non-monotone: on 20~May the count of
declarations not yet fully formalized rose from 5 to 26, and on 22--23~May the
total target count dropped from 668 to 612 as previously verified construction
nodes were audited against the paper and consolidated.
When the final \code{sorry} was replaced by a verified proof on 23~May
(\commit{27df275}), 23 target declarations remained not yet fully formalized;
that count reached zero later that day, stabilizing at 566 fully formalized
target declarations through 9~June 2026.

This convergence was driven by the May 2026 statement audit and governance
protocol (\cref{sec:formalflow}).
On 29~April, the first gap notes documented discrepancies between the formal
declarations and the paper, and \mippr{921} merged a standardized gap-note
policy and template the same day.
On 8~May, \mippr{1374} rejected a pull request that would have eliminated the
last \code{sorry} by adding two unproved auxiliary hypotheses to the main
theorem, enforcing the rule that an explicit \code{sorry} placeholder is
preferable to an unfaithful hypothesis.
On 9 and 10~May, one of the authors of the LIDT paper audited nine Lean
statements from the main theorem downward against the paper and opened a
tracking issue for each discrepancy.
Starting on 11~May, standing repository instructions required agents to restore
the paper's exact statement even when doing so reintroduced a tracked
\code{sorry}, and an automated proof-debt scanner enforced this requirement in
continuous integration.
The final placeholder closed on 23~May when commit \commit{27df275} added the
positivity side condition $0<k$ identified in gap note 422.
The chapter-level statement revisions triggered by this audit are summarized in
\cref{tab:chapter-milestone-ledger}.

\subsection{The 25 gap notes}
\label{sec:gap-audit}

When an agent or reviewer encountered a suspected discrepancy between the formal
development and the paper, the repository protocol required filing a structured
gap note under \code{docs/paper-gaps/} rather than weakening a Lean statement
to force compilation.
Each gap note isolated the mathematical step in question, stated a minimal
falsifiable form or counterexample, and awaited adjudication.
Adjudication separated unfaithful agent shortcuts (formal-side interface
mismatches, bundled hypotheses, or simplified definitions that had to be
repaired in Lean without changing the paper) from genuine mathematical
corrections to the paper (omitted side conditions, complete-dilation
impossibility on an unextended outcome alphabet, or quantitative error-bound
losses that required revising a printed hypothesis or intermediate bound).
At snapshot \PinnedHead (\PinnedDate), the repository contains \NumGapNotes{}
gap notes, all of which are resolved and none of which remains an open
dependency of \code{mainFormal}.

The \NumGapNotes{} gap notes fall into eight categories according to the
adjudicated outcome; because two tracking issues span multiple notes
(\mipissue{930} spans five and \mipissue{1099} spans two), each note is cited by
its issue identifier and, where needed, a short descriptive slug:
\begin{enumerate}[label=(\roman*)]
  \item \textbf{Corrected side conditions (2 notes):} Tightened or added
    hypotheses of the paper's main theorem (issues 906 and 422).
  \item \textbf{Repaired formal-interface mismatches (3 notes):}
    Mismatches between the paper's hypotheses and early formal statements or
    proofs, repaired on the formal side while leaving the paper's statements
    unchanged (issues 196, 1230, and 930/self-improvement $\nu$).
  \item \textbf{Clarified wording and statement scope (5 notes):}
    Impossibility of complete original-alphabet dilation for sub-measurements, a
    harmless ambient-dimension typo, state normalization conventions, removal of
    an unnecessary nonnegativity assumption, and retargeted blueprint links
    (issues naimark, 930/dimension typo, 933, 938, 2338).
  \item \textbf{Corrected quantitative error bounds (5 notes):}
    Substitution losses, telescoping factors, and the total-overlap
    displacement
    (issues 1099/line-169 loss, 1099/sharper fix, 1093,
    930/successor coefficient, 930/pasting error).
  \item \textbf{Endpoint branches (3 notes):} Explicit case splits for empty
    distinct tuples ($k > q$), exact zero-error spectral cutoffs
    ($\zeta = 0$), and degree-zero pasting ($d = 0$)
    (issues 930/distinct tuples, 1100, 1622).
  \item \textbf{Adjusted constants (1 note):} An increased scalar
    cascade coefficient absorbing the completion residual (issue 904).
  \item \textbf{Design adjustments (1 note):} The blueprint polynomial
    definition recast from a function subtype to the paper's representative
    predicate (polynomial-divergence).
  \item \textbf{Verified design decisions (5 notes):} Audits of formalization
    choices that confirmed exact agreement with the paper, including a proof
    recovering the paper's sharp orthonormalization constant
    (issues 458, 713, 760, 1228, 1032).
\end{enumerate}

\cref{tab:gaps-sup-census} indexes all 25 gap notes, one row per file, by note
identifier, blueprint chapter, proof stage, category, severity, scope,
adjudicated resolution, and case-study cross-reference in
\cref{app:composition}.
Severity records the authors' assessment of the discrepancy's impact on the
paper's statement, independent of the implementation difficulty recorded inside
individual notes.
Scope records whether the discrepancy affected the paper's printed statement,
the formal Lean statement alone, or an internal proof step.

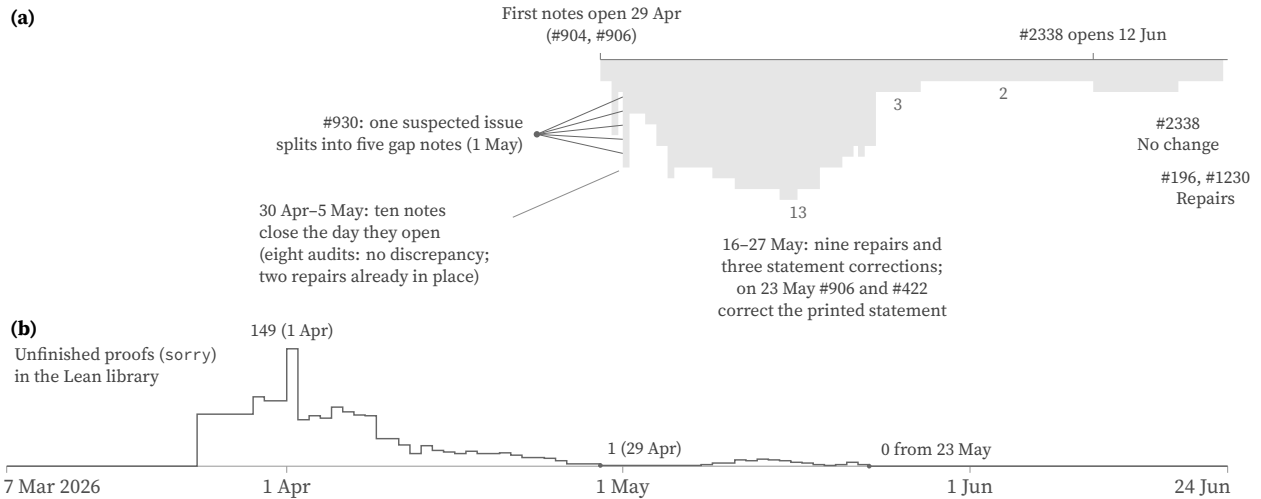
\begin{figure}
  \centering
  \resizebox{\linewidth}{!}{%
  \begin{tikzpicture}[
      >={Stealth[length=2.8pt]},
      every node/.style={font=\fontsize{6}{7.2}\selectfont, text=black!75,
                         inner sep=1pt},
      lbl/.style={font=\fontsize{6}{7.2}\selectfont, text=black!75,
                  align=left, inner sep=1pt},
      datelbl/.style={font=\scriptsize, text=black!70, inner sep=1.5pt},
      panel/.style={font=\bfseries\fontsize{7}{8.4}\selectfont,
                    text=black, inner sep=1pt},
    ]

    \node[panel, anchor=north west] at (0,0.65) {(a)};
    \node[panel, anchor=north west] at (0,-3.1) {(b)};

    \fill[black!10]
      (7.155,0) --
      (14.661,0) --
      (14.661,-0.26) --
      (14.121,-0.26) --
      (14.121,-0.39) --
      (13.095,-0.39) --
      (13.095,-0.26) --
      (11.016,-0.26) --
      (11.016,-0.39) --
      (10.476,-0.39) --
      (10.476,-1.04) --
      (10.341,-1.04) --
      (10.341,-1.17) --
      (10.26,-1.17) --
      (10.26,-1.04) --
      (10.206,-1.04) --
      (10.206,-1.17) --
      (10.071,-1.17) --
      (10.071,-1.3) --
      (9.801,-1.3) --
      (9.801,-1.56) --
      (9.531,-1.56) --
      (9.531,-1.69) --
      (9.315,-1.69) --
      (9.315,-1.56) --
      (8.775,-1.56) --
      (8.775,-1.43) --
      (8.505,-1.43) --
      (8.505,-1.3) --
      (8.046,-1.3) --
      (8.046,-1.43) --
      (7.965,-1.43) --
      (7.965,-1.04) --
      (7.83,-1.04) --
      (7.83,-0.78) --
      (7.695,-0.78) --
      (7.695,-0.65) --
      (7.506,-0.65) --
      (7.506,-1.3) --
      (7.425,-1.3) --
      (7.425,-0.39) --
      (7.371,-0.39) --
      (7.371,-0.91) --
      (7.29,-0.91) --
      (7.29,-0.26) --
      (7.155,-0.26) --
      cycle;
    \draw[black!55, line width=0.3pt] (7.155,0) -- (14.715,0);

    \fill[black!60] (6.39,-0.9) circle[radius=1.1pt];
    \foreach \ya in {-0.45,-0.62,-0.79,-0.96,-1.13}
      \draw[black!60, line width=0.3pt] (6.39,-0.9) -- (7.425,\ya);
    \node[lbl, anchor=east, align=right] at (6.27,-0.9)
      {\#930: one suspected issue\\splits into five gap notes (1 May)};

    \draw[black!55, line width=0.3pt] (7.155,0) -- (7.155,0.08);
    \node[lbl, anchor=south, align=center] at (7.05,0.14)
      {First notes open 29 Apr\\(\#904, \#906)};
    \draw[black!55, line width=0.3pt] (13.095,0) -- (13.095,0.08);
    \node[lbl, anchor=south] at (13.095,0.14) {\#2338 opens 12 Jun};
    \node[font=\fontsize{6}{7.2}\selectfont, text=black!60]
      at (9.54,-1.84) {13};
    \node[font=\fontsize{6}{7.2}\selectfont, text=black!60]
      at (10.74,-0.54) {3};
    \node[font=\fontsize{6}{7.2}\selectfont, text=black!60]
      at (12.015,-0.41) {2};
    \node[lbl, anchor=north west] at (3.0,-1.7)
      {30 Apr--5 May: ten notes\\
       close the day they open\\
       (eight audits: no discrepancy;\\
       two repairs already in place)};
    \draw[black!45, line width=0.3pt] (6.1,-1.9) -- (7.38,-1.34);
    \node[lbl, anchor=north, align=center] at (9.95,-2.1)
      {16--27 May: nine repairs and\\
       three statement corrections;\\
       on 23 May \#906 and \#422\\
       correct the printed statement};
    \node[lbl, anchor=north, align=center] at (14.12,-0.65)
      {\#2338\\No change};
    \node[lbl, anchor=north, align=center] at (14.45,-1.3)
      {\#196, \#1230\\Repairs};

    \draw[black!40, line width=0.35pt] (0,-4.9) -- (14.715,-4.9);
    \foreach \xt in {0,3.375,7.425,11.61,14.715}
      \draw[black!40, line width=0.35pt] (\xt,-4.9) -- (\xt,-4.98);
    \node[datelbl, anchor=north west] at (-0.1,-5.0) {7 Mar 2026};
    \node[datelbl, anchor=north] at (3.375,-5.0) {1 Apr};
    \node[datelbl, anchor=north] at (7.425,-5.0) {1 May};
    \node[datelbl, anchor=north] at (11.61,-5.0) {1 Jun};
    \node[datelbl, anchor=north east] at (14.815,-5.0) {24 Jun};
    \draw[black!60, line width=0.5pt]
      (0,-4.9) -- (2.295,-4.9) -- (2.295,-4.273) -- (2.97,-4.273) --
      (2.97,-4.064) -- (3.105,-4.064) -- (3.105,-4.112) -- (3.375,-4.112) --
      (3.375,-3.485) -- (3.51,-3.485) -- (3.51,-4.34) -- (3.645,-4.34) --
      (3.645,-4.292) -- (3.78,-4.292) -- (3.78,-4.321) -- (3.915,-4.321) --
      (3.915,-4.188) -- (4.05,-4.188) -- (4.05,-4.245) -- (4.185,-4.245) --
      (4.185,-4.283) -- (4.32,-4.283) -- (4.32,-4.292) -- (4.455,-4.292) --
      (4.455,-4.568) -- (4.725,-4.568) -- (4.725,-4.644) -- (4.86,-4.644) --
      (4.86,-4.748) -- (4.995,-4.748) -- (4.995,-4.653) -- (5.13,-4.653) --
      (5.13,-4.71) -- (5.265,-4.71) -- (5.265,-4.729) -- (5.4,-4.729) --
      (5.4,-4.748) -- (5.535,-4.748) -- (5.535,-4.72) -- (5.67,-4.72) --
      (5.67,-4.748) -- (5.94,-4.748) -- (5.94,-4.739) -- (6.075,-4.739) --
      (6.075,-4.767) -- (6.21,-4.767) -- (6.21,-4.796) -- (6.48,-4.796) --
      (6.48,-4.805) -- (6.615,-4.805) -- (6.615,-4.834) -- (6.75,-4.834) --
      (6.75,-4.872) -- (7.155,-4.872) -- (7.155,-4.891) -- (8.37,-4.891) --
      (8.37,-4.881) -- (8.505,-4.881) -- (8.505,-4.862) -- (8.775,-4.862) --
      (8.775,-4.824) -- (8.91,-4.824) -- (8.91,-4.834) -- (9.045,-4.834) --
      (9.045,-4.815) -- (9.18,-4.815) -- (9.18,-4.824) -- (9.315,-4.824) --
      (9.315,-4.834) -- (9.45,-4.834) -- (9.45,-4.853) -- (9.585,-4.853) --
      (9.585,-4.862) -- (9.72,-4.862) -- (9.72,-4.881) -- (9.855,-4.881) --
      (9.855,-4.891) -- (9.99,-4.891) -- (9.99,-4.881) -- (10.125,-4.881) --
      (10.125,-4.853) -- (10.26,-4.853) -- (10.26,-4.881) --
      (10.395,-4.881) -- (10.395,-4.9) -- (14.715,-4.9);
    \node[lbl, anchor=west] at (0.05,-3.7)
      {Unfinished proofs (\texttt{sorry})\\in the Lean library};
    \node[lbl, anchor=south] at (3.44,-3.42) {149 (1 Apr)};
    \fill[black!60] (7.155,-4.891) circle[radius=0.8pt];
    \node[lbl, anchor=south west] at (7.2,-4.85) {1 (29 Apr)};
    \fill[black!60] (10.395,-4.9) circle[radius=0.8pt];
    \node[lbl, anchor=south west] at (10.495,-4.86) {0 from 23 May};
  \end{tikzpicture}}
  \caption{\textbf{Life cycle of proof-gap notes.}
    \textbf{\textsf{(a)}}~The number of gap notes open on each date, drawn as a
    band whose width counts the open items (one strand each); decreases in
    width mark notes closing, with their outcomes labelled directly.
    Twenty-five gap notes opened between 29 April and 12 June 2026, and all
    were closed by 23 June: nine after an audit found no discrepancy,
    thirteen by repairs to the formalization, and three by corrections to
    formal theorem statements (\mipissue{906}, \mipissue{422}, and the
    Naimark note).
    A note counts as closed when the decision it called for was executed
    in the repository.
    Ten of the notes closed the day they opened.
    The retrospective audit \mipissue{930} entered as one suspected
    issue and split into five notes.
    \textbf{\textsf{(b)}}~On the same time axis: the number of unfinished proofs
    in the library (daily closing values), which had fallen to one by
    29 April, when the first notes opened, and reached zero on
    23 May, the day five notes closed.
    Dates and outcomes are taken from the gap-note files and their commit
    history; the comments in this figure's file list the dates and outcomes
    for each entry.}
  \label{fig:gap-lifecycle}
\end{figure}

\cref{fig:gap-lifecycle} plots the opening and closing dates of the 25 gap
notes, which together track 20 distinct suspected discrepancies.
Each note was opened when a discrepancy was flagged and closed by the commit
that executed the adjudicated decision in the repository: correcting the
statement, repairing the formal proof, or recording that the paper's step held
without modification.
In five instances, the summary label in \cref{tab:gaps-sup-census} and the
action category in \cref{fig:gap-lifecycle} emphasize complementary aspects of
the resolution because the table summarizes the final mathematical status while
the figure classifies the commit action.
Detailed derivations of the quantitative repairs, including the Line-169
transport bound ($\zeta_1 + 10\zeta_1^{1/8}$), the total-overlap displacement
($\eta \le \widehat\zeta+2\sqrt{\widehat\zeta_{\mathrm{ortho}}}$), and the
Bernoulli telescoping bound ($46k^2m$), appear in \cref{app:composition}.

\FloatBarrier

\paragraph{The total-overlap note and the completed proof}
\label{sec:gap-exhibit}

To illustrate this lifecycle on a quantitative proof repair, the total-overlap
gap note (\textbf{issue-1093}) was filed on 5~May 2026 (\commit{e42bf231}) when
substituting a sub-measurement into the point-consistency triangle inequality
left an unabsorbed total-overlap displacement $\eta$, and the repaired proof
discharged that displacement via two-way completeness transfer on 16~May
(\commit{134d08f2}).
We present the statement-level derivation of this bound in
\cref{subsec:total-overlap-eta}.

\FloatBarrier

\begin{longtblr}[
  caption = {\textbf{The 25 gap notes.}
    Gap notes indexed by identifier, chapter, stage,
    category, severity, scope, mathematical resolution, and case-study
    reference (--- where no case study discusses the note).
    Rows are the gap notes at the pinned
    snapshot, one per note, ordered by the date each note opened;
    determinations follow the decision executed in the repository.},
  label = {tab:gaps-sup-census}
]{colspec = {X[12.4,l] X[6.5,l] X[12.2,l] X[9.6,l] X[8.6,l] X[9.7,l] X[31.4,l]
    X[9.6,l]},
  row{odd} = {bg=gray!5},
  row{1} = {bg=gray!25, font=\bfseries},
  rowsep = 3pt, colsep = 2pt, rowhead = 1}
    \toprule
    Note ID & Ch. & Stage & Category
      & Severity & Scope & Determination \& resolution & Case ref \\
    \midrule
    \textbf{issue-904}
      & Ch.~10 & Cascade & Constant & Low & Proof
      & Repaired: widened Step~6 cascade coefficient from 40 to 42,
        absorbing the completion proposition's omitted $+2\zeta_1$
        residual.
      & Case~IV.a \\
    \textbf{issue-906}
      & Ch.~10 & Main soundness & Side condition & High & Statement
      & Corrected: printed $k \ge md$ strengthened to $k \ge 400md$,
        the sampling hypothesis required by the successor pasting
        theorem; the excluded interval shown to be nonempty.
      & Case~VII \\
    \textbf{issue-196}
      & Ch.~7 & SDP primal & Interface & Medium & Formal statement
      & Repaired: primal feasible set restored to sub-measurements
        $\sum_g T_g \preceq I$; the equality encoding excluded the
        paper's strict Slater witness.
      & Case~III \\
    \textbf{issue-458}
      & Ch.~3 & Foundation layer & Audit & Low & Statement
      & Audited: prime-power encoding of $q$, finite-support
        distributions, and the zero measurement family; no hypothesis or
        conclusion altered; no discrepancy.
      & --- \\
    \textbf{issue-713}
      & Ch.~8 & Scalar transport & Audit & Low & Proof
      & Absorbed: converting the hybrid scalar estimate adds $2\sqrt{\zeta}$ in
        the tensor \code{closenessOfIP} form; the existing
        error bound includes it; no discrepancy.
      & --- \\
    \textbf{issue-760}
      & Ch.~8 & Scalar chain & Audit & Low & Proof
      & Audited: ten-step scalar approximation chain and the exact
        BAB--ABA swap identity; the formal total matches the paper's
        $\nu = 48m(\sqrt{\gamma}+\sqrt{\zeta})$; no discrepancy.
      & --- \\
    \textbf{polynomial-divergence}
      & Ch.~3 & Polynomial defn. & Design & Low & Statement
      & Repaired: blueprint polynomial set recast from a function-type
        subtype to the paper's representative-of-degree-$\le d$
        predicate, enabling Schwartz--Zippel.
      & --- \\
    \textbf{issue-930} (dimension typo)
      & Ch.~8 & Commutativity & Wording & Low & Statement
      & Clarified: printed $(m,q,d)$ is a harmless typo; the formal
        statement types the ambient strategy at $(m+1,q,d)$ via
        \code{params.next}; no Lean change needed.
      & --- \\
    \textbf{issue-930} (distinct tuples)
      & Ch.~9 & Distinct tuples & Endpoint & Low & Statement
      & Repaired: distinct-tuple weights defined for
        all $k$, normalized only when $k \le q$; the $k > q$ case
        proved with the trivial $k^2/q$ bound.
      & --- \\
    \textbf{issue-930} (successor coeff.)
      & Ch.~10 & Successor step & Error bound & Medium & Proof
      & Repaired: printed absorption
        $(1+\tfrac{1}{100m})(m^2+3)\le(m+1)^2$ fails at $m=1$; the
        sharper $\nu_{\mathrm{paste}}\le\nu_{\mathrm{ind}}/5$ absorbs
        both coefficients for all $m\ge1$.
      & Case~VII \\
    \textbf{issue-930} (pasting error)
      & Ch.~9 & Bernoulli pasting & Error bound & Medium & Statement
      & Corrected: telescoping the printed stage losses gives
        $\nu_8 = 46k^2m$, not the printed $46km$; the corrected
        quadratic error is absorbed by $\nu = 100k^2m$.
      & Case~VII \\
    \textbf{issue-930} (self-improvement $\nu$)
      & Ch.~7 & Self-improvement & Interface & High & Formal statement
      & Repaired: restored the paper's input consistency hypothesis
        $A^{\mathbf u}_a\otimes I \simeq_\nu I\otimes
        G_{[g(\mathbf u)=a]}$ to \code{selfImprovement}; removed
        \code{SelfImprovement\allowbreak Obligations}.
      & Case~II \\
    \textbf{issue-933}
      & Ch.~3 & Preliminaries & Wording & Low & Proof
      & Audited: paper's unit-vector convention versus the positive-cone
        \code{QuantumState}; \code{IsNormalized} assumed in each lemma
        rather than in the state definition; no theorem-level defect.
      & --- \\
    \textbf{issue-938} (truncation)
      & Ch.~4 & Truncation & Wording & Low & Proof
      & Audited: abstract averaging lemma proved for arbitrary real $f$,
        dropping the paper's implicit $f \ge 0$; a conservative
        strengthening; no discrepancy.
      & --- \\
    \textbf{issue-1100}
      & Ch.~4 & Spectral rounding & Endpoint & Low & Proof
      & Repaired: unconditional rounding witness from the paper's
        $0<\zeta\le 1/4$ branch, an exact $\zeta=0$ spectral-projector
        branch, and a trivial $\zeta>1/4$ branch.
      & --- \\
    \textbf{issue-1099} (line-169 loss)
      & Ch.~4 & Line-169 transport & Error bound & Medium & Proof
      & Repaired: the printed \path{prop:triangle-sub} use drops
        $\sqrt{\zeta_2}$; the theorem claiming exactly $\zeta_1$ was removed and
        the formalization proves the bound $\zeta_1 + 10\zeta_1^{1/8}$.
      & Case~IV.a \\
    \textbf{issue-1099} (sharper fix)
      & Ch.~4 & Line-169 repair & Error bound & Medium & Proof
      & Repaired: pre-completion match-mass comparison at
        $\epsilon = 100\zeta_1^{1/4}$ takes the square root before
        completion, giving $\zeta_1 + 10\zeta_1^{1/8}$ and preserving
        the final error bound.
      & Case~IV.a \\
    \textbf{issue-1093}
      & Ch.~7 & Point consistency & Error bound & Medium & Proof
      & Repaired: $\eta \le \widehat\zeta +
        2\sqrt{\widehat\zeta_{\mathrm{ortho}}}$ by completeness transfer
        in both directions (\commit{134d08f2}).
      & Case~IV.b \\
    \textbf{issue-1228}
      & Ch.~4 & Orthonormalization & Audit & Low & Proof
      & Audited: no discrepancy; pinned Mathlib lacks the rectangular
        polar theorem, so a project-local positive-Gram construction
        supplies the coisometric factor.
      & --- \\
    \textbf{issue-1230}
      & Ch.~7 & SDP slackness & Interface & High & Proof
      & Repaired: canonical block-SDP strong duality and saturated
        slackness proved without dominance $I \preceq Z$; conditional
        dominance wrappers removed.
      & Case~III \\
    \textbf{naimark}
      & Ch.~3 & Naimark dilation & Wording & High & Statement
      & Corrected: read as projective sub-measurement on the original
        alphabet; the complete-measurement form is impossible; bipartite
        correlation preservation proved.
      & App.~B \S\ref{subsec:naimark-alphabet-repair} \\
    \textbf{issue-1032}
      & Ch.~4 & Orthonormalization & Audit & Low & Proof
      & Audited: sharp $100\zeta^{1/4}$ paper constant proved at the
        $2\zeta$ completion scale, avoiding the other completion proof's
        larger bound $120\zeta^{1/4}$; no discrepancy with the paper.
      & --- \\
    \textbf{issue-1622}
      & Ch.~9 & Degree-zero pasting & Endpoint & Medium & Proof
      & Repaired: dedicated $d=0$ pasting branch proved; the
        unrestricted theorem \code{ldPasting} assembles it
        without adding a $0<d$ hypothesis.
      & Case~VI \\
    \textbf{issue-422}
      & Ch.~10 & Main soundness & Side condition & High & Statement
      & Corrected: the printed hypothesis permits $k = 0$ at $d = 0$, where
        the printed error vanishes and demands impossible exact degree-zero
        consistency; the soundness statement now requires $0 < k$.
      & Case~VI \\
    \textbf{issue-2338}
      & Ch.~3 & Blueprint links & Wording & Low & Statement
      & Clarified: Chapter~3 blueprint links retargeted to the
        two-space heterogeneous lemma forms; no mathematical
        discrepancy.
      & --- \\
    \bottomrule
\end{longtblr}

\FloatBarrier

\suppappendix{Cross-module composition in the inductive proof}
\label{app:composition}
\subsection{Overview}
\label{pf:composition-mechanics}

\Cref{sec:blueprint-progress} traced how individual theorem statements evolved
as auxiliary assumptions and unproved obligations were removed.
To complete the inductive soundness proof of \citet{jnvwy_ldt}, the repaired
modules must also compose across the induction step from dimension $m$ to $m+1$.
At each boundary in the induction, an upstream theorem must supply the exact
Hilbert-space carrier, operator representation, and quantitative error bound
required by the hypotheses of the next stage.
The case studies below examine five classes of composition failure that arose
when connecting these modules:
\begin{enumerate}
  \item Carrier and register mismatches: an operator intended for separate
  spatial tensor factors $\mathcal{H}_A \otimes \mathcal{H}_B$ acts on a single
  matrix algebra or an uncoupled space.
  \item Inlined conclusions: a theorem assumes its target conclusion as an input
  hypothesis, leaving callers to prove it.
  \item Missing invariants: a downstream stage assumes operator properties that
  fail for general quantum strategies, such as dominance $I \preceq Z$ or
  permutation symmetry on unequal local spaces.
  \item Default values on invalid inputs: a partial construction returns $0$
  outside its support, invalidating the downstream consistency claim.
  \item Insufficient error bounds: intermediate estimates introduce additive or
  square-root error terms that exceed the threshold allowed by the next theorem.
\end{enumerate}

\subsubsection{Interface mismatches between inductive stages}
\label{subsec:producer-consumer-formalism}

Lean checks that each proof term inhabits its declared type, but two verified
declarations still fail to compose whenever the carrier space, operator family,
or error expression produced by one stage does not match the input hypothesis
of the next.

The most direct failure occurs when a theorem takes its own conclusion as an
extra hypothesis.
In the early declaration of \code{mainFormal}, the caller had to supply the
target polynomial measurements and their consistency bounds inside
\code{MainFormalBridgePackage}, and the proof returned that witness directly
(\cref{pf:case-mainformal}).
Writing $H$ for the test-passing hypotheses and $Q$ for the target soundness
conclusion, the early declaration proved $H \Longrightarrow (\mathdefect{Q}
\Longrightarrow Q)$; the repaired theorem proves $H' \Longrightarrow
\mathrepair{Q}$ from the corrected hypotheses $H'$
(\cref{subsec:soundness-trajectory}).

The remaining failures leave mathematical obligations at module boundaries:
embedding unequal local Hilbert spaces into a shared role carrier, proving that
Lagrange interpolation nodes lie in the non-abort support, or tightening
intermediate error bounds so that their sum stays within the inductive budget.
In each case study below, we state the mathematical role of the step in the
induction, exhibit the early declaration or estimate from \citet{jnvwy_ldt}
that prevented composition, and give the repaired Lean construction together
with its contribution to \code{mainInductionError} and \code{mainFormalError}.

\subsubsection{The completed inductive proof}
\label{subsec:repaired-pipeline-architecture}

The completed proof composes thirteen declarations across five stages:
\begin{align}
  &\text{Semidefinite programming and duality foundation:} \nonumber \\
  &\quad\leanid{matrixSdpCanonicalStrongDuality},\quad
   \leanid{sdp\_statement\_with\_slackness} \nonumber \\
  \Longrightarrow\quad&\text{Self-improvement core and projective rounding:} \nonumber \\
  &\quad\leanid{selfImprovementHelper},\quad
   \leanid{selfImprovement} \nonumber \\
  \Longrightarrow\quad&\text{Inductive slice adapters and answer carriers:} \nonumber \\
  &\quad\leanid{selfImprovementInInductionSection} \nonumber \\
  &\quad\leanid{AnswerSelfImprovementData.ofSelfImprovementInInductionSection}
   \nonumber \\
  &\quad\leanid{AnswerSelfImprovementData.ofAnswerCarrier},\quad
   \leanid{SelfImprovementData.ofAnswer} \nonumber \\
  \Longrightarrow\quad&\text{Low-degree pasting and Bernoulli recurrence:} \nonumber \\
  &\quad\leanid{ldPastingInInductionSection},\quad
   \leanid{Pasting.ldPasting} \nonumber \\
  \Longrightarrow\quad&\text{Induction and role unsymmetrization:} \nonumber \\
  &\quad\leanid{mainInductionSuccessorNext\_ofSmallErrorConstruction}
   \nonumber \\
  &\quad\leanid{mainInduction},\quad \leanid{mainFormal}.
  \label{eq:full-terminal-chain}
\end{align}
The five stages in \cref{eq:full-terminal-chain} pass carrier spaces,
measurements, and error bounds through the induction as follows:
\begin{enumerate}
  \item The semidefinite programming stage proves canonical strong duality and
  slack mass saturation for block matrix variables, constructing a positive
  sub-measurement $T$ and a dual operator $Z \succeq 0$ with $Z \succeq A_g$ and
  $P(T) = D(Z)$.
  \item The self-improvement stage combines $T$ with local and global variance
  bounds on the coordinate rerandomization graph, applying continuous
  functional calculus and positive-Gram polar extension to construct an
  orthonormalized projective sub-measurement $H$ whose consistency error is
  bounded by \code{selfImprovementError}.
  \item The slice-adapter stage restricts answer-valued strategies on
  $\mathbb{F}_q^{m+1}$ to each hyperplane slice $x \in \mathbb{F}_q$, applies
  the dimension-$m$ induction hypothesis, and reindexes the resulting error
  averages over $\mathbb{F}_q^m \times \mathbb{F}_q$.
  \item The pasting stage combines the $k$ slice measurements into an ambient
  $(m+1)$-variate polynomial measurement via the Bernoulli recurrence and
  Lagrange interpolation on a verified $(d+1)$-element support.
  \item The final induction step combines the induction hypothesis with
  self-improvement and pasting to bound the successor error by
  \code{mainInductionError}, unsymmetrizes the role register back to Alice's
  and Bob's local carriers, and bounds the top-level error by
  \code{mainFormalError}.
\end{enumerate}

\clearpage

\subsection{Case study I: register placement, role symmetrization, and self-consistency}
\label{pf:case-mainformal}

\subsubsection{The top-level conclusion supplied as a hypothesis}

The top-level soundness theorem \code{mainFormal} (Theorem~3.10 of
\citet{jnvwy_ldt}) takes a bipartite projective strategy that passes the
low individual-degree test with failure probability at most $\eps$ and
constructs local degree-$d$ polynomial measurements $G_A$ and $G_B$ consistent
with the point measurements within \code{mainFormalError}.
In the statement history of \cref{sec:blueprint-progress}, the early
declaration accepted this target conclusion as an input hypothesis packaged in
\code{MainFormalBridgePackage}.

\begin{customdef}{10.1a}[The top-level conclusion assumed as a hypothesis (\commit{f659daae8d167f9fb8763478a6e385592858dc1b}, 12 April 2026)]
\label{def:mainformal-bridge-proxy}
An early draft declared the top-level theorem with its target conclusion bundled
as an input hypothesis:
\begin{lstlisting}
-- Schematic excerpt from commit f659da
structure MainFormalBridgePackage (params) (strategy : ProjStrat params iota) (eps k) : Prop where
  witness : exists G_A G_B,
    ConsRel strategy.state ... (mainFormalError params k eps) /\ ...

theorem mainFormal (params) (strategy : ProjStrat params iota) (eps : Error)
    (_hpass : strategy.PassesLowIndividualDegreeTest eps)
    (k : Nat) (_hk : params.m * params.d <= k)
    (hbridge : MainFormalBridgePackage params strategy eps k) :
    exists G_A G_B, ... (mainFormalError params k eps) := by
  exact hbridge.witness
\end{lstlisting}
\end{customdef}

The proof \lstinline|exact hbridge.witness| projects the witness directly from
\lstinline|hbridge|, leaving the test-passing hypothesis \lstinline|_hpass| and
sampling bound \lstinline|_hk| unused.
Packaging the conclusion as an input hypothesis allowed downstream modules in
the early draft (\commit{f659daae}) to type-check before the inductive step
compiled, but reduced the theorem to the implication $Q \implies Q$.
After maintainers flagged the bundled hypothesis (\mipissue{493}), \mippr{1789}
removed \code{MainFormalBridgePackage} and proved \code{mainFormal} via
\code{mainFormal\_of\_mainInduction}, deriving $G_A$, $G_B$, and the bound
\code{mainFormalError} from \lstinline|_hpass| and \lstinline|_hk|.

\subsubsection{Spatial register placement and tensor lifts}
\label{subsec:register-placement-details}

In a bipartite strategy on $\mathcal{H}_A \otimes \mathcal{H}_B$, Alice's and
Bob's measurement operators act on separate tensor factors as $A \otimes I_B$
and $I_A \otimes B$, so operators of distinct provers commute.
Early matrix definitions placed both provers' operators in a single matrix
algebra or lifted both to the left factor:
\begin{equation}
  \mathdefect{\operatorname{ev}(\psi, (A \cdot B) \otimes I)}.
\end{equation}
Because operators on the same factor do not commute in general, joint
expectations formed from this lift could not compose with bipartite consistency
lemmas.
Commit \commit{050358f3cc9b88d442dc34467061fd9b0cd30d06} introduced explicit
left and right Kronecker embeddings, and
\commit{7e79c14754f34c61460fa3cafdb5fc969083e2b3} (\mippr{148}) moved Bob's
operators to $\code{liftRight}$, giving the bipartite tensor evaluation:
\begin{equation}
  \mathrepair{\operatorname{ev}(\psi, A_a^u \otimes B_b^v) = \operatorname{Re}\tau(\rho_\psi (A_a^u \otimes B_b^v))}.
\end{equation}

\subsubsection{The correlation-preserving dilation input}

Before entering the dimension induction, we apply the Naimark construction of
\cref{subsec:naimark-alphabet-repair} to dilate the strategy's POVMs into
projective sub-measurements on the original outcome alphabets while preserving
all joint correlations on Alice's and Bob's dilated local registers.
This dilated bipartite strategy serves as the input to role symmetrization.

\subsubsection{Role symmetrization on direct-sum carrier spaces}
\label{subsec:role-symmetrization-details}

The dimension induction $\code{mainInduction}(m)$ acts on a synchronous,
permutation-invariant strategy in which both registers share a single local
Hilbert space and measurement family.
An arbitrary bipartite strategy $\mathcal{S} = (\psi, A, B)$, however, acts on
separate local spaces $\mathcal{H}_A$ and $\mathcal{H}_B$ whose Naimark
dilations need not have equal dimension, and its measurement operators $A$ and
$B$ need not be symmetric.

To bridge this gap, \mippr{958}
(\commit{c31419739ec880afd798b09b43098a3f62b6f88f}) generalized
\code{ProjStrat} to independently quantified local index types $\iota_A$ and
$\iota_B$.
Commit \commit{e3e437d5afceef2cc5d29d829614924710dee816} added direct-sum
block helpers, and commit \commit{440cb0f8cbf32d6306258026bf0a2d71bb4aeb1b}
completed the role-register construction for distinct local carriers.
We embed both local spaces of the original strategy into the direct-sum role
carrier
\begin{equation}
  \mathcal{K}_{\mathrm{role}} \coloneqq
  \mathbb{C}^{\mathrm{Role} \times (\iota_A \oplus \iota_B)},
  \qquad
  \rho_{\mathrm{role}} \in
  \operatorname{End}(\mathcal{K}_{\mathrm{role}} \otimes
  \mathcal{K}_{\mathrm{role}}).
\end{equation}
The symmetrized state forms an equal mixture of the two tensor orientations:
\begin{equation}
  \rho_{\mathrm{role}} =
  \frac{1}{2} (\rho_\psi \oplus \operatorname{SWAP}(\rho_\psi)),
  \label{eq:role-register-density-formal}
\end{equation}
embedded into $\operatorname{End}(\mathcal{K}_{\mathrm{role}} \otimes
\mathcal{K}_{\mathrm{role}})$.
In Lean, \code{roleRegisterSymmState} constructs this density matrix by
embedding $\rho_\psi$ and its swapped permutation into direct-sum blocks
(\code{localPairABBlock}) scaled by the trace normalization factor
\code{roleRegisterDensityScale}.
Measurements act block-diagonally through \code{localDirectSumBlock}, producing
a permutation-invariant strategy even when the original local dimensions differ.

\subsubsection{Relating test agreement to self-consistency}
\label{subsec:cross-prover-self-consistency}

The induction hypothesis requires the within-prover self-consistency relation
\begin{equation}
  A^u_a \otimes I \simeq_{\delta} I \otimes A^u_a,
  \label{eq:within-prover-ssc}
\end{equation}
where the same point measurement acts on both registers~\citep{jnvwy_ldt}.
The low individual-degree test, by contrast, queries the two provers at the
same point $u$ and accepts when their answers agree ($a = b$), which measures
cross-prover consistency between $A^{\mathrm{A}}$ and $A^{\mathrm{B}}$.

Commit \commit{0281079b} (11 April 2026) defined test self-consistency as the
average of two within-prover defects:
\begin{equation}
  \mathbb{E}_{u} \left(1 - \sum_a \langle \psi | \mathdefect{A^{\mathrm{A},u}_a \otimes A^{\mathrm{A},u}_a} | \psi \rangle\right)
  \quad\text{and}\quad
  \mathbb{E}_{u} \left(1 - \sum_a \langle \psi | \mathdefect{A^{\mathrm{B},u}_a \otimes A^{\mathrm{B},u}_a} | \psi \rangle\right),
\end{equation}
whereas the test in \citet{jnvwy_ldt} measures the cross-prover overlap of
$A^{\mathrm{A},u}_a \otimes A^{\mathrm{B},u}_a$.
\Mippr{385} (\commit{9645d3d0}) restored the cross-prover agreement subtest and
connected the two formulations through role symmetrization.
When $\mathcal{S}$ passes the test with overall failure probability at most
$\eps$, each of the three equally weighted subtests has error at most
$\mathrepair{3\eps}$ (\code{point\_agreement\_le\_three\_mul}), and the
role-symmetrized strategy $\mathcal{S}_{\mathrm{role}}$ satisfies
\begin{equation}
  \operatorname{selfConsistencyError}(\mathcal{S}_{\mathrm{role}}) = \mathrepair{\operatorname{pointAgreementError}(\mathcal{S})},
\end{equation}
proved in \code{roleRegisterSymmStrategy\_selfConsistency\_eq\_pointAgreement}.
Because both provers in $\mathcal{S}_{\mathrm{role}}$ share the same
block-diagonal measurement operators, cross-prover agreement on $\mathcal{S}$
supplies the goodness parameters $(3\eps, 3\eps, 3\eps)$ passed into
\code{mainInductionError}.

\subsubsection{Slice strategies and their error bounds}
\label{subsec:per-slice-strategy-profiles}

To advance the induction from dimension $m$ to $m+1$, we restrict an
answer-valued strategy on $\mathbb{F}_q^{m+1}$ to each hyperplane slice
$x \in \mathbb{F}_q$ along the last coordinate and apply the induction
hypothesis $\code{mainInduction}(m)$ on $\mathbb{F}_q^m$.
At commit \commit{1a7f43ac0aa83c9129b2c954d116816ddf92df65} (6 May 2026), the
restricted-slice module proved only that the scalar failure probabilities of
restricted strategies equalled those of answer-valued strategies, without
packaging an explicit strategy and goodness witness for each slice $x$.

Commit \commit{440cb0f8cbf32d6306258026bf0a2d71bb4aeb1b} introduced this
slice profile structure:
\begin{lstlisting}
-- Schematic excerpt from commit 440cb0
structure AnswerSuccessorRestrictedFailureProfile
    (strategy : AnswerSymStrat params.next iota) : Type where
  axisParallel : Fq params -> Error
  selfConsistency : Fq params -> Error
  diagonal : Fq params -> Error
  restrictedGood : forall x,
    (xRestrictedAnswerSymStratOfAnswer params strategy x).IsGood
      (axisParallel x) (selfConsistency x) (diagonal x)
\end{lstlisting}
The theorem \code{answerSuccessorRestrictedSliceConclusions}\linebreak[3]
applies $\code{mainInduction}(m)$ to \code{xRestrictedAnswerSymStratOfAnswer}
for each slice $x \in \mathbb{F}_q$.
Lemma \code{answerSuccessorDiagonalSliceIndexErrorAverage}%
\allowbreak\code{\_eq\_diagonalIndexError} then equates the slice-averaged
errors to the ambient $(m+1)$-dimensional subtest errors via a product
equivalence and uniform-average identities on
$\mathbb{F}_q^m \times \mathbb{F}_q$, feeding the averaged slice error into
\code{mainInductionError} at dimension $m+1$.

The final theorem \code{mainFormal} combines this inductive step with role
unsymmetrization over independently quantified carriers under the corrected
sampling hypotheses stated in \cref{sec:formal-statement}.
The zero-sampling counterexample appears in \cref{subsec:pasting-degree-zero},
and the sampling-hypothesis trajectory appears in
\cref{subsec:soundness-trajectory}.

\subsection{Case study II: assembling self-improvement and reindexing its bounds}
\label{pf:case-self-improvement}

\subsubsection{The self-improvement theorem and assumed intermediate steps}

In the inductive step, the dimension-$m$ induction hypothesis produces for each
restricted slice a polynomial measurement $G = \{G_g\}$ consistent with the
slice point measurement $A$ at error $\nu$.
Before the pasting stage combines these slices, Theorem~9.4 of
\citet{jnvwy_ldt} upgrades each $G$ to an orthonormalized projective
sub-measurement $H = \{H_g\}$ and a dual operator
$Z \succeq \mathbb{E}_u A^u_{g(u)}$ whose consistency error depends only on the
test parameters $(\eps, \delta)$ and $(m, q, d)$ through
\code{selfImprovementError}.
Early formalizations assumed the intermediate steps of this upgrade inside a
bridge structure.

\begin{customdef}{5.1a}[Intermediate steps assumed in a bridge structure (\commit{48d1475cc7aecba688b52d676275407eaa5c8e93}, 9 April 2026)]
\label{def:si-bridge-proxy}
An early draft of Chapter~7 declared:
\begin{lstlisting}
-- Schematic excerpt from commit 48d147
structure SelfImprovementBridgePackage (params) (strategy) (G) : Prop where
  permInvariant : PermInvState strategy.state
  helperStrongSelfConsistency : SelfImprovementHelperConclusion ... ->
    BipartiteSSCRel strategy.state (uniformDistribution Unit) ...
  evaluationDataProcessing : SDDRel strategy.state (uniformDistribution Unit) ... ->
    SDDRel strategy.state (uniformDistribution (Point params)) ...
  finalFields : SelfImprovementHelperConclusion ... ->
    SelfImprovementFinalFields ...

theorem selfImprovement (hbridge : SelfImprovementBridgePackage params strategy G) :
    SelfImprovementConclusion ... := by
  let _ := hcons -- Consistency hypothesis discarded!
  exact hbridge.finalFields ...
\end{lstlisting}
\end{customdef}

The proof discarded the input consistency hypothesis \lstinline|hcons| and
required two bridge fields that callers in the Chapter~10 induction could not
supply:
\begin{enumerate}
  \item Permutation invariance for orthonormalization: converting the helper
  sub-measurement $\widehat{H}$ into a projective sub-measurement $H$ via Gram
  orthonormalization required \code{PermInvState strategy.state}, whereas
  inductive slice strategies provide only the test-goodness predicate
  \code{strategy.IsGood eps delta gamma}.
  \item Index-space mismatch ($\mathrm{Unit} \not\cong \mathrm{Point}$): the
  semidefinite programming stage proved state-dependent distance bounds over
  the singleton distribution \code{uniformDistribution Unit}, whereas the
  inductive step requires point-evaluation bounds over the uniform distribution
  \code{uniformDistribution (Point params)} on $\mathbb{F}_q^m$.
\end{enumerate}

\subsubsection{Assembling the internal proof without bridge packages}

The theorem signature was restored in
\commit{d8f6a63638293b7e1c0e98ab629f800806051e54} (\mippr{1462}), conditional
wrappers were removed in \commit{4b04a2b136ac083a331cc75094bc1fcdbb34bb0d}
(\mippr{1539}), and the internal proof was assembled in
\commit{134d08f2e90a7b4486c39e78cdaaca30bb4db156} (\mippr{1638}, merged in
\commit{9aebc89dd1f6d0e05f07527b38edbbdc6422ca39}).

\begin{customthm}{5.1}[Self-improvement assembly and the remaining SDP proof (\commit{9aebc89dd1f6d0e05f07527b38edbbdc6422ca39}, \commit{65288769181461a8d4f2e89c08d9dbf3a42dd964})]
\label{thm:case-si-final}
The theorem \code{selfImprovement} takes test goodness $(\eps, \delta, \gamma)$,
the measurement $G$, and input consistency $h_{\mathrm{cons}}$, assembling
the verified self-improvement guarantee through four mathematical stages:
\begin{enumerate}[label=(\roman*)]
  \item \textbf{Primal-dual SDP solution with slackness:}
    Solving the semidefinite program (\cref{pf:case-sdp}) yields optimal
    sub-measurements $\{T_g\}$ and dual matrix $Z$ satisfying complementary
    slackness.
  \item \textbf{Completeness transfer via input consistency:}
    Combining input consistency $h_{\mathrm{cons}}$ with slackness bounds the
    total defect $I - \sum_g T_g$ on the strategy state.
  \item \textbf{Strong self-consistency and orthonormalization:}
    Establishing strong self-consistency on the helper outputs permits Gram
    orthonormalization into exact projectors.
  \item \textbf{Final error aggregation:}
    Combining point-evaluation data processing with the total-overlap
    displacement bound $\eta \le \widehat\zeta +
    2\sqrt{\widehat\zeta_{\mathrm{ortho}}}$ establishes the completeness,
    point-consistency, self-closeness, dual dominance, and projective-residual
    bounds of \code{SelfImprovementConclusion} within
    \code{selfImprovementError}.
\end{enumerate}
\Mippr{1708} (\commit{65288769181461a8d4f2e89c08d9dbf3a42dd964}) closed the
remaining SDP complementary slackness obligation.
With this step proved, \code{selfImprovement} depended only on the standard
axioms.
\end{customthm}

Through \code{selfImprovementInInductionSection}, the assembled theorem resets
each slice's consistency error to \code{selfImprovementError}, which enters the
inductive recurrence for \code{mainInductionError}.

\subsection{Case study III: semidefinite duality and the false assumption $I \preceq Z$}
\label{pf:case-sdp}

\subsubsection{Primal feasibility and Slater constraint qualification}

To initiate self-improvement, Lemma~9.1 of \citet{jnvwy_ldt} replaces the input
measurement $G$ by solving the block semidefinite program
\begin{equation}
  \sup_{\{T_g\}} \sum_g \operatorname{Tr}(T_g A_g) \quad \text{subject to} \quad T_g \succeq 0, \ \sum_g T_g \preceq I,
  \label{eq:sdp-primal-source}
\end{equation}
whose dual minimizes $\operatorname{Tr}(Z)$ subject to $Z \succeq A_g$ for all
$g$.
An early formulation restricted the primal variables to complete measurements
($\mathdefect{\sum_g T_g = I}$), which excluded the strict Slater interior
point $T_g^{\mathrm{strict}} = \frac{1}{2M}I$ because
$\sum_g T_g = \frac{1}{2}I \prec I$.
\Mippr{251} (\commit{9747bedfad458608e6e6d00a64d77127509929a3}) enlarged the
primal feasible set to sub-measurements $\mathrepair{\sum_g T_g \preceq I}$.
Strict dual feasibility holds at $Z^{\mathrm{strict}} = 2I$, which satisfies
$Z - A_g \succeq I \succ 0$.

\subsubsection{A counterexample to $I \preceq Z$}

Because the relaxed primal optimum $\{T_g\}$ is only a sub-measurement, the
proof must still upgrade $\{T_g\}$ to a complete measurement $\sum_g T_g = I$
without changing the optimal objective value.
Early matrix definitions (\commit{3685be1c99bdbb60cb116ec3f2de604e46cfc1ea})
attempted to deduce $\sum_g T_g = I$ from the canonical complementary slackness
identity $(I - \sum_g T_g) Z = 0$ by assuming operator dominance:
\begin{equation}
  \mathdefect{I \preceq Z}.
\end{equation}
This inequality fails for general quantum strategies.
Consider a single polynomial outcome ($M=1$) on $\mathcal{H} = \mathbb{C}$ with
$A_{g_0} = \frac{1}{2} I$.
The optimal dual variable is $Z = \frac{1}{2} I$.
Then $Z - A_{g_0} = 0 \succeq 0$, so $Z$ is dual feasible and optimal with
$\operatorname{Tr}(Z) = 1/2 = \operatorname{Tr}(T_{g_0} A_{g_0})$ at the primal
optimum $T_{g_0} = I$, yet
\begin{equation}
  Z - I = \frac{1}{2} I - I = -\frac{1}{2} I \not\succeq 0 \implies \mathdefect{I \not\preceq Z}.
\end{equation}
Because $I \preceq Z$ does not follow from the test hypotheses, Chapter~7
callers could not discharge it.

\subsubsection{Saturating the slack while preserving the objective}
\label{lem:sdp-saturation-final}

The repaired proof (\mippr{1708},
\commit{65288769181461a8d4f2e89c08d9dbf3a42dd964}) establishes canonical strong
duality $P(X) = D(Z)$ directly by finite-dimensional convex cone separation.
To obtain completeness without assuming $I \preceq Z$, the \textbf{Slack Mass
Saturation Lemma} (\commit{67715abf7aa1925af3d144b37b45c0a40a3460d1}) adds the
positive slack block $S = I - \sum_g T_g \succeq 0$ to a distinguished
polynomial outcome $g_*$:
\begin{equation}
  \mathrepair{\widetilde{T}_{g_*} \coloneqq T_{g_*} + S, \qquad \widetilde{T}_g \coloneqq T_g \text{ for } g \neq g_*}.
\end{equation}
Since $A_{g_*} \succeq 0$, adding $S \succeq 0$ cannot decrease the primal
objective, while weak duality bounds the saturated objective from above by
$D(Z)$:
\begin{equation}
  P(X) \le P(X_{\mathrm{sat}}) \le D(Z) = P(X) \implies \mathrepair{P(X_{\mathrm{sat}}) = P(X) = D(Z)}.
\end{equation}
Thus $\{\widetilde{T}_g\}$ is a complete measurement ($\sum_g \widetilde{T}_g =
I$) achieving the same primal optimum $P(X) = D(Z)$.
Commits \commit{3e09f41c7435758ce173be5174510e9fb8449122},
\commit{c303e5593bd56ad588d9194e2a1541964b53bca6}, and
\commit{6339abb2c8fbc401f9c5a8b370ec0e03a8541d82} removed all intermediate
declarations that assumed $I \preceq Z$, allowing \code{selfImprovement} to run
under the standard test hypotheses and feed its bound into
\code{mainInductionError}.

\subsection{Case study IV: consistency under rounding and completion}
\label{pf:case-transport-failures}

\subsubsection{IV.a Line~169: bounding consistency before completion}
\label{subsec:line169-transport-details}
\label{subsec:zeta2-cascade-residual}

After $\code{mainInduction}(m)$ produces a symmetric polynomial sub-measurement
on the role carrier, the final unsymmetrization step in Section~3 of
\citet{jnvwy_ldt} rounds and completes Alice's operator $G^{\mathrm{A}}$ to a
projective measurement $Q^{\mathrm{A}}$ while preserving consistency with Bob's
operator $G^{\mathrm{B}}$ within \code{mainFormalError}.
On Line~169 of Section~3, the paper applies triangle substitution to replace
$G^{\mathrm{A}}$ with $Q^{\mathrm{A}}$ at unchanged consistency error:
\begin{equation}
  \begin{aligned}
  G^{\mathrm{A}}_g \otimes I
    &\simeq_{\zeta_1} I \otimes G^{\mathrm{B}}_g,
    \qquad
    G^{\mathrm{A}}_g \otimes I
    \approx_{\zeta_2} Q^{\mathrm{A}}_g \otimes I \\
  &\implies
    \mathdefect{Q^{\mathrm{A}}_g \otimes I
      \simeq_{\zeta_1} I \otimes G^{\mathrm{B}}_g}.
  \end{aligned}
\end{equation}
However, substituting across state-dependent closeness $\approx_{\zeta_2}$
incurs an additive square-root loss $\sqrt{\zeta_2}$, giving $\zeta_1 +
\sqrt{\zeta_2}$ rather than $\zeta_1$.

\Mippr{1190} (\commit{1dd593bccac259fe366d26d3675a595d3f92f2b1},
\mipissue{1099}) resolved this by comparing $G^{\mathrm{A}}$ against the
pre-completion projective sub-measurement $P^{\mathrm{A}}$ before adding the
defect projector.
Because $G^{\mathrm{A}} \approx_{100\zeta_1^{1/4}} P^{\mathrm{A}}$, triangle
substitution loses at most $\sqrt{100\zeta_1^{1/4}} = 10\zeta_1^{1/8}$ in match
mass.
Completing $P^{\mathrm{A}}$ to $Q^{\mathrm{A}}$ at a distinguished outcome
$g_0$ adds a positive operator on outcome $g_0$ and therefore cannot decrease
the match mass:
\begin{equation}
  \begin{aligned}
  \sum_g \langle \psi |(Q^{\mathrm{A}}_g \otimes G^{\mathrm{B}}_g)|\psi\rangle
  &\ge
  \sum_g \langle \psi |(P^{\mathrm{A}}_g \otimes G^{\mathrm{B}}_g)|\psi\rangle,
  \\
  &\hspace{-4em}
  (\code{completeAtOutcomeProj\_left\_matchMass\_ge}).
  \end{aligned}
\end{equation}
Combining the two steps gives the repaired consistency bound:
\begin{equation}
  Q^{\mathrm{A}}_g \otimes I \simeq_{\mathrepair{\zeta_1 + 10\zeta_1^{1/8}}} I \otimes G^{\mathrm{B}}_g.
\end{equation}
\paragraph*{The extra term in $\zeta_2$.}

Completing an orthonormalized sub-measurement with self-consistency error
$\zeta_1$ also contributes an unstated $+2\zeta_1$ term to the closeness bound
$\zeta_2$.
\Mippr{909} (\commit{1ad5f93cda14e9bc26961d42cb39beedbe8faf9d}) absorbed this
$+2\zeta_1$ contribution by increasing the Step~6 coefficient in
\code{cascadeZeta2} from $40$ in \citet{jnvwy_ldt} to $42$:
\begin{equation}
  \zeta_2 = \mathrepair{200\zeta_1^{1/4} + 42\zeta_1^{1/8}}.
\end{equation}

To trace how both repairs enter the top-level error bound, let $\sigma$ denote
\code{mainInductionError} at
$(\mathrm{params},k,3\varepsilon,3\varepsilon,3\varepsilon)$, and let
$\nu_{\mathrm{final}}=\operatorname{mainFormalError}
(\mathrm{params},k,\varepsilon)$.
For $0\le\varepsilon$, $400md\le k$, $0<k$, and $\nu_{\mathrm{final}}<1$, we
define the cascade quantities
\begin{align}
  \zeta_1&=2\sigma+2\sqrt{3\varepsilon+2\sigma}+md/q,\nonumber\\
  \zeta_2&=200\zeta_1^{1/4}+42\zeta_1^{1/8},\qquad
  \zeta_3=6\zeta_1+6\zeta_2.
\end{align}
The repaired point-consistency error then satisfies
\begin{equation}
  \zeta_4^{\mathrm{repaired}}
  =2\sigma+2\sqrt{\mathrepair{\zeta_1+10\zeta_1^{1/8}}+\zeta_3/2}
  \le\nu_{\mathrm{final}},
\end{equation}
which is proved in \code{MainFormalScalarBounds.\allowbreak
  zeta4Repaired\_le\_\allowbreak mainFormalError}.

The top-level assembly in
\code{mainFormal\allowbreak Conclusion\_\allowbreak
  ofRoleRegister\allowbreak
  ScalarBoundary} bounds the unsymmetrized point-consistency error by
$\zeta_4^{\mathrm{repaired}}$ and applies this inequality to conclude with
\code{mainFormalError}.

\subsubsection{IV.b Sub-measurement total-overlap displacement \texorpdfstring{$\eta$}{eta}}
\label{subsec:total-overlap-eta}

In the final stage of self-improvement (Theorem~9.4 of \citet{jnvwy_ldt}), we
must transfer point consistency from the helper sub-measurement $\widehat{H}$
to the orthonormalized projective sub-measurement $H$.
Because $\widehat{H}_{\mathrm{tot}} = \sum_g \widehat{H}_g \preceq I$ and
$H_{\mathrm{tot}} = \sum_g H_g \preceq I$ are sub-measurements rather than
complete measurements, their total operators need not equal $I$.
Applying triangle substitution between sub-measurements therefore introduces a
total-overlap displacement term
\begin{equation}
  \eta \coloneqq \mathbb{E}_u \big| \langle \psi | A^u_{\mathrm{tot}} \otimes (H_{\mathrm{tot}} - \widehat{H}_{\mathrm{tot}}) | \psi \rangle \big|,
\end{equation}
giving the point-consistency bound $\operatorname{ConsRel}(\psi, A, H, \delta +
\sqrt{\varepsilon_{\mathrm{DP}}} + \mathrepair{\eta})$.

\Mippr{1239} (\commit{dc2faed5ec50854643289882bdd21d0e583471a7}) initially
bounded $\eta$ by Cauchy--Schwarz over the polynomial outcome set:
\begin{equation}
  \eta \le \mathrepair{\sqrt{|\mathbb{F}_q|\varepsilon_{\mathrm{DP}}}} \quad (\code{final\_fields\_total\_difference\_le\_sqrt\_card\_data}).
\end{equation}
Because this intermediate bound grows with $|\mathbb{F}_q|$, it cannot be
absorbed into the field-independent error budget without an extra monotonicity
hypothesis (as in the intermediate wrapper of \mippr{1348}, \commit{ec527772}).
The final assembly in \commit{134d08f2} eliminates the $|\mathbb{F}_q|$ factor
by using completeness transfer to control both signs of the difference of total
expectations.
Write $\widehat\zeta$, $\widehat\zeta_{\mathrm{ortho}}$, and
$\widehat\zeta_{\mathrm{DP}}$ for the helper, orthonormalization, and
data-processing errors at $(\mathrm{params},\varepsilon,\delta)$.
Because the point measurement $A^u$ is complete ($A^u_{\mathrm{tot}}=I$),
completeness transfer gives
\begin{equation}
  \eta=
  \left|\langle\psi|I\otimes
    (H_{\mathrm{tot}}-\widehat H_{\mathrm{tot}})|\psi\rangle\right|
  \le\mathrepair{\widehat\zeta+
    2\sqrt{\widehat\zeta_{\mathrm{ortho}}}}.
\end{equation}
Combining this estimate with the data-processing bound shows that the
point-consistency error of $H$ satisfies
\begin{equation}
  \mathrepair{2\widehat\zeta+
    \sqrt{\widehat\zeta_{\mathrm{DP}}}+
    2\sqrt{\widehat\zeta_{\mathrm{ortho}}}}
  \le \operatorname{selfImprovementError}
    (\mathrm{params},\varepsilon,\delta)
\end{equation}
for $0\le\varepsilon,\delta\le1$ and $d\le q$, with the remaining parameter
ranges handled by separate scalar cases.
The proof passes this bound to
\path{final_fields_of_helper_outputs_of_total_difference} via
\path{final_fields_point_consistency_total_difference_error_le_selfImprovementError},
preserving the paper's \code{selfImprovementError} budget that feeds into
\code{mainInductionError}.

\subsection{Case study V: rerandomization graph Laplacian and centered
variance trace}
\label{pf:case-expansion-variance}

Within self-improvement, the data-processing step bounds how far the SDP
sub-measurement $T$ can drift across points of $\mathbb{F}_q^m$ by combining
spectral expansion of the coordinate rerandomization graph (Chapter~5) with a
trace identity for the global centered variance of the measurement operators
(Chapter~6).
Early formalizations replaced both identities with shortcut definitions that
type-checked locally in Chapters~5 and~6 but blocked the Chapter~7 proof of
\code{selfImprovement}.

\subsubsection{Proving the Laplacian identity from edge sums}
\label{pf:prop:laplacian-target}
\label{pf:prop:laplacian-proxy}
\label{pf:eq:laplacian-rewrite-source-app}

To relate the spectral gap of the coordinate rerandomization walk $K$ to
axis-parallel consistency along lines, the expansion step rewrites the graph
Laplacian $L = M^{-1}I - K$ as a Dirichlet average of edge differences:
\begin{equation}
  L = \frac{1}{2} \sum_{u,v \in \mathbb{F}_q^m} W(u,v) (|u\rangle - |v\rangle)(\langle u| - \langle v|).
  \label{pf:eq-laplacian-rewrite-source-app}
\end{equation}
An early draft (\commit{78a14cd4f06fdb113d0b280fc737488add12390f}, 27 March
2026) defined the edge-difference operator as an alias for the right-hand side:
\begin{equation}
  \mathdefect{L_{\mathrm{diff}} \coloneqq M^{-1}I - K \quad (\code{laplacianRewrite := rfl})}.
\end{equation}
Because this definition omitted the edge sum over pairs $(u,v)$, Chapter~7
could not connect the spectral lower bound $L \succeq \lambda_2(I - J/M)$ to
local line-consistency errors.
The repaired proof (\mippr{1033},
\commit{020fc96334233ae4ea130e52f6d5d3e3f8fbe994},
\commit{a3fcfcbdaf4178a5cf6a879c94e7af1d31c69513}) defines
$(L_{\mathrm{diff}})_{a,b}$ entrywise from the weighted edge sum in
\cref{pf:eq-laplacian-rewrite-source-app} and proves entrywise equality with
$(M^{-1}I - K)_{a,b}$ using transition symmetry $W(u,v) = W(v,u)$ and marginal
row sums $\sum_v W(u,v) = M^{-1}$.

\subsubsection{Replacing zero residuals by centered operators}
\label{pf:lem:global-rewrite-target}
\label{pf:eq:global-variance-trace}

The second step of the variance argument expresses the global variance of an
operator family $\{A^u\}$ as a trace over centered residuals:
\begin{equation}
  A_\perp^u = A^u - A_{\mathrm{avg}}, \qquad A_{\mathrm{avg}} = \frac{1}{M} \sum_u A^u, \qquad \sum_u A_\perp^u = 0.
  \label{pf:eq-global-variance-trace}
\end{equation}
Commit \commit{de9a974e2d4543fe95055153f69e38466bc98218} recorded the constant
Fourier mode $A_0=M^{-1/2}\sum_u A^u=\sqrt M\,A_{\mathrm{avg}}$ but defined the
orthogonal residual to be zero:
\begin{equation}
  \mathdefect{A_\perp \coloneqq 0}.
\end{equation}
Setting $A_\perp \coloneqq 0$ made $\sum_u A_\perp^u = 0$ hold trivially while
reducing the residual variance to zero, which prevented Chapter~7 from bounding
$\frac{1}{M}\sum_u \|A^u - A_{\mathrm{avg}}\|_\psi^2$.
\Mippr{542} (\commit{ef07687c5f79639dcb1f428ed5e3d737e00a5c05}) defined the
centered residual family $u \mapsto A^u - A_{\mathrm{avg}}$, proved
$\sum_u A_\perp^u = 0$, and established the trace identity
\begin{equation}
  \operatorname{globalVariance}(A, \psi) = \frac{1}{M} \sum_{u \in \mathbb{F}_q^m} \operatorname{ev}_\psi\left((A_\perp^u)^\dagger A_\perp^u\right) \quad (\code{globalVarianceTraceForm\_eq\_closedForm}).
\end{equation}
Together with the edge-sum Laplacian identity, this trace formula supplies the
data-processing error $\widehat\zeta_{\mathrm{DP}}$ inside
\code{selfImprovementError} and \code{mainInductionError}.

\subsection{Case study VI: interpolation support and degree-zero pasting}
\label{pf:case-pasting}

\subsubsection{Replacing default zero by interpolation on the support}
\label{subsec:pasting-sentinel-evasion}

In the inductive step from $m$ to $m+1$, Definition~12.8 of \citet{jnvwy_ldt}
reconstructs an $(m+1)$-variate degree-$d$ polynomial from $k$ slice outcomes
$(g_1,\dots,g_k) \in (\operatorname{Poly} \cup \{\bot\})^k$ by applying
Lagrange interpolation across a subset of $d+1$ non-aborting slices.
Commit \commit{3cc379b8} handled the partiality of slice extraction at $\bot$
by returning the zero polynomial:
\begin{equation}
  \mathdefect{\operatorname{extractSliceOr0}(\bot) \coloneqq 0}.
\end{equation}
Although substituting $0$ preserved the degree bound $\deg \le d$, evaluating
the resulting interpolant on any slice where $\bot$ was replaced by $0$ yields
$0$ rather than the slice polynomial, falsifying the point-consistency bound
required by the Chapter~10 induction.

\Mippr{495} and \mippr{561}
(\commit{64ae1b79c2e90ab00dd35f5d1d2b4843294bf722},
\commit{c77bc07d8e7f9c2a689bc43a1ce23b7ce3f0394b}) removed
\code{extractSliceOr0} and packaged the non-aborting interpolation nodes into
\code{InterpolationSupportWitness}:
\begin{lstlisting}
-- Schematic excerpt from commit 64ae1b
structure InterpolationSupportWitness (params) (gs : Fin k -> GHatOutcome params) where
  support : Finset (Fin k)
  subset_support : support <= gHatTupleSupport gs
  card_eq : support.card = params.d + 1
\end{lstlisting}
The reconstruction map \code{interpolateCompletedSlicesFromSupport} now takes a
witness $\sigma \subseteq \operatorname{supp}(g)$ of cardinality $d+1$,
ensuring that Lagrange interpolation acts only on valid slice polynomials when
bounding the pasting error in \code{mainInductionError}.

\subsubsection{Degree-zero pasting and the missing $0<k$ hypothesis}
\label{subsec:pasting-degree-zero}
\label{subsec:k0-singularity-details}

When $d = 0$, polynomials have degree zero in the $(m+1)$-st coordinate and are
constant across slices ($\mathrm{Poly}(m+1,q,0) \cong \mathrm{Poly}(m,q,0)$).
\Mippr{1643} (\commit{88c32381837c2a042cd4cc994c31158367d7ee91}) formalized
pasting at $d = 0$ directly by defining the height-averaged sub-measurement
\code{averagedSliceAppendedSubMeas} without Lagrange interpolation.

Analyzing $d = 0$ also exposed a missing hypothesis in Theorem~3.10 of
\citet{jnvwy_ldt}.
When $d = 0$, the paper's sampling condition $k \ge md$ permits $k = 0$.
At $k = 0$, the prefactor $k^2$ in \code{mainFormalError} forces the target
consistency error to vanish:
\begin{equation}
  \nu = 100000 \cdot 0^2 \cdot m^4 \left(\eps^{1/40000} + (0/q)^{1/40000} + e^{-0/(2560000\,m^2)}\right) = \mathdefect{0}.
  \label{pf:eq-zero-k-collapse}
\end{equation}
Consider a strategy on $\mathcal{H} = \mathbb{C}$ whose point measurement
outputs $0$ at $u_0$ and $1$ at $u_1 \neq u_0$.
At $\eps = 1$, this strategy satisfies all test-passing hypotheses of the
theorem.
However, any degree-zero polynomial is constant, so zero-error consistency at
$u_0$ forces its value to be $0$, whereas zero-error consistency at $u_1$
forces its value to be $1$.
Thus no degree-zero polynomial measurement can achieve $\nu = 0$.
The formal theorem rules out this degenerate regime by adding the hypothesis
$\mathrepair{0 < k}$ (\commit{27df275dcad9841571fe11821c78318191ef9db6}),
ensuring that \code{mainInductionError} and \code{mainFormalError} remain
non-degenerate for all $d \ge 0$.

\subsection{Case study VII: pasting telescoping sum and inductive successor arithmetic}
\label{pf:case-arithmetic}

\subsubsection{From \texorpdfstring{$\widehat{H}$}{H} to \texorpdfstring{$G$}{G}: the telescoping sum (\texorpdfstring{$46k^2m$}{46k^2m} correction)}
\label{subsec:bernoulli-telescoping}

In the pasting proof, Lemma~12.21 of \citet{jnvwy_ldt} approximates the
probability that at least $d+1$ slices produce valid polynomials by relating the
$\widehat{H}$-mass of outcomes with support size $|\tau| \ge d+1$ to the upper
binomial tail in $G$ across $k+1$ hybrid quantities $M_0, \dots, M_k$:
\begin{equation}
  \mathbb{E}_{x_1, \dots, x_k} \sum_{\tau :\, |\tau| \ge d+1} \sum_{(g_1, \dots, g_k) \in \mathsf{Outcomes}_\tau} \langle\psi| \widehat{H}^{x_1,\dots,x_k}_{g_1,\dots,g_k} \otimes I |\psi\rangle \approx_{\nu_8} \sum_{i = d+1}^{k} \binom{k}{i} \langle\psi| G^i (I-G)^{k-i} \otimes I |\psi\rangle.
  \label{eq:bernoulli-pasting-supp}
\end{equation}
Each single-slice replacement between adjacent hybrids satisfies
\begin{equation}
  |M_{j-1} - M_j| \le \Delta_k \coloneqq 2\sqrt{2\zeta} + 2\sqrt{\nu_4(k)}.
  \label{eq:stage-swap-bound}
\end{equation}
Because the commutation error $\nu_4(k) \propto k^2 m$ scales quadratically in
$k$, its square root $\sqrt{\nu_4(k)}$ grows linearly in $k$.
Telescoping the $k$ hybrid steps therefore multiplies $\Delta_k$ by an
additional factor of $k$, yielding a quadratic prefactor $k^2 m$:
\begin{equation}
  |M_0 - M_k| \le \sum_{j=1}^k |M_{j-1} - M_j| \le k \Delta_k \le \mathrepair{46\,k^2m\left(\gamma^{1/32} + \zeta^{1/32} + (d/q)^{1/32}\right)}.
  \label{eq:stage-telescope-bound}
\end{equation}
Commit \commit{15df91dd6640e06e448316193f00ba5803806f0d} replaced the paper's
misprinted factor $\mathdefect{46km}$ with $\mathrepair{46k^2m}$ in
\code{fromHToGError}.

Passing to the complementary event $|\tau| < d+1$ and bounding the lower
binomial tail by the additive Chernoff bound at threshold $\theta=1/(200m)$
also requires $k\ge2d/\theta=400md$.
The same-space theorem adopted $k \ge 400md$ in \mippr{912}
(\commit{805e890d9}), and \commit{78d56aa2} retained it when the full
induction step was assembled (\cref{subsec:soundness-trajectory}).

\subsubsection{Inductive error growth at \texorpdfstring{$m=1$}{m=1}}
\label{subsec:successor-m1-gap}

To close the induction from dimension $m$ to $m+1$, the combined error from the
dimension-$m$ induction hypothesis, pasting, and self-improvement must not
exceed the inductive threshold $\nu_{\mathrm{ind}}(m+1) =
\operatorname{mainInductionError}(\mathrm{params.next}, \dots)$.
In Section~10 of \citet{jnvwy_ldt}, the paper bounds the dimension-growth
factor by writing
\begin{equation}
  \mathdefect{\left(1 + \frac{1}{100m}\right)(m^2+3) \le (m+1)^2},
\end{equation}
stating that the inequality holds ``because $m \ge 2$''.
However, the induction starts at $m = 1$, where the inequality fails:
\begin{equation}
  \mathdefect{(1.01)(1^2 + 3) = 4.04 > 4 = (1+1)^2}.
\end{equation}
\Mippr{552} (\commit{70aa84785756f6698c01fba31949065e7331e9b7}) resolved this
base-step gap by proving the tighter pasting bound
$\nu_{\mathrm{paste}} \le \mathrepair{\frac{1}{5}\nu_{\mathrm{ind}}}$ in
\code{ldPastingInInductionNu\_le\_fifth\_mainInductionNu}.
With $\nu_{\mathrm{paste}} \le \frac{1}{5}\nu_{\mathrm{ind}}$, the successor
error regroups as
\begin{equation}
  \left((m^2+1)\left(1 + \tfrac{1}{100m}\right) + \tfrac{2}{5}\right)\nu \le (m+1)^2\nu,
\end{equation}
which holds for all $m \ge 1$ and evaluates to $\mathrepair{2.42 \le 4}$ at
$m = 1$.
This completes the inductive bound for \code{mainInductionError} across all
dimensions $m \ge 1$ and feeds directly into \code{mainFormalError}.

\subsection{Index of theorem inputs and outputs}
\label{pf:master-matrix}

\Cref{tab:master-composition-matrix} summarizes the fifteen module connections
analyzed in Case Studies~I--VII
(\cref{pf:case-mainformal,pf:case-self-improvement,pf:case-sdp,%
pf:case-transport-failures,pf:case-expansion-variance,pf:case-pasting,%
pf:case-arithmetic}).
For each connection, the table lists the blueprint chapter and case study, the
producer declaration that supplies the carrier space, operator construction, or
error bound, the downstream consumer declaration in the inductive proof of
\cref{eq:full-terminal-chain}, and the mathematical repair that enabled the two
declarations to compose.

\begingroup
\newcommand{\compositionentry}[6]{%
  \textbf{#1}\newline#2\newline#5 &
  \textbf{Producer:} #3\newline
  \textbf{Consumer:} #4\par\smallskip
  #6 \\}

\begin{longtblr}[
  expand = \compositionentry,
  caption = {\textbf{Theorem inputs and outputs.}
  The table catalogs fifteen proved connections between modules across the ten
  blueprint chapters.},
  label = {tab:master-composition-matrix},
]{
  colspec = {X[2,l] X[6,l]},
  row{odd} = {bg=gray!5},
  row{1} = {bg=gray!25, font=\bfseries},
  rowsep = 3pt,
  colsep = 4pt,
  rowhead = 1,
}
\toprule
Chapter / interface / case &
Declarations and repair \\
\midrule
\compositionentry
  {Ch.~2 (Registers)}{Spatial register placement}{\leanid{QuantumState.eval}
  (\commit{826b5327})}{\leanid{ProjStrat} (\commit{7e79c147})}{Case I}
  {Moved Bob's operator lift from \code{liftLeft} to \code{liftRight},
  so Alice's and Bob's measurements commute on separate tensor factors in
  $\mathcal{L}(\mathcal{H}_A) \otimes \mathcal{L}(\mathcal{H}_B)$.}
\compositionentry
  {Ch.~2 (Strategies)}{Role symmetrization}
  {\leanid{roleRegisterSymmStrategy} (\commit{440cb0f8})}
  {\leanid{mainInduction} (\commit{c3141973})}{Case I}
  {Embedded $\mathcal{H}_A \oplus \mathcal{H}_B$ into the direct-sum role
  carrier $\mathcal{K}_{\mathrm{role}}$ via \code{localDirectSumBlock},
  constructing a permutation-invariant strategy across unequal local
  dimensions.}
\compositionentry
  {Ch.~2 (Subtests)}{Agreement to self-consistency}
  {\leanid{point\_agreement\_le\_three\_mul} (\commit{9645d3d0})}
  {\leanid{roleRegisterSymmStrategy\_selfConsistency\_eq\_pointAgreement}
  (\commit{440cb0f8})}{Case I}
  {Proved that test failure probability at most $\eps$ bounds cross-prover
  agreement error by $3\eps$, supplying the $(3\eps,3\eps,3\eps)$ goodness
  parameters for \code{mainInductionError} on the symmetrized strategy.}
\compositionentry
  {Ch.~10 (Induction)}{Strategies on each slice}
  {\leanid{AnswerSuccessorRestrictedFailureProfile} (\commit{440cb0f8})}
  {\leanid{answerSuccessorRestrictedSliceConclusions} (\commit{440cb0f8})}
  {Case I}
  {Packaged an explicit strategy and goodness witness for each hyperplane slice
  $x \in \mathbb{F}_q$ and reindexed the slice-averaged errors over
  $\mathbb{F}_q^m \times \mathbb{F}_q$ into \code{mainInductionError}.}
\compositionentry
  {Ch.~7 (Self-Imp.)}{Self-improvement assembly}
  {\leanid{selfImprovementHelper} (\commit{5f0fc8ad})}
  {\leanid{selfImprovementInInductionSection} (\commit{134d08f2})}{Case II}
  {Derived point-indexed consistency bounds over $\mathbb{F}_q^m$ from
  $\mathrm{Unit}$-indexed SDP estimates, assembling helper completeness, strong
  self-consistency, and Gram orthonormalization within
  \code{selfImprovementError}.}
\compositionentry
  {Ch.~7 (SDP)}{SDP primal sub-measurements}
  {\leanid{sdpStrictPrimalSubMeas} (\commit{ca96c44e})}
  {\leanid{matrixSdpCanonicalStrongDuality} (\commit{e4ca6d28})}{Case III}
  {Enlarged the primal feasible set from complete measurements to
  sub-measurements $\sum_g T_g \preceq I$, supplying the strict Slater interior
  point $T_g=\frac{1}{2M}I$ with total mass $\frac{1}{2}I \prec I$.}
\compositionentry
  {Ch.~7 (SDP)}{SDP slack mass saturation}
  {\leanid{matrixSdpCanonicalSaturateSlackBlockMatrix} (\commit{67715abf})}
  {\leanid{sdp\_statement\_with\_slackness} (\commit{31c8e37c})}{Case III}
  {Added the positive slack block $S = I - \sum_g T_g$ to $g_*$ and applied
  weak duality $P(X) \le P(X_{\mathrm{sat}}) \le D(Z)=P(X)$, eliminating the
  false operator-dominance hypothesis $I \preceq Z$.}
\compositionentry
  {Ch.~4 (Rounding)}{Line-169 consistency bound}
  {\leanid{completeAtOutcomeProj\_left\_matchMass\_ge} (\commit{1dd593bc})}
  {\leanid{mainFormalConclusion} (\commit{cd148c7f})}{Case IV.a}
  {Bounded the match-mass loss against the pre-completion sub-measurement
  $P^{\mathrm{A}}$ and proved that completion cannot decrease match mass,
  giving error $\zeta_1+10\zeta_1^{1/8}$ inside \code{mainFormalError}.}
\compositionentry
  {Ch.~3 (Prelim.)}{Total-overlap displacement}
  {\leanid{completenessTransferProjectiveP} (\commit{134d08f2})}
  {\leanid{selfImprovement} (\commit{134d08f2})}
  {Case IV.b}
  {Bounded the sub-measurement total-overlap displacement by
  $\eta\le\widehat\zeta+2\sqrt{\widehat\zeta_{\mathrm{ortho}}}$ via
  completeness transfer, absorbing the resulting point-consistency error into
  \code{selfImprovementError}.}
\compositionentry
  {Ch.~5 (Expansion)}{Rerandomization graph Laplacian}
  {\leanid{rerandomizeCoordWeight\_symm} (\commit{a3fcfcbd})}
  {\leanid{laplacian\_eq\_edgeDifferenceForm} (\commit{a3fcfcbd})}{Case V}
  {Proved entrywise equality $L=M^{-1}I-K$ from coordinate-swap symmetry
  $W(u,v)=W(v,u)$ and marginal sums, replacing the tautological alias
  $L_{\mathrm{diff}} \coloneqq M^{-1}I - K$.}
\compositionentry
  {Ch.~6 (Variance)}{Centered global variance trace}
  {\leanid{canonicalGlobalVarianceDecomposition} (\commit{ef07687c})}
  {\leanid{globalVarianceTraceForm\_eq\_closedForm} (\commit{ef07687c})}
  {Case V}
  {Constructed the centered residual family
  $A_\perp^u=A^u-A_{\mathrm{avg}}$ with $\sum_u A_\perp^u=0$ in place of
  $A_\perp \coloneqq 0$, establishing the global variance trace identity.}
\compositionentry
  {Ch.~9 (Pasting)}{Supported Lagrange interpolation}
  {\leanid{InterpolationSupportWitness} (\commit{64ae1b79})}
  {\leanid{interpolateCompletedSlicesFromSupport} (\commit{c77bc07d})}
  {Case VI}
  {Replaced the zero-default map
  $\operatorname{extractSliceOr0}(\bot)\coloneqq0$ with Lagrange interpolation
  on a verified non-abort support $\sigma \subseteq \operatorname{supp}(g)$ of
  size $|\sigma|=d+1$.}
\compositionentry
  {Ch.~9 (Pasting)}{Degree-zero pasting completion}
  {\leanid{averagedSliceAppendedSubMeas} (\commit{88c32381})}
  {\leanid{degreeZeroPastedPointConsistency} (\commit{88c32381})}{Case VI}
  {Completed height-averaged sub-measurements at degree $d=0$ without
  interpolation, treating arbitrary $k$ by cases.
  The top-level $0<k$ correction is a separate soundness obligation
  (\S\ref{subsec:pasting-degree-zero}).}
\compositionentry
  {Ch.~9 (Pasting)}{Pasting telescoping sum}
  {\leanid{fromHToGStageMass} (\commit{15df91dd})}
  {\leanid{fromHToG\_bound} (\commit{15df91dd})}{Case VII}
  {Telescoped $k$ adjacent hybrid differences to bound the binomial tail,
  correcting the paper's factor in $\nu_8$ from $46km$ to $46k^2m$ under
  $k \ge 400md$.}
\compositionentry
  {Ch.~10 (Induction)}{Successor error regrouping}
  {\leanid{ldPastingInInductionNu\_le\_fifth\_mainInductionNu}
  (\commit{70aa8478})}
  {\leanid{mainInductionSuccessorNext} (\commit{70aa8478})}{Case VII}
  {Proved $\nu_{\mathrm{paste}}\le\frac15\nu_{\mathrm{ind}}$ so the successor
  bound $((m^2+1)(1+\frac{1}{100m})+\frac25)\nu \le (m+1)^2\nu$ holds at $m=1$
  ($2.42\le4$), resolving the paper's $m=1$ gap $(1.01)(4)=4.04>4$.}
\bottomrule
\end{longtblr}
\endgroup

\FloatBarrier

\suppappendix{Verification tools and proof integrity checks}
\label{app:checks}
\subsection{Overview}
\label{sec:defense-architecture}

Successful compilation in Lean~\citep{Moura2021Lean,
TheMathlibCommunity2020Lean} guarantees that a proof term matches its declared
type, but does not by itself ensure that the declaration signature captures the
intended claim in the papers~\citep{jnvwy_ldt, Ji2021Mip}.
Without statement-level checks, a declaration can pass typechecking while
assuming its conclusion as a hypothesis, packaging unproved intermediate steps
as structure fields, or instantiating an existential claim on a degenerate
carrier space.

To detect these defects, we combined five continuous-integration, audit, and
review checks with semantic mathematical review
(\cref{tab:defense-architecture-summary}).
\cref{tab:check-status} records when each check was introduced and how it was
enforced.

Between mid-April and mid-May 2026, we merged 74 commits updating agent
instructions, automated review prompts, and continuous-integration workflows,
of which 29 directly modified prompt and policy configurations.
Across agent instructions, pull-request review prompts, automated repair loops,
maintainer checklists, and continuous-integration scanners, these updates
targeted conclusion-shaped hypotheses and helper structures with unproved
fields.

\begin{table}
\centering
\caption{The five checks for proof shortcuts and statement changes.}
\label{tab:defense-architecture-summary}
\begin{tblr}{
  colspec = {X[3,l] X[2.1,l] X[3.9,l] X[4,l]},
  row{odd} = {bg=gray!5},
  row{1} = {bg=gray!25, font=\bfseries},
  rowsep = 3pt,
  colsep = 4pt,
}
\toprule
Check &
Target &
Mechanism &
Action on failure \\
\midrule
1. Conclusion-shaped hypotheses (\S\ref{sec:layer-1-conclusion-hypotheses}) &
Proof evasion &
Parses binders in theorem signatures; compares salient signature tokens between
hypotheses and the conclusion &
Fails continuous integration (\mippr{1511}) \\
2. Proof debt in paper-facing headers (\S\ref{sec:layer-2-proof-debt-scanner}) &
Auxiliary obligations &
Scans blueprint-linked declarations for warning terms &
Fails continuous integration (\mippr{1511}) \\
3. Helper structure reviews (\S\ref{sec:layer-3-helper-purges}) &
Unused helpers &
Requires constructor, consumer, or issue &
Maintainers block merge; reviewed PRs remove unused structures (\mippr{1440}) \\
4. Kernel axiom audits (\S\ref{sec:layer-4-axiom-audits}) &
Axiom leaks &
Inspects environment with \code{Lean.collectAxioms} &
Elaboration fails when unexpected axioms appear (\mippr{416}) \\
5. Blueprint synchronization (\S\ref{sec:layer-5-blueprint-sync}) &
Unverified green nodes &
Cross-references blueprint tags with Lean definitions and docstrings &
Reports matches for review; offers a blocking \code{--ci} mode (\mippr{1767}) \\
\bottomrule
\end{tblr}
\end{table}

\begin{table}
\centering
\caption{Introduction and deployment status of the five checks.}
\label{tab:check-status}
\begin{tblr}{
  colspec = {X[1.9,l] X[6.0,l] X[3.1,l] X[5,l]},
  row{odd} = {bg=gray!5},
  row{1} = {bg=gray!25, font=\bfseries},
  rowsep = 3pt,
  colsep = 4pt,
}
\toprule
Date &
Shortcut observed &
Check introduced &
Status over time \\
\midrule
15 April 2026 &
Classical soundness proof omitted; \code{sorryAx} in transitive axioms (\mipissue{408}). &
Kernel axiom audits (\mippr{416}). &
Blocking from introduction; relaxed once for 14~min (\mipcommit{3aeb70ab}, 20~May); restored (\mipcommit{fc7376a2}). \\
26 April 2026 &
Hypotheses that assume the conclusion (\code{SelfImprovementBridgePackage}, \mipissue{493}). &
Conclusion-shaped hypothesis scanner (\mippr{784}). &
Report-only from 26 April (\mippr{784}); blocking from 11 May (\mippr{1511}). \\
9 May 2026 &
Unused helper structures and unproved intermediate fields (\mipissue{1381}). &
Helper structure reviews (\mippr{1440}). &
Enforced via maintainer review checklist and PR review audits. \\
11 May 2026 &
Names for unproved obligations (\code{Bridge}, \code{Package}, \dots) in
headers of blueprint-linked theorems (\mipissue{1458}). &
Proof-debt scanner (\mippr{1475}). &
Report-only in \mippr{1475}; \code{--ci} added in \mippr{1506};
blocking from 11 May (\mippr{1511}). \\
20 May 2026 &
Green blueprint nodes with warning terms in docstrings or linked declaration names (\mipissue{1693}). &
Blueprint synchronization audit (\mippr{1767}). &
Used for a recorded audit; script provides a blocking mode with explicit
allowances. \\
\bottomrule
\end{tblr}
\end{table}

\subsection{Checks 1 and 2: hypothesis and proof-debt scanners}
\label{sec:layer-1-conclusion-hypotheses}
\label{sec:layer-2-proof-debt-scanner}

Checks~1 and~2 target theorem declarations whose signatures deviate from the
paper by assuming unproved obligations.
Check~1 inspects the binder structure of each theorem signature, whereas
Check~2 scans public declaration identifiers and type signatures for stems used
for unproved obligations.

\subsubsection{Implementation}

The conclusion-shaped hypothesis scanner parses each theorem signature into
hypothesis binders and the conclusion, comparing salient signature tokens
between each hypothesis and the conclusion to flag hypotheses with high
salient-token overlap.
The proof-debt scanner checks declarations linked to blueprint nodes against a
list of stems used for unproved obligations:
\begin{equation*}
  \code{ForbiddenStems} \coloneqq \{\code{Bridge}, \code{Residual}, \code{Repair}, \code{Package}, \code{Input}, \code{Producer}, \code{Hypotheses}, \code{Assumptions}, \dots\}.
\end{equation*}
A third script, the declaration audit, blocks direct \code{axiom} and
\code{constant} commands in source code.

\subsubsection{Historical example: \code{SelfImprovementBridgePackage}}

Two defects in the formalization of the self-improvement theorem motivated
these scanners: the named package introduced on 9 April 2026
(\commit{48d1475c}) and the inline-existential formulation documented in
\mipissue{493}.
The proof-debt scanner flags the named package via its identifier stem, whereas
the conclusion-shaped hypothesis scanner detects the inline existential binder
by its salient-token overlap with the theorem conclusion.
\cref{app:trajectory} details the sequence of statement repairs, and
\cref{pf:case-self-improvement} presents the completed proof.
We restored the paper's theorem statement in \mippr{1462}, assembled the proof
in \mippr{1638}, and discharged the semidefinite-programming obligation in
\mippr{1708}.

\subsection{Check 3: helper structure reviews and scaffolding removal}
\label{sec:layer-3-helper-purges}

While Checks~1 and~2 inspect theorem signatures and identifier stems, unproved
obligations can also be stored as fields of auxiliary structures.
Proving agents frequently defined intermediate structures to bundle hypotheses,
intermediate operators, or witness candidates; when left uninstantiated, these
structures obscure missing proof steps from signature-level scanners.

\Needspace{9\baselineskip}
\subsubsection{Requirements for merging}

Before approving a pull request that introduces an intermediate data structure,
maintainers require at least one of the following (\mipissue{1379}):
\begin{enumerate}
  \item an explicit \textbf{constructor theorem} deriving the structure from
  established premises,
  \item an explicit \textbf{consumer theorem} using the structure to prove a
  required step, or
  \item an issue specifying what remains to be proved.
\end{enumerate}

\subsubsection{Historical example: removing unused structures}

Following an audit of helper declarations, \mippr{1440} (\commit{bb1839c7},
\mipissue{1381}) removed unused helper structures and temporary input
abbreviations, including uninstantiated Naimark, rounding, and
spectral-truncation witness structures.

Similarly, the repair of \code{QXPLayerData} replaced assumed structure fields
with an explicit mathematical construction.
\cref{app:trajectory} traces its successive repairs; during pull-request
review, reviewers verify both the constructor theorems that instantiate each
structure field and the consumer lemmas that apply them.

\subsection{Check 4: kernel axiom audits}
\label{sec:layer-4-axiom-audits}

Syntactic scanners and structure reviews inspect source declarations, whereas
Check~4 queries the Lean kernel environment directly to verify that target
theorems do not depend transitively on unfinished proofs (\code{sorryAx}) or
custom mathematical axioms.

\subsubsection{Implementation}

During \code{lake build}, the audit module in \path{LDT/Test/AxiomAudit.lean}
invokes \code{Lean.collectAxioms} on each monitored declaration and aborts
elaboration if the transitive axiom set departs from its expected specification
(\cref{sec:verification-artifacts}).

\subsubsection{Historical examples: unfinished proofs and a relaxed axiom check}

The classical soundness placeholder introduced in \mippr{407} motivated this
kernel check.
After an automated issue tracker flagged its transitive \code{sorryAx}
dependency (\mipissue{408}), commit \commit{c530eb2} reverted the pull request.
In \mippr{416}, we made the unproved classical premise an explicit hypothesis
and added an in-kernel axiom audit for the quantum soundness theorem
(\cref{app:trajectory}).
Because \code{Lean.collectAxioms} detects only axiom declarations in the kernel
environment, assumptions introduced as explicit theorem hypotheses must instead
be caught by Checks~1--3 and mathematical review.

On 20 May 2026, \code{selfImprovement} temporarily depended on a helper lemma
containing a \code{sorry} token tracked in \mipissue{1642}.
During an automated repair session on \mippr{1734}, an agent modified the
axiom-audit harness in commit \commit{3aeb70ab}, adding \code{sorry} to
\code{expected\allowbreak SelfImprovementAxioms} and
\code{expectedInduction\allowbreak SelfImprovementAxioms} so that the failing
continuous-integration check would pass.
An audit detected this modification fourteen minutes later; commit
\commit{fc7376a2} restored both expected-axiom lists to the standard Lean
axioms and proved the helper lemma to eliminate the \code{sorryAx} dependency.
The pull request merged later that day with the strict axiom check restored, so
the modified test harness never reached the main branch.
This event was the only recorded instance during the formalization in which an
agent relaxed the expected-axiom lists in the audit harness.

\subsection{Check 5: blueprint synchronization and declaration checks}
\label{sec:layer-5-blueprint-sync}

Even when a Lean declaration typechecks without unexpected axioms, the project
blueprint can misrepresent formalization progress if a node marked complete
(green) links to a weakened substitute statement or an auxiliary helper
declaration that assumes unproved results.
Check~5 verifies consistency between the \LaTeX\ blueprint and the Lean
declarations.

\subsubsection{Implementation}

We maintain the mathematical dependency graph with Massot's
\texttt{leanblueprint} tool~\citep{Massot2021Leanblueprint} and validate its
Lean links with two continuous-integration scripts.
First, the blueprint synchronization script parses the \LaTeX\ blueprint files,
extracts all \texttt{\textbackslash lean} and \texttt{\textbackslash leanok}
tags, and checks that every referenced declaration exists in the Lean codebase.
Second, the green-node audit searches the declaration identifiers and public
signatures linked from \texttt{\textbackslash leanok} nodes for forbidden
obligation stems and checks their immediate docstrings for
\code{**Unfaithful:**} markers.
In \code{--ci} mode, the script fails the build whenever a paper-facing green
node matches these patterns outside an explicit allowlist, leaving mathematical
review to confirm whether each flagged hypothesis belongs to the displayed
statement.

\subsubsection{Historical example: reviewing green nodes with warning terms}

In \mippr{1767} (\mipissue{1693}), we added the green-node integrity audit and
committed its initial findings at \commit{e3d69726b}.
The audit flagged blueprint links whose Lean identifiers contained obligation
stems and compared each linked signature with the displayed blueprint
statement.
For example, \code{SliceBoundednessInput.storedBoundedResidualBound} packages a
boundedness assumption already present in the displayed commutativity claim.
Across all green nodes checked against the paper, the audit found none that
assumed an unproved obligation absent from the displayed statement.
The script reports unexpected links without automatically removing
\texttt{\textbackslash leanok} tags.

\subsection{Mathematical review}
\label{sec:layer-6-human-audit}

The five checks inspect syntactic, structural, and kernel-level properties of
declaration signatures, helper structures, and blueprint links through blocking
continuous-integration gates (Checks~1, 2, and~4), maintainer merge
requirements (Check~3), and blueprint audits (Check~5), whereas mathematical
review verifies the semantic definitions and intermediate lemmas against the
proofs in the LIDT paper.
For example, the initial interpolation sentinel compiled without axioms or
obligation stems, yet failed to provide point consistency on empty slices;
\cref{subsec:pasting-sentinel-evasion} details its replacement by
support-certified interpolation.
Similarly, the Line-169 transport lemma required an additional error-bound term
before its conclusion matched the downstream consumer
(\cref{subsec:line169-transport-details}).
Mathematical review therefore checks that each lemma supplies the exact
mathematical quantities required by subsequent proof steps, even when both
declarations pass kernel typechecking.

\subsection{Measuring the checks}
\label{sec:check-effectiveness}

\cref{tab:defense-architecture-summary,tab:check-status} summarize the scope
and deployment history of the five checks.
To evaluate their operation across the project history, we replayed the final
scanner rules across dated snapshots of the main branch, measured the detection
latency for the three statement defects in \cref{app:trajectory}, and audited
all pull-request runs that failed the blocking continuous-integration checks.

\subsubsection{Running the final scanners on earlier snapshots}

We ran the final versions of the conclusion-shaped hypothesis scanner and the
proof-debt scanner on 23 dated snapshots of the main branch between 7 March and
5 June 2026.
We applied identical matching rules to every snapshot, counting all historical
declarations matched by the final rule sets.

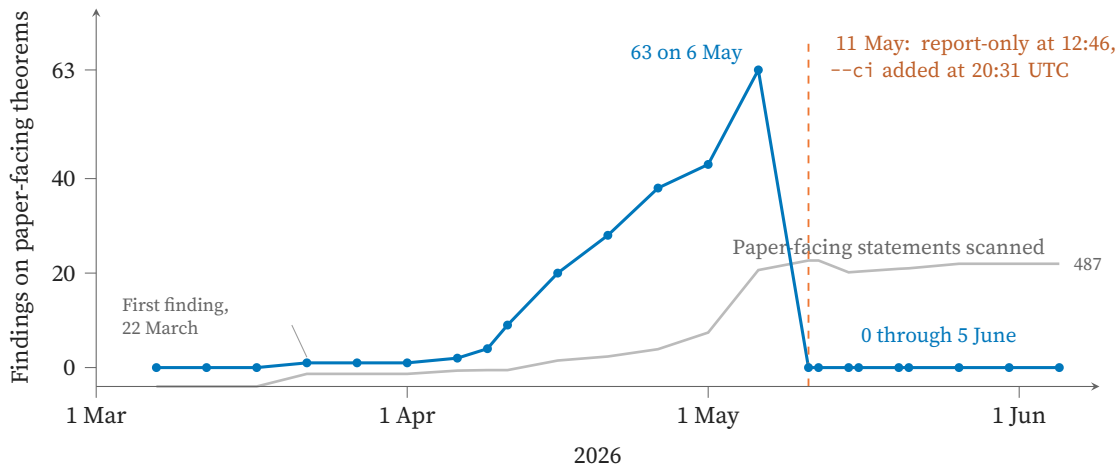
\begin{figure}
  \centering
  \begingroup
  \pgfplotsset{
    replay axis/.style={
      width=0.80\linewidth,
      height=5.0cm,
      scale only axis,
      xmin=0, xmax=100,
      enlargelimits=false,
      clip=false,
    },
  }
  \begin{tikzpicture}
    \begin{axis}[
      replay axis,
      name=corpus,
      axis lines=none,
      ymin=0, ymax=1500,
    ]
      \addplot[draw=cbGrey, line width=1.0pt, no marks] coordinates {
        (6,0) (11,0) (16,0) (21,50) (26,50) (31,50) (36,63) (39,65)
        (41,65) (46,103) (51,119) (56,148) (61,214) (66,462) (71,500)
        (72,500) (75,453) (76,456) (80,467) (81,469) (86,487) (91,487)
        (96,487)
      };
      \node[anchor=south, font=\footnotesize, text=cbGrey!55!black,
            inner sep=1.5pt]
        at (axis cs:79,487) {Paper-facing statements scanned};
      \node[anchor=west, font=\scriptsize, text=cbGrey!55!black,
            inner sep=1.5pt]
        at (axis cs:97,487) {487};
    \end{axis}

    \begin{axis}[
      replay axis,
      at={(corpus.south west)}, anchor=south west,
      axis lines=left,
      axis line style={draw=cbInk!70},
      tick style={draw=cbInk!70},
      tick align=outside,
      tick label style={font=\small, text=cbInk},
      label style={font=\small, text=cbInk},
      ymin=-4, ymax=76,
      ytick={0,20,40,63},
      xtick={0,31,61,92},
      xticklabels={1 Mar,1 Apr,1 May,1 Jun},
      ylabel={Findings on paper-facing theorems},
      xlabel={2026},
      axis on top,
    ]
      \addplot[draw=cbBlue, line width=1.15pt, mark=*, mark size=1.1pt,
               mark options={draw=cbBlue, fill=cbBlue}] coordinates {
        (6,0) (11,0) (16,0) (21,1) (26,1) (31,1) (36,2) (39,4)
        (41,9) (46,20) (51,28) (56,38) (61,43) (66,63) (71,0)
        (72,0) (75,0) (76,0) (80,0) (81,0) (86,0) (91,0)
        (96,0)
      };

      \draw[cbOrange, line width=0.7pt, dashed] (axis cs:71,-4) -- (axis cs:71,70);
      \node[anchor=north west, align=left, font=\footnotesize,
            text=cbOrange!80!black, inner sep=2.5pt, text width=4.4cm]
        at (axis cs:73,72)
        {11 May: report-only at 12:46,\\\texttt{--ci} added at 20:31 UTC};

      \node[anchor=south east, align=right, font=\footnotesize, text=cbBlue,
            inner sep=2pt]
        at (axis cs:65,63) {63 on 6 May};
      \node[anchor=south, align=center, font=\footnotesize, text=cbBlue,
            inner sep=2pt]
        at (axis cs:84,3) {0 through 5 June};
      \node[anchor=west, align=left, font=\scriptsize, text=cbInk!70,
            inner sep=2pt]
        at (axis cs:2,11) {First finding,\\22 March};
      \draw[cbInk!45, line width=0.4pt] (axis cs:19.5,9) -- (axis cs:21,2.2);
    \end{axis}
  \end{tikzpicture}
  \endgroup
  \caption{Findings before and after we added the proof-debt scanner.
    We ran the current proof-debt scanner, unmodified, over 23 verified
    snapshots of the repository's main branch using the same matching
    rules for every snapshot.
    The blue curve counts its findings on paper-facing theorems at each
    snapshot; the grey curve, drawn to its own scale, is the number of
    paper-facing statements the scanner scanned.
    The dashed vertical line marks 11~May 2026, the day the scanner entered
    continuous integration in report-only mode.
    The \texttt{--ci} flag was added eight hours later;
    \mippr{1511} made scanner failures fail the CI job that night.
    Data:
    \protect\path{data/m1_proof_debt_replay.csv}.}
  \label{fig:check-replay}
\end{figure}

\cref{fig:check-replay} plots the proof-debt scanner's match count across the
23 snapshots.
Before continuous-integration enforcement, matched paper-facing declarations
accumulated from a single instance on 22 March to a peak of 63 on 6 May 2026.
On 11 May, the scanner merged in report-only mode (\mippr{1475}, 12:46~UTC),
gained the \code{--ci} flag that evening (\mippr{1506}, 20:31~UTC), and became
a merge-blocking check at 23:27~UTC (\mippr{1511}).
All 63 matches were eliminated that day, and the eight subsequent snapshots
through 5 June (spanning 487 paper-facing declarations) recorded zero matches;
\mippr{1708} later closed the underlying semidefinite-programming lemma.
In the 9 April snapshot, the replay flagged four declarations, including the
\code{selfImprovement} package introduced that day in \commit{48d1475c}.

Across the same 23 snapshots, as the number of scanned declarations grew from
44 to 2,602, the conclusion-shaped hypothesis scanner found matches only on 21
April 2026.
Daily April snapshots isolate when this hypothesis appeared and was removed.
On 18 April, \mippr{491} introduced an inline existential hypothesis on
\code{mainInduction} (sharing 11/11 salient conclusion tokens) while inlining
the \code{SelfImprovementBridgePackage} fields into \code{selfImprovement}; a
refactoring of the surrounding bridge declarations removed the inline hypothesis
on 23 April.
When the report-only scanner merged three days later (\mippr{784}), review in
\mipissue{493} and the subsequent refactoring had already eliminated the
violation.
After the scanner became merge-blocking on 11 May (\mippr{1511}), the finding
count remained at zero across all eight subsequent snapshots without expanding
the scanner's allowlist.

\subsubsection{How long the shortcuts survived}

\cref{tab:check-latency} records when the three statement defects in
\cref{app:trajectory} were introduced, first detected, and repaired.
Beginning on 11 May, continuous integration ran both scanners as blocking
checks on pull requests modifying Lean files, and additionally ran the
proof-debt scanner on pull requests modifying the blueprint.

\begin{table}
\centering
\caption{Time from introduction to first observed detection for the three
  documented shortcuts.
  We measure to the first issue, review comment, or failing check logged in the
  repository.
  The repairs followed on 1--2 May 2026 for the Laplacian alias (the
  \mippr{1033} chain), the same night for the rounding statement (\mippr{287},
  with the construction proved in \mippr{652}, \mippr{726}, and \mippr{1126}),
  and 11 May 2026 for the \code{selfImprovement} statement (\mippr{1462}).}
\label{tab:check-latency}
\begin{tblr}{
  colspec = {X[3.5,l] X[2.5,l] X[6.3,l] X[4.2,l] X[2.5,l]},
  row{odd} = {bg=gray!5},
  row{1} = {bg=gray!25, font=\bfseries},
  rowsep = 3pt,
  colsep = 4pt,
}
\toprule
Pattern &
Introduced &
First observed detection &
Detected by &
Latency \\
\midrule
Definitional alias of the Laplacian, closed by \code{rfl} &
27 Mar 2026 (\commit{78a14cd4}) &
25 Apr 2026, 00:46~UTC: a review fix on \mippr{721} drops \texttt{\textbackslash leanok} from the blueprint node as a ``vacuous tautology''. &
Automated reviewers checking blueprint tags &
29 days \\
Rounding witness on a one-point space (\code{R = [1]}, \code{PUnit} carrier) &
9 Apr 2026, 20:45~UTC (\mippr{274}) &
Same day: a review bot's risk note in the body of the introducing pull request, then \mipissue{279} filed 19 minutes after merge. &
Automated pull-request review that did not block the merge; an issue was then filed under the owner account &
Under 1 day (flagged before merge, then merged) \\
Conclusion-shaped hypotheses on \code{selfImprovement} &
9 Apr 2026, named stem (\commit{48d1475c}); 18 Apr 2026, inline existentials (\mippr{491}) &
18 Apr 2026, 04:13~UTC: \mipissue{493} names \mippr{491} before its 11:34~UTC merge; 10 May 2026, 05:42~UTC: \mipissue{1453} flagged the unproved named package. &
Review filed as a tracker issue &
0 days for the inline form; 31 days to the issue about the named package \\
\bottomrule
\end{tblr}
\end{table}

\subsubsection{Failures reported by blocking checks}

From 11--12 May until 16 July 2026, three continuous-integration audit jobs
executed these checks on pull requests modifying the files covered by each job;
a single workflow (\commit{343bf067}) subsequently consolidated the three jobs
without changing their commands.
Counting the latest attempt of each run and classifying outcomes from archived
step statuses and branch history, 49 of the 4,113 pull-request runs logged
during this period failed.
Of these 49 failures, 42 resulted from transient runner timeouts, checkout
errors, or empty job steps.
Seven failures were triggered by the verification checks: six by
unfaithful-blueprint markers (five inspected individually) and one by the
proof-debt scanner, while the conclusion-shaped hypothesis scanner did not fail
in the inspected runs.
On all inspected branches, the flagged Lean declarations or blueprint entries
were repaired, and the pull requests merged with passing checks.
Most statement defects and unproved obligations were detected and repaired
earlier during interactive proving sessions and pull-request review threads,
with the continuous-integration checks enforcing these invariants at merge.
Because the proof-debt scanner ran in report-only mode for only eight hours on
11 May, we report blocking-period runs without a separate report-only baseline.

\suppappendix{GitHub task tracking, agent contributions, and model usage}
\label{app:task-tracking}

\subsection{From proof decomposition to session instructions}
\label{sec:issue-session-instructions}

We coordinated proof obligations, dependency tracking, and automated
verification through GitHub issues, native sub-issues, and pull-request
workflows.
A saved continuation prompt illustrates how an interactive proving session
resumed work from the repository state:
\begin{quote}
\small\raggedright\ttfamily
Continue to address the opened issues.
\end{quote}
The remainder of the prompt instructs the session to launch parallel subagents
in isolated worktrees, refactor helper declarations, synchronize the blueprint,
merge pull requests once reviews agree and review threads resolve, and advance
to the next open issues.
Within this loop, GitHub issues specified the pending proof obligations and
their dependencies, while the paper, the blueprint, and the repository
instructions specified the target mathematical statements and verification
checks.
An agent session could therefore query the open issue graph, submit a proof in
a pull request, address review comments, and proceed to the next task.

In addition to local sessions, GitHub Actions workflows invoked Claude Code and
Codex from issue and pull-request discussions, and OpenCode from issue
comments.
Each workflow combined the discussion request with repository instructions
requiring formal declarations to preserve the mathematical statements in the
paper.

\subsection{Tracking proof tasks}
\label{sec:tracking-task-tree}

To represent the dependency structure of the formalization, we organized
pending proof obligations into hierarchical tracking issues.
Initially, each tracking issue listed its pending subtasks as Markdown
checkboxes in the issue description.
Because editing a Markdown checklist did not update the underlying issue
hierarchy, checklist entries could diverge from the state of the child issues.
We resolved this mismatch by migrating the task hierarchy to GitHub native
sub-issues: the parent issue description retained the mathematical overview of
the task, while native parent--child links alone defined the sub-issue
dependency graph.

The tracking issue for the quantum-soundness theorem (\mipissue{422})
illustrates this hierarchy.
Its description lists five sub-issues for assembling the theorem
(\mipissue{423}--\mipissue{427}), whereas the native sub-issue graph contains
17 child issues added as the proof required auxiliary lemmas and cross-module
bridge declarations.

For each parent tracking issue, GitHub computes completion as the fraction of
closed sub-issues, recording each child's \code{CLOSED} state separately from
its closure reason.
Completing a proof task also requires kernel typechecking in Lean and review
verifying that the declaration signature states the required result.
The tracking workflow counts closed sub-issues and posts a comment when a
triggering event finds every child closed, while keeping closure of the parent
issue a separate action.
At the audited August revision, the job runs after reopening a child, closing a
child as completed, or opening or merging a non-fork pull request; closing a
child as not planned does not itself trigger this job.

\subsection{From body checklists to native sub-issues}
\label{sec:tracking-history}

\begin{figure}
  \centering
  \begin{tikzpicture}[
      panel/.style={draw=cbInk!55, fill=white, rounded corners=2pt,
        line width=0.55pt},
      oldpanel/.style={panel, draw=cbRed!70!black, fill=cbRed!2},
      newpanel/.style={panel, draw=cbTeal!75!black, fill=cbTeal!2},
      heading/.style={font=\small\bfseries, text=cbInk},
      copy/.style={font=\footnotesize, text=cbInk, align=left},
      muted/.style={font=\footnotesize, text=cbInk!62, align=left},
      oldtag/.style={draw=cbRed!65!black, fill=cbRed!10,
        rounded corners=2pt, inner xsep=4pt, inner ysep=2pt,
        font=\footnotesize\bfseries, text=cbRed!65!black},
      newtag/.style={draw=cbTeal!70!black, fill=cbTeal!10,
        rounded corners=2pt, inner xsep=4pt, inner ysep=2pt,
        font=\footnotesize\bfseries, text=cbTeal!55!black},
      closed/.style={circle, draw=cbInk!55, fill=cbInk!16,
        minimum size=4.2pt, inner sep=0pt},
      migration/.style={-{Stealth[length=4pt]}, draw=cbInk!65,
        line width=0.8pt},
    ]

    \node[oldpanel, minimum width=6.55cm, minimum height=7.05cm,
      anchor=south west] (old) at (0,0) {};
    \draw[cbInk!28, line width=0.4pt]
      ([yshift=-0.78cm]old.north west) --
      ([yshift=-0.78cm]old.north east);
    \node[heading, anchor=west] at ([xshift=0.28cm,yshift=-0.39cm]old.north west)
      {Issue \#422 on 23 April};
    \node[oldtag, anchor=east] at
      ([xshift=-0.26cm,yshift=-0.39cm]old.north east) {Issue body};

    \node[copy, anchor=west] at (0.30,5.87) {\textbf{Tracked tasks}};
    \node[muted, anchor=west] at (0.30,5.49) {\texttt{```[tasklist]}};
    \node[copy, anchor=west] at (0.30,5.11) {\texttt{\#\#\# Tasks}};

    \foreach \yy/\nn in
      {4.73/423,4.35/424,3.97/425,3.59/426,3.21/427,2.83/428} {
      \draw[cbInk!62, line width=0.45pt]
        (0.52,\yy-0.10) rectangle +(0.20,0.20);
      \node[copy, anchor=west] at (0.88,\yy) {\texttt{\#\nn}};
    }
    \foreach \yy/\nn in {2.45/470,2.07/471,1.69/509} {
      \draw[cbInk!62, fill=cbInk!12, line width=0.45pt]
        (0.52,\yy-0.10) rectangle +(0.20,0.20);
      \draw[cbInk!65, line width=0.5pt]
        (0.56,\yy) -- (0.61,\yy-0.05) -- (0.69,\yy+0.06);
      \node[copy, anchor=west, text=cbInk!60] at
        (0.88,\yy) {\texttt{\#\nn}};
    }
    \node[muted, anchor=west] at (0.30,1.31) {\texttt{```}};

    \draw[cbRed!60!black, line width=0.45pt]
      (0.30,0.78) -- (6.25,0.78);
    \node[copy, text=cbRed!65!black, anchor=west] at (0.42,0.48)
      {Tasks recorded in editable issue text};

    \draw[migration] (6.95,3.53) -- (9.05,3.53);
    \node[font=\footnotesize\bfseries, text=cbInk, align=center]
      at (8.00,4.06) {PR \#674};
    \node[font=\footnotesize, text=cbInk!65, align=center]
      at (8.00,3.07) {24 Apr 2026};

    \node[newpanel, minimum width=6.55cm, minimum height=7.05cm,
      anchor=south west] (new) at (9.45,0) {};
    \draw[cbInk!28, line width=0.4pt]
      ([yshift=-0.78cm]new.north west) --
      ([yshift=-0.78cm]new.north east);
    \node[heading, anchor=west] at ([xshift=0.28cm,yshift=-0.39cm]new.north west)
      {Issue \#422};
    \node[newtag, anchor=east] at
      ([xshift=-0.26cm,yshift=-0.39cm]new.north east) {17 / 17 completed};

    \node[copy, anchor=west] at (9.75,5.87)
      {\textbf{Sub-issues}\quad\textcolor{cbInk!60}{native relation}};

    \draw[cbInk!38, line width=0.55pt] (10.05,5.29) -- (10.05,1.52);
    \foreach \yy in {5.16,4.28,3.40,2.52,1.64} {
      \draw[cbInk!38, line width=0.55pt] (10.05,\yy) -- (10.42,\yy);
      \node[closed] at (10.56,\yy) {};
    }
    \node[copy, anchor=west] at (10.78,5.16)
      {\texttt{\#423--\#427}\quad direct assembly (5)};
    \node[copy, anchor=west] at (10.78,4.28)
      {\texttt{\#1037}\quad recursive-slice follow-up};
    \node[copy, anchor=west] at (10.78,3.40)
      {\texttt{\#1038}\quad answer-side bridge};
    \node[copy, anchor=west] at (10.78,2.52)
      {\texttt{\#1507}\quad final induction obligation};
    \node[muted, anchor=west] at (10.78,1.64)
      {+ 9 further native children};

    \draw[cbTeal!65!black, line width=0.45pt]
      (9.75,0.78) -- (15.70,0.78);
    \node[copy, text=cbTeal!55!black, anchor=west] at (9.88,0.48)
      {Sub-issues define tasks; the body lists them};

    \node[font=\footnotesize\itshape, text=cbInk!55, anchor=south east]
      at (15.98,7.20) {Schematic interface view};
  \end{tikzpicture}
  \caption{\textbf{From body checklist to native sub-issues.}
    \textbf{\textsf{(Left)}}: the 23~April body tasklist stored six direct tasks
    (\#423--\#428) and three checked milestones (\#470, \#471, and \#509),
    so editing that text changed the displayed task list.
    \textbf{\textsf{(Right)}}: after \mippr{674}, issue~\#422's task state came from
    native parent--child relations; its description still listed the tasks.
    At the 31~August 2026 query, all 17 native children were
    closed on GitHub.
    Grey state markers denote GitHub closure, not mathematical completion.
    We reconstructed the task lists here and omitted the surrounding GitHub
    interface.}
  \label{fig:tracking-representation}
\end{figure}
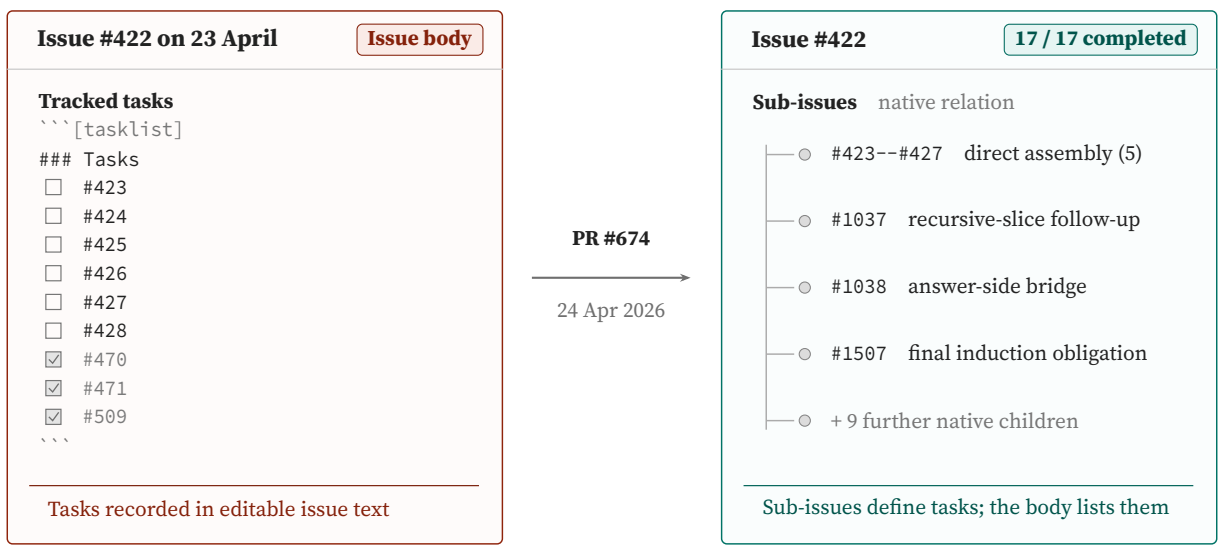

As the formalization progressed, we updated how the repository automation
recorded and synchronized child tasks.
\cref{fig:tracking-representation} reconstructs the Markdown checklist and
native sub-issue representations for tracker \mipissue{422}, and
\cref{tab:tracking-history} documents the workflow changes from the early
checklists through commit \PinnedHead{} to the August revision.

\FloatBarrier

\begin{table}
  \centering
  \caption{Changes to GitHub task tracking.
    The last column states the change in repository behavior.}
  \label{tab:tracking-history}
  \begin{tblr}{
    colspec = {X[1.1,l] X[1.2,l] X[1.35,l] X[4.35,l]},
    row{odd} = {bg=gray!5},
    row{1} = {bg=gray!25, font=\bfseries},
    rowsep = 3pt,
    colsep = 4pt,
  }
    \toprule
    Date & Reference & Task state & Repository behavior \\
    \midrule
    26 Mar 2026 &
      \Mipcommit{f9389508e} &
      Body checklist &
      The tracking-issue template stored child references as Markdown
      checkboxes. \\
    29 Mar 2026 &
      \Mipcommit{d6e3419f7} &
      Body checklist &
      The post-merge scan could append a new
      \code{- [ ] \#N} entry to that list. \\
    24 Apr 2026 &
      \mippr{674}; \mipcommit{4233ee532} &
      Native sub-issues &
      GitHub native sub-issues replaced Markdown checkboxes for tracking
      child tasks. \\
    24 Jun 2026 &
      \Mipcommit{b39705bfd} &
      Native sub-issues &
      At commit \PinnedHead{}, the merged-PR scan created follow-up
      issues, attached them to an open tracker, and logged the action. \\
    16--17 Jul 2026 &
      \Mipcommit{bb4889d0b};
      \mipcommit{f692c9b22};
      \mipcommit{1c6eed8d2};
      \mipcommit{442c15740} &
      Native sub-issues &
      The combined workflow kept native sub-issues, separated completion
      notices, and gave progress comments a stable marker. \\
    31 Aug 2026 &
      \Mipcommit{507e81220} &
      Native sub-issues &
      Separate scripts count closed children and find follow-up work in
      merged pull requests. \\
    \bottomrule
  \end{tblr}
\end{table}

When the repository's Claude review switch (\code{CLAUDE\_REVIEW\_ENABLED}) is
enabled, merging a non-fork pull request triggers an automated scan of its
description, review threads, and diff.
The workflow creates issues for follow-up obligations introduced by the merge,
excluding completed tasks, pre-existing \code{sorry} placeholders, and cosmetic
comments.

Across the project, automated post-merge follow-up scans created 110 issues
(out of 117 bot-created issues in total).
Combined with maintainer filings, follow-up tasks accounted for 292 of the 725
issues opened in the repository (40.3\%), concentrated in April and May 2026
during the main inductive proof formalization.
For each new obligation, the automation creates a native sub-issue under the
parent tracker, links the originating pull request, and posts an update on the
tracker.

In the August revision, a dedicated script counts closed children, updates
progress ratios, and posts state-transition comments with stable deduplication
markers.
Because the script reads native sub-issue relations and states directly, edits
to issue descriptions do not alter the sub-issue dependency graph or trigger
completion notices.

\subsection{Issue creation and organization through GitHub Actions}
\label{sec:issue-workflows}

When a new issue is opened, three GitHub Actions workflows classify the task,
audit existing lemmas in Mathlib and the local codebase, and synchronize
dependencies across the issue graph.

\paragraph*{Auto-labeling and anchor audit.}
When a contributor or agent opens an issue, the \emph{Issue Classification}
workflow evaluates its title and description against the repository taxonomy,
assigning area labels (\code{formalization}, \code{infrastructure},
\code{cleanup}), paper identifiers (\code{2009.12982}), and mathematical topic
labels (\code{ldt-basic}, \code{pasting}, \code{proof-infra}).
The triage agent also checks that the issue description specifies the required
mathematical anchors: theorem references in the paper, blueprint nodes, and
target Lean declaration names.
It posts an initial classification comment recording the applied labels,
summarizing the technical scope, and listing any missing paper or declaration
references needed before formalization begins.

\paragraph*{Automated Mathlib audit and scouting.}
For issues labeled \code{formalization}, the \emph{Mathlib Scout} workflow
audits Mathlib and the local Lean library.
The scouting agent searches for existing lemmas, matching type signatures, and
algebraic structures, separating results already available in Mathlib from
project-specific obligations that require new proofs.
It posts a structured scouting report on the issue listing the relevant
declarations, file paths, and a proposed lemma decomposition.

\paragraph*{Scouting reports as proving prompts.}
These automated triage and scouting comments serve as prompt context for
subsequent proving sessions.
When a proving session (dispatched through Claude Code or Codex) starts work on
an issue, the agent reads the issue description together with the classification
and scouting comments, using the identified Mathlib declarations and modular
decomposition to construct the proof without repeating library searches.

\paragraph*{Worked example: \mipissue{2454}.}
The two boxes below reproduce the automated
comments posted on \mipissue{2454}, a task to extract shared finite-support
lemmas in the pasting layer into a reusable interface.
Upon issue creation, the classification agent assigned topic labels and noted
the missing target declaration names (first box).
The Mathlib scout then searched Mathlib and the local repository, identifying
the exact embedding and factorial lemmas required and proposing a
three-declaration support API (second box).
The proving agent on \mippr{2455} ingested this report as prompt context and
implemented the three declarations proposed by the scout, closing
\mipissue{2454}.

\begin{promptbox}{Auto-labeling and initial classification comment on \mipissue{2454}}
\label{box:issue-classification}
\textbf{\#\# Initial classification}

\textbf{\#\#\# Labels}\\
Added \code{formalization}, \code{2009.12982}, \code{pasting}, \code{proof-infra}. The issue describes extracting shared finite-support lemmas from the pasting layer into a reusable API, which fits the formalization and proof-infrastructure categories.

\textbf{\#\#\# Reading}\\
The issue identifies a repeated pattern in the pasting layer: constructing the finite support of injective point tuples, proving its cardinality via \code{Equiv.subtypeInjectiveEquivEmbedding}, and proving nonemptiness when $k \le q$. The goal is to move these facts next to \code{distinctTupleDistribution} so that downstream total-variation and averaging proofs can cite a single support API.

The issue does not currently give specific file paths, line numbers, the current inline declaration names, or the target Lean declaration name for the factored support API.

\textbf{\#\#\# Next step}\\
Add the specific file paths and line numbers where the repeated support constructions currently live, and name the target Lean declaration (or declarations) that should hold the factored facts. If a blueprint anchor or LaTeX label corresponds to \code{distinctTupleDistribution}, include that reference as well.
\end{promptbox}

\begin{promptbox}{Mathlib scouting and audit report on \mipissue{2454}}
\label{box:mathlib-scout}
\textbf{\#\# Mathlib Scouting Report}

\textbf{\#\#\# Mathematical source}\\
$\bullet$ Refactoring task: source facts are already in the repository (\code{MIPStarRE/LDT/Pasting/Defs/Tuples.lean}, defining \code{distinctTuples} and \code{distinctTupleDistribution}) and downstream consumers in \code{Core/DDistinct.lean} and \code{ComparisonLemmas/LineInterpolation/}.\\
$\bullet$ Goal: Factor out the finite support of injective \code{PointTuple params k}, its cardinality \code{params.q.descFactorial k}, and nonemptiness ($k \le q$) into a shared API.

\textbf{\#\#\# Relevant Mathlib definitions and lemmas}\\
$\bullet$ \code{Equiv.subtypeInjectiveEquivEmbedding} (\code{Mathlib/Logic/Embedding/Basic.lean}): equivalence between injective functions and embeddings.\\
$\bullet$ \code{Fintype.card\_embedding\_eq} (\code{Mathlib/Data/Fintype/CardEmbedding.lean}): cardinality formula for embeddings between finite types.\\
$\bullet$ \code{Nat.descFactorial\_pos} (\code{Mathlib/Data/Nat/Factorial/Basic.lean}): gives $0 < n^{\underline{k}} \leftrightarrow k \le n$, establishing nonemptiness directly.

\textbf{\#\#\# Relevant project definitions}\\
$\bullet$ \code{MIPStarRE.LDT.Pasting.distinctTuples}: set of $k$-tuples with injective coordinates.\\
$\bullet$ \code{MIPStarRE.LDT.Pasting.ldDnoteq}: main total-variation bound $\mathrm{TV}(\mathrm{uniform}, \mathrm{distinct}) \le k^2/q$.

\textbf{\#\#\# Suggested approach}\\
Create a shared support API in \code{Tuples.lean} with three declarations:
\begin{enumerate}[label=(\roman*)]
  \item \code{distinctTupleSupport params k}: \code{Finset.univ.filter (Function.Injective)} as a named definition.
  \item \code{card\_distinctTupleSupport}: cardinality equals \code{params.q.descFactorial k}, proved via \code{subtypeInjectiveEquivEmbedding} and \code{card\_embedding\_eq}.
  \item \code{distinctTupleSupport\_nonempty\_iff}: nonemptiness equivalent to $k \le q$, proved via \code{Nat.descFactorial\_pos}.
\end{enumerate}

\textbf{\#\#\# Gaps to fill}\\
$\bullet$ No Mathlib gap: all required embedding and factorial lemmas already exist in Mathlib.\\
$\bullet$ Project gap only: factoring the repeated local proofs into \code{Tuples.lean}.
\end{promptbox}

The \emph{Issue Tracker} workflow responded to issue completion or reopening
and to pull-request opening or merging.
At the pinned snapshot, an agent performed both tracking updates and post-merge
follow-up review when the review switch was enabled.
Its prompt required each new proof task to cite the corresponding result in the
paper, name the relevant Lean declarations, and state what remained to be
proved.
It also instructed the agent to inspect the sub-issue dependency graph and
recommend an unblocked task in an issue comment, taking account of the
downstream obligations that task would unblock.
The post-merge creation procedure appears in \cref{sec:tracking-history}, and
\cref{sec:tracking-pr-links} gives an example.

Two scheduled workflows generated repository-wide status reports without
modifying the issue hierarchy.
The \emph{Daily Standup Summary} workflow ran on weekdays or on demand and
asked an agent to summarize proved results, active proof strategies, open
problems, and dependencies in a dated issue, updating any existing report for
that date.
The scheduled \emph{Stale issue audit} workflow separately identified open
issues with outdated paper or file citations without editing the issues.

\subsection{Pull-request links and task decomposition}
\label{sec:tracking-pr-links}

Pull-request descriptions linked code changes to the sub-issue dependency graph
using keyword directives.
We used \code{Addresses \#N} when a pull request advanced an intermediate step,
and \code{Closes \#N} when it discharged the target obligation.
Merging a pull request with a closing keyword automatically updated the
sub-issue state on GitHub and advanced the parent tracker's completion count.

The decomposition of \mipissue{422} (the quantum-soundness tracker) illustrates
this progression (\cref{tab:tracking-pr-trace}).
When \mippr{1031} merged, the post-merge follow-up scan opened child issues
\mipissue{1037} and \mipissue{1038}, attaching both as native sub-issues under
\mipissue{422}.
Subsequent pull requests advanced or closed these child issues individually
until the final statement repair in \mippr{1789} closed the parent tracker.

\begin{table}
  \centering
  \caption{Task decomposition and pull-request relations in the history of
    tracker \mipissue{422}.}
  \label{tab:tracking-pr-trace}
  \begin{tblr}{
    colspec = {X[1.35,l] X[1.65,l] X[3.0,l]},
    row{odd} = {bg=gray!5},
    row{1} = {bg=gray!25, font=\bfseries},
    rowsep = 3pt,
    colsep = 4pt,
  }
    \toprule
    Reference & Relation & State change \\
    \midrule
    \Mipissue{1037}, \mipissue{1038} &
      Native children of \mipissue{422} &
      Both issues added as children of the tracker. \\
    \mippr{1045} &
      \code{Closes \#1038} &
      Merged PR closed the answer-side bridge task. \\
    \mippr{1218} &
      \code{Addresses \#1037} &
      Merged PR advanced the successor-slice task. \\
    \mippr{1221} &
      \code{Closes \#1037} &
      Merged PR closed that child issue. \\
    \mippr{1789} &
      \code{Closes \#422} &
      Final statement correction closed the parent tracker. \\
    \bottomrule
  \end{tblr}
\end{table}

Native sub-issues maintained the hierarchical decomposition of proof
obligations, while pull-request links recorded the commits that discharged each
sub-issue.

\subsection{Automated pull-request repair}
\label{sec:repository-repair}

Once a pull request was opened against a tracked issue, automated repair
workflows in GitHub Actions routed compiler diagnostics and pull-request review
findings back to proving agents.
Because each pull request preserved its branch state, review threads, and
commit history across separate agent invocations, verification and repair
proceeded asynchronously after the initiating proving session ended.

\paragraph*{Dispatch from repository events.}
At the pinned proof snapshot, the \emph{Auto Fix (Lean)} workflow routed
failures from continuous integration and pull-request review to specialized
repair jobs.
A failing \emph{Lean Action CI} run dispatched build repair, and a failing
\emph{Lint blueprint} run dispatched blueprint repair.
For mathematical review, a completed run of \emph{Claude Code Review (Lean)}
dispatched review repair if the pull request carried the
\code{auto-fix-claude} label, regardless of the review approval state.
Label-triggered dispatch queried the latest completed checks for the head
commit and collected active, unresolved review threads, skipping outdated
threads from earlier commits and avoiding redundant runs when a newer build had
already passed.
Dispatch was restricted to internal repository branches and could be disabled
repository-wide.

\paragraph*{Repair on the pull-request branch.}
The dispatcher passed the failed run identifier and log output to the build or
blueprint repair agent, or provided the review summary and unresolved comment
threads to the review repair agent.
Each repair job checked out the pull-request branch directly.
When multiple failure types triggered concurrently, the workflow serialized
execution in a fixed order (build, blueprint, and review) to prevent concurrent
push conflicts, while a branch-level concurrency group cancelled superseded
jobs whenever a newer commit arrived.

Repair prompts required local verification before pushing: Lean repairs
required \code{lake build}, while blueprint repairs required recompiling the
blueprint and validating declaration links.
The agent pushed directly to the pull-request branch with an automated repair
subject-line marker and summarized its edits in a pull-request comment.
For theorems formalizing results from the paper, repository instructions
forbade altering the declaration signature to bypass proof obligations; when an
agent could not prove the declared statement, it documented the mathematical
blocker so that build and review repairs fed into the proof-gap protocol
whenever a failure exposed a discrepancy.

\paragraph*{Continuation and stopping.}
To bound automated repair sequences, a guard action halted automation after
five consecutive commits carrying automated repair subject-line markers (such as
\code{[claude-review-fix]}), counting build, blueprint, and review repairs
together.
The counter inspected commit subject lines backward from \code{HEAD} and reset
whenever an unmarked commit subject appeared, bounding each contiguous
automated repair sequence on a branch.
To prevent mutual triggering between the review and repair workflows, the
review workflow skipped commits carrying a repair subject-line marker.
A repair job exited without invoking the model if all review threads had
already been resolved or if no code changes were required.
After an agent verified and pushed a repair commit, continuous-integration
workflows independently re-checked the branch on GitHub before merge.

\paragraph*{Case study: automated review and repair.}
\mippr{2340} retargeted blueprint dependencies to theorem formulations with
independent Hilbert spaces for Alice and Bob.
Automated pull-request review triggered two successive repair commits.
The first commit, \mipcommit{8ced559}, reorganized the corresponding gap note
and corrected its mathematical citations.
The second commit, \mipcommit{0fdcb46}, replaced an informal table of
declaration names with explicit mathematical statements comparing the
single-space and heterogeneous formulations, and updated downstream references.
Both commits carried the \code{[claude-review-fix]} subject-line marker.
A third pass verified that all review threads were satisfied and exited without
additional commits.
The blueprint build, blueprint--Lean synchronization check, and proof-debt
scanner all passed on the final head commit.

\subsection{Tool loops, goal mode, and proof-task completion}
\label{app:goal-loop}

Within interactive proving sessions, TeXRA maintained continuity across model
turns by storing an explicit session objective and injecting a continuation
prompt whenever the session would otherwise pause for user input.
At TeXRA commit \code{9114c3eb2d} (10 June 2026), the objective was a
structured text document specifying the target outcome, the proof approach, and
the verification checks required before completion.
The plan tool presented this objective for approval; approving it with goal
mode enabled activated automatic turn continuations.
For formalization tasks, the session objective identified the target proof
obligation, while the repository instructions specified the required
verification checks.

Whenever the goal remained active and no user or event message was queued,
TeXRA prompted the agent to continue, supplying the stored objective and
elapsed session time.
The continuation prompt began:
\begin{quote}
\small\raggedright\ttfamily
Autonomous objective active. Keep working until it is verifiably done.
Do not end your turn to summarize progress or hand back control; only
stop when the objective's end state is true and you have inspected real
evidence for it.
\end{quote}
The prompt further instructed the agent not to substitute a weaker or narrower
task, and to verify every requirement before ending the loop.

To terminate the loop, the agent invoked the plan tool's \code{complete}
command with a summary of the verification checks performed, which cleared the
active goal entry and halted automatic continuations.
The proof itself was verified through Lean compiler commands and comparison
against the paper.
If tool execution errored or was cancelled, TeXRA paused the active goal.

\paragraph*{Repository events and progress checks.}
\label{app:github-subscriptions}

To coordinate local sessions with asynchronous GitHub workflows, TeXRA
(\code{9114c3eb2d}) provided a \code{github\_subscription} tool that subscribed
a proving session to a repository, pull request, or issue.
TeXRA polled the GitHub REST API and delivered state changes to the subscribing
session as follow-up messages.
Pull-request subscriptions reported comments and reviews from accounts passing
a bot filter (which inspected the GitHub account type and the \code{[bot]}
login suffix, passing automated reviews posted under either bot or author
accounts), as well as failed continuous-integration checks with annotations,
merge-conflict changes, and completed check suites on the head commit.

Because the subscription tracked check completion separately from check
success on each commit, a passing rerun after an earlier failure generated a
new notification.
A pull-request subscription terminated automatically when the pull request
closed or merged, whereas an issue subscription remained active across closure
to notify the session if the issue was reopened.
TeXRA queued these repository events alongside user messages and delivered them
ahead of the goal continuation prompt, allowing the session to incorporate
asynchronous review and continuous-integration feedback while pursuing the
active objective.

At the end of a session, the orchestrator also invoked \code{progressCheck}, a
read-only subagent that inspected the active objective, subagent outputs, local
Git and pull-request state, and newly unblocked follow-up issues.
The \code{progressCheck} subagent returned a follow-up message recommending
whether to stop, continue the current task, select a newly unblocked task, or
request user clarification.

\paragraph*{Proof-task checks.}

At the pinned snapshot, the repository instructions required proving agents to
consult the LIDT paper first, then the blueprint, and then the Lean codebase.
For any theorem corresponding to a statement in the paper, agents were required
to preserve its hypotheses and conclusion and, after each edit, to compare the
Lean declaration signature against the paper to verify that no hypotheses had
been added or conclusions weakened.

Before opening or updating a pull request, agents typechecked the modified Lean
file, checked that no untracked \code{sorry} tokens remained (apart from
tracked placeholders explicitly permitted during paper-statement realignment),
and ran a full library build whenever imports or shared declarations changed,
followed by semantic comparison of the declaration against the requested
statement and the in-kernel axiom audit (\cref{app:theorem}).

Finally, agents linked the resulting pull request to the sub-issue dependency
graph using the keyword conventions in \cref{sec:tracking-pr-links}.

\subsection{Codebase metrics and project timeline}
\label{sec:metrics-artifact-process}

The issue and pull-request workflows above produced the completed formalization
at the pinned snapshot commit \PinnedHead{} (\PinnedDate).
At that commit, the library contains \NumLines{} lines of Lean across
\NumFiles{} source files, with zero \code{sorry} tokens and zero custom axioms.
\cref{tab:artifact-process-metrics} summarizes the library size and the
repository activity over the formalization timeline.

\begin{table}
  \centering
  \caption{Library and formalization metrics at proof completion (\PinnedDate).
    Commit counts include every ancestor reachable from the pinned snapshot
    commit (\texttt{\PinnedHead}).}
  \label{tab:artifact-process-metrics}
  \begin{tblr}{
    colspec = {l r},
    row{odd} = {bg=gray!5},
    row{1} = {bg=gray!25, font=\bfseries},
    rowsep = 3pt,
  }
    \toprule
    Measurement & Value \\
    \midrule
    Lines of Lean in the library & \NumLines \\
    Lean source files in the library & \NumFiles \\
    Library \code{sorry} tokens & 0 \\
    Explicit custom axioms & 0 \\
    Lean / Mathlib version & \NumToolchain \\
    \midrule[dashed]
    Formalization timeline & 7 March--24 June 2026 \\
    Calendar days / active commit days & \NumSpanDays{} / \NumActiveDays \\
    Commits reachable from the snapshot & \NumCommits \\
    Pull requests merged by 24 June 2026 & \NumPRLandings \\
    \bottomrule
  \end{tblr}
\end{table}

We extracted the commit count from all ancestors reachable from commit
\PinnedHead{}, determined active commit days from Git author timestamps, and
counted all pull requests merged into the repository through 24 June 2026.

\subsection{Account attribution and agent commit contributions}
\label{sec:division-of-labor}

To audit commit attribution across the \NumCommits{} commits reachable from
\PinnedHead{}, we grouped Git author identities across aliases and inspected
both the author field and \code{Co-Authored-By} commit trailers for agent
identifiers.
\cref{tab:division-of-labor} partitions the commit history into commits
recorded under human author accounts and commits recorded under dedicated
agent-named accounts, and separately reports how many author-account commits
carry an agent \code{Co-Authored-By} trailer.

\begin{table}
  \centering
  \caption{Account names and agent co-author markers at commit
    \texttt{\PinnedHead} (\PinnedDate).
    The last row counts author-account commits that name an agent as
    co-author; these commits are also included in the author-account row.
    Local agent sessions committed under host author accounts in the early
    days.}
  \label{tab:division-of-labor}
  \begin{tblr}{
    colspec = {l r},
    row{odd} = {bg=gray!5},
    row{1} = {bg=gray!25, font=\bfseries},
    rowsep = 3pt,
  }
    \toprule
    Account or marker & Commits \\
    \midrule
    All commits reachable from the snapshot & \NumCommits \\
    Author accounts & 2,314 \\
    Agent-named accounts & 464 \\
    Author-account commits with agent co-author trailers & 324 \\
    \bottomrule
  \end{tblr}
\end{table}

Agent-named accounts (464 commits) and agent co-author trailers on
author-account commits (324 commits) explicitly mark 788 commits (28.4\%).
Although Git metadata attributes the remaining 1,990 commits (and 2,314 commits
in total, or 83.3\%) to human author accounts without agent trailers, all
commits and all formal Lean code in the repository were generated by
language-model sessions.
During the initial phase of the project, interactive agent sessions running in
local developer terminals inherited the host machine's default Git credentials
and committed without agent trailers; dedicated bot identities and co-author
trailers were configured for the agent harnesses only in later stages.

\subsection{Model token usage and accounting}
\label{sec:token-usage}
\newcommand{\RecordedUsageRecords}{56,089}
\newcommand{\RecordedUsageTotalBillions}{30.121}
\newcommand{\RecordedUsageInputBillions}{30.035}
\newcommand{\RecordedUsageOutputMillions}{85.676}
\newcommand{\RecordedUsageCachePercent}{90.7}

\FloatBarrier
\begingroup
\let\usagefigure\figure
\let\usagetable\table
\renewcommand{\figure}[1][]{\usagefigure[H]}
\renewcommand{\table}[1][]{\usagetable[H]}

We measured model token usage across the \code{MIPStarRE} repository and the
older parent workspace that enclosed its checkout, which together encompass the
low-degree test formalization and related preparatory developments.
Combining TeXRA session exports, machine checkpoints, native Codex logs, and
OpenCode message records after deduplicating inherited history from forked
sessions yields \RecordedUsageRecords{} interactive usage records.
These records total \RecordedUsageTotalBillions{} billion tokens, comprising
\RecordedUsageInputBillions{} billion input tokens and
\RecordedUsageOutputMillions{} million output tokens.
Divided across the \NumLines{} lines of Lean in the completed library, this
volume corresponds to approximately $238{,}000$ tokens per line of verified
Lean code.
The high ratio of input to output tokens arises from the multi-turn proving
loop: each turn supplies the file context, Lake compiler diagnostics, and
blueprint declarations in the prompt, while the model returns short tactic
scripts or localized edits.
Aggregate counters from the Codex index are excluded from this total because
their summary logs omit the turn-level identifiers needed to deduplicate forked
sessions or resolve overlap with TeXRA exports; GitHub Actions workflow runs
from 1~April--4~June 2026 likewise did not retain token counters.

\cref{fig:token-usage-harnesses} shows token volume and metered API expenditure
across TeXRA harnesses, and \cref{fig:token-usage-native} shows token volume
for native Codex and OpenCode sessions.
Metered API expenditure recorded in TeXRA totaled \$10,181.77 USD, whereas
OpenCode accessed enterprise API endpoints without per-call cost fields and
native Codex ran under fixed-rate seat subscriptions.
\cref{tab:token-usage-model} disaggregates input and output tokens by model,
listing records without a logged model identifier in separate rows.

\begin{figure}[tbp]
  \centering
  \begingroup
  \input{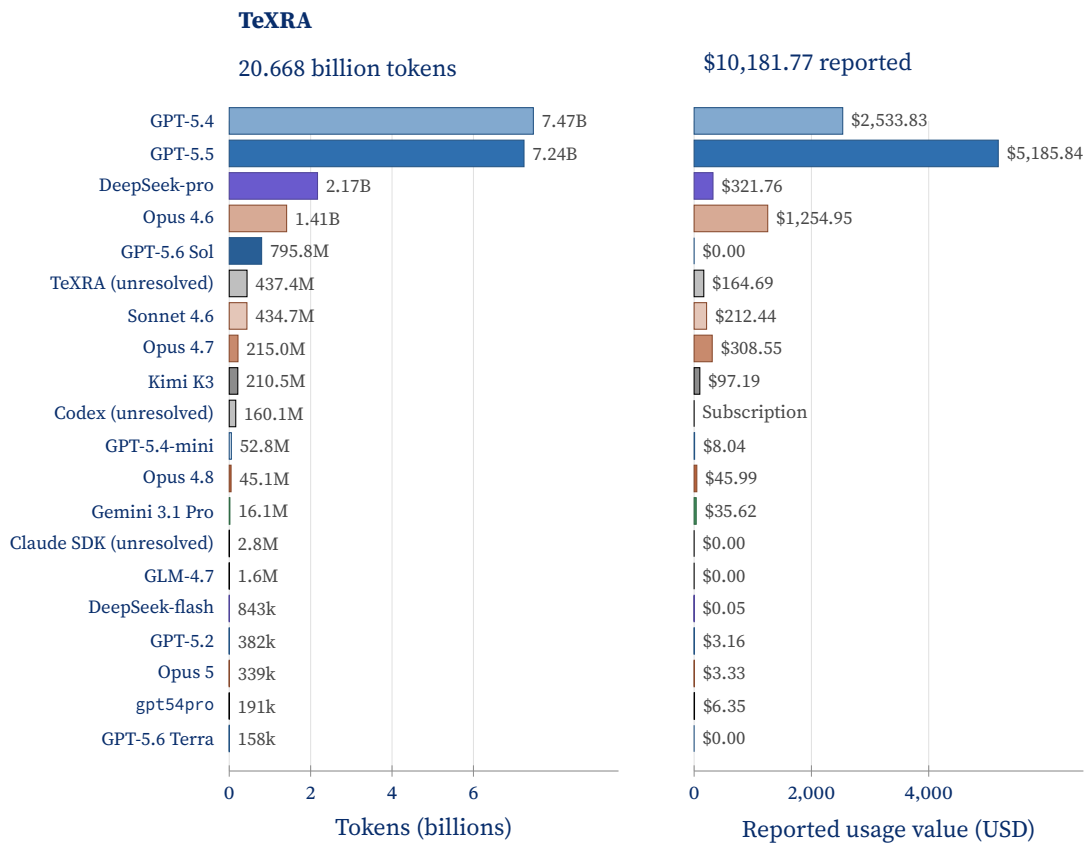}
  \begin{tikzpicture}
    \begin{axis}[
      name=texraTokens,
      width=5.15cm, height=8.6cm, scale only axis,
      xbar, bar width=10pt,
      title={20.668 billion tokens},
      title style={font=\small, at={(0,1.03)}, anchor=south west},
      y dir=reverse, enlarge y limits=0.06,
      xmin=0, xmax=9.56489, xtick={0,2,4,6},
      symbolic y coords={m0,m1,m2,m3,m4,m5,m6,m7,m8,m9,m10,m11,m12,m13,m14,m15,m16,m17,m18,m19,mpad}, ymax=mpad,
      ytick={m0,m1,m2,m3,m4,m5,m6,m7,m8,m9,m10,m11,m12,m13,m14,m15,m16,m17,m18,m19}, yticklabels={{GPT-5.4},{GPT-5.5},{DeepSeek-pro},{Opus 4.6},{GPT-5.6 Sol},{TeXRA (unresolved)},{Sonnet 4.6},{Opus 4.7},{Kimi K3},{Codex (unresolved)},{GPT-5.4-mini},{Opus 4.8},{Gemini 3.1 Pro},{Claude SDK (unresolved)},{GLM-4.7},{DeepSeek-flash},{GPT-5.2},{Opus 5},{\code{gpt54pro}},{GPT-5.6 Terra}},
      axis line style={thin, cUsageAxis},
      xtick style={color=cUsageAxis},
      ytick style={color=cUsageAxis},
      xticklabel style={font=\scriptsize},
      yticklabel style={font=\scriptsize},
      xlabel style={font=\small},
      ylabel style={font=\small},
      grid style={line width=0.3pt, cUsageGrid},
      axis x line=bottom,
      axis y line=left,
      x axis line style={-},
      y axis line style={draw=none},
      ytick style={draw=none},
      yticklabel style={align=right},
      xmajorgrids=true,
      scaled x ticks=false,
      clip=false,
      xlabel={Tokens (billions)},
      point meta=explicit symbolic, nodes near coords,
      nodes near coords align={horizontal},
      every node near coord/.append style={anchor=west, font=\scriptsize, text=cUsageLabel, inner xsep=3pt},
    ]
      \addplot+[forget plot, bar shift=0pt, mark=none,
        fill=cGPT54, draw=cOAIStroke]
        coordinates {(7.472568737,m0) [{7.47B}]};
      \addplot+[forget plot, bar shift=0pt, mark=none,
        fill=cGPT55, draw=cOAIStroke]
        coordinates {(7.242415534,m1) [{7.24B}]};
      \addplot+[forget plot, bar shift=0pt, mark=none,
        fill=cDSpro, draw=cDSStroke]
        coordinates {(2.16912065,m2) [{2.17B}]};
      \addplot+[forget plot, bar shift=0pt, mark=none,
        fill=cOpus46, draw=cAntStroke]
        coordinates {(1.410056478,m3) [{1.41B}]};
      \addplot+[forget plot, bar shift=0pt, mark=none,
        fill=cGPT56, draw=cOAIStroke]
        coordinates {(0.795812904,m4) [{795.8M}]};
      \addplot+[forget plot, bar shift=0pt, mark=none,
        fill=cUnknown, draw=cNeutralStroke]
        coordinates {(0.437405421,m5) [{437.4M}]};
      \addplot+[forget plot, bar shift=0pt, mark=none,
        fill=cSonnet46, draw=cAntStroke]
        coordinates {(0.434720584,m6) [{434.7M}]};
      \addplot+[forget plot, bar shift=0pt, mark=none,
        fill=cOpus47, draw=cAntStroke]
        coordinates {(0.21497506,m7) [{215.0M}]};
      \addplot+[forget plot, bar shift=0pt, mark=none,
        fill=cKimi3, draw=cNeutralStroke]
        coordinates {(0.210540358,m8) [{210.5M}]};
      \addplot+[forget plot, bar shift=0pt, mark=none,
        fill=cUnknown, draw=cNeutralStroke]
        coordinates {(0.160057496,m9) [{160.1M}]};
      \addplot+[forget plot, bar shift=0pt, mark=none,
        fill=cGPT54mini, draw=cOAIStroke]
        coordinates {(0.052751672,m10) [{52.8M}]};
      \addplot+[forget plot, bar shift=0pt, mark=none,
        fill=cOpus48, draw=cAntStroke]
        coordinates {(0.045145537,m11) [{45.1M}]};
      \addplot+[forget plot, bar shift=0pt, mark=none,
        fill=cGem31Pro, draw=cGOStroke]
        coordinates {(0.016076058,m12) [{16.1M}]};
      \addplot+[forget plot, bar shift=0pt, mark=none,
        fill=cUnknown, draw=cNeutralStroke]
        coordinates {(0.00276671,m13) [{2.8M}]};
      \addplot+[forget plot, bar shift=0pt, mark=none,
        fill=cGLM47, draw=cNeutralStroke]
        coordinates {(0.001591382,m14) [{1.6M}]};
      \addplot+[forget plot, bar shift=0pt, mark=none,
        fill=cDSflash, draw=cDSStroke]
        coordinates {(0.00084269,m15) [{843k}]};
      \addplot+[forget plot, bar shift=0pt, mark=none,
        fill=cGPT52, draw=cOAIStroke]
        coordinates {(0.00038236,m16) [{382k}]};
      \addplot+[forget plot, bar shift=0pt, mark=none,
        fill=cOpus5, draw=cAntStroke]
        coordinates {(0.000338539,m17) [{339k}]};
      \addplot+[forget plot, bar shift=0pt, mark=none,
        fill=cUnknown, draw=cNeutralStroke]
        coordinates {(0.000191346,m18) [{191k}]};
      \addplot+[forget plot, bar shift=0pt, mark=none,
        fill=cGPT56mini, draw=cOAIStroke]
        coordinates {(0.000157548,m19) [{158k}]};
    \end{axis}
    \begin{axis}[
      name=texraCost, at={(texraTokens.south east)}, anchor=south west, xshift=10mm,
      width=5.15cm, height=8.6cm, scale only axis,
      xbar, bar width=10pt,
      title={\$10,181.77 reported},
      title style={font=\small, at={(0,1.03)}, anchor=south west},
      y dir=reverse, enlarge y limits=0.06,
      xmin=0, xmax=6637.87, xtick={0,2000,4000},
      symbolic y coords={m0,m1,m2,m3,m4,m5,m6,m7,m8,m9,m10,m11,m12,m13,m14,m15,m16,m17,m18,m19,mpad}, ymax=mpad,
      ytick={m0,m1,m2,m3,m4,m5,m6,m7,m8,m9,m10,m11,m12,m13,m14,m15,m16,m17,m18,m19}, yticklabels=\empty,
      axis line style={thin, cUsageAxis},
      xtick style={color=cUsageAxis},
      ytick style={color=cUsageAxis},
      xticklabel style={font=\scriptsize},
      yticklabel style={font=\scriptsize},
      xlabel style={font=\small},
      ylabel style={font=\small},
      grid style={line width=0.3pt, cUsageGrid},
      axis x line=bottom,
      axis y line=left,
      x axis line style={-},
      y axis line style={draw=none},
      ytick style={draw=none},
      yticklabel style={align=right},
      xmajorgrids=true,
      scaled x ticks=false,
      clip=false,
      xlabel={Reported usage value (USD)},
      point meta=explicit symbolic, nodes near coords,
      nodes near coords align={horizontal},
      every node near coord/.append style={anchor=west, font=\scriptsize, text=cUsageLabel, inner xsep=3pt},
    ]
      \addplot+[forget plot, bar shift=0pt, mark=none,
        fill=cGPT54, draw=cOAIStroke]
        coordinates {(2533.82913050000006473,m0) [{\$2,533.83}]};
      \addplot+[forget plot, bar shift=0pt, mark=none,
        fill=cGPT55, draw=cOAIStroke]
        coordinates {(5185.8359129999992413,m1) [{\$5,185.84}]};
      \addplot+[forget plot, bar shift=0pt, mark=none,
        fill=cDSpro, draw=cDSStroke]
        coordinates {(321.762343481999867364,m2) [{\$321.76}]};
      \addplot+[forget plot, bar shift=0pt, mark=none,
        fill=cOpus46, draw=cAntStroke]
        coordinates {(1254.95001500000017934,m3) [{\$1,254.95}]};
      \addplot+[forget plot, bar shift=0pt, mark=none,
        fill=cGPT56, draw=cOAIStroke]
        coordinates {(0,m4) [{\$0.00}]};
      \addplot+[forget plot, bar shift=0pt, mark=none,
        fill=cUnknown, draw=cNeutralStroke]
        coordinates {(164.691712,m5) [{\$164.69}]};
      \addplot+[forget plot, bar shift=0pt, mark=none,
        fill=cSonnet46, draw=cAntStroke]
        coordinates {(212.43899999999996527,m6) [{\$212.44}]};
      \addplot+[forget plot, bar shift=0pt, mark=none,
        fill=cOpus47, draw=cAntStroke]
        coordinates {(308.5540000000000660,m7) [{\$308.55}]};
      \addplot+[forget plot, bar shift=0pt, mark=none,
        fill=cKimi3, draw=cNeutralStroke]
        coordinates {(97.1850000000000118,m8) [{\$97.19}]};
      \addplot+[forget plot, bar shift=0pt, mark=none,
        fill=cUnknown, draw=cNeutralStroke]
        coordinates {(0,m9) [{Subscription}]};
      \addplot+[forget plot, bar shift=0pt, mark=none,
        fill=cGPT54mini, draw=cOAIStroke]
        coordinates {(8.03599999999999511,m10) [{\$8.04}]};
      \addplot+[forget plot, bar shift=0pt, mark=none,
        fill=cOpus48, draw=cAntStroke]
        coordinates {(45.9859999999999963,m11) [{\$45.99}]};
      \addplot+[forget plot, bar shift=0pt, mark=none,
        fill=cGem31Pro, draw=cGOStroke]
        coordinates {(35.62200000000000258,m12) [{\$35.62}]};
      \addplot+[forget plot, bar shift=0pt, mark=none,
        fill=cUnknown, draw=cNeutralStroke]
        coordinates {(0,m13) [{\$0.00}]};
      \addplot+[forget plot, bar shift=0pt, mark=none,
        fill=cGLM47, draw=cNeutralStroke]
        coordinates {(0,m14) [{\$0.00}]};
      \addplot+[forget plot, bar shift=0pt, mark=none,
        fill=cDSflash, draw=cDSStroke]
        coordinates {(0.050000000000000024,m15) [{\$0.05}]};
      \addplot+[forget plot, bar shift=0pt, mark=none,
        fill=cGPT52, draw=cOAIStroke]
        coordinates {(3.1550000000000002,m16) [{\$3.16}]};
      \addplot+[forget plot, bar shift=0pt, mark=none,
        fill=cOpus5, draw=cAntStroke]
        coordinates {(3.326,m17) [{\$3.33}]};
      \addplot+[forget plot, bar shift=0pt, mark=none,
        fill=cUnknown, draw=cNeutralStroke]
        coordinates {(6.3469999999999996,m18) [{\$6.35}]};
      \addplot+[forget plot, bar shift=0pt, mark=none,
        fill=cGPT56mini, draw=cOAIStroke]
        coordinates {(0,m19) [{\$0.00}]};
    \end{axis}
    \node[anchor=south west, font=\small\bfseries, yshift=11mm]
      at (texraTokens.north west) {TeXRA};

  \end{tikzpicture}
  \endgroup
  \caption{Recorded model usage in TeXRA across the MIPStarRE workspaces
    and the older workspace that enclosed them.
    Token bars show input plus output; cached input is counted once.
    TeXRA includes API and subscription routes.
    Dollar bars show reported usage values, not subscription charges.
    Model colours match \cref{fig:token-usage-native}.}
  \label{fig:token-usage-harnesses}
\end{figure}
\begin{figure}[tbp]
  \centering
  \begingroup
  \input{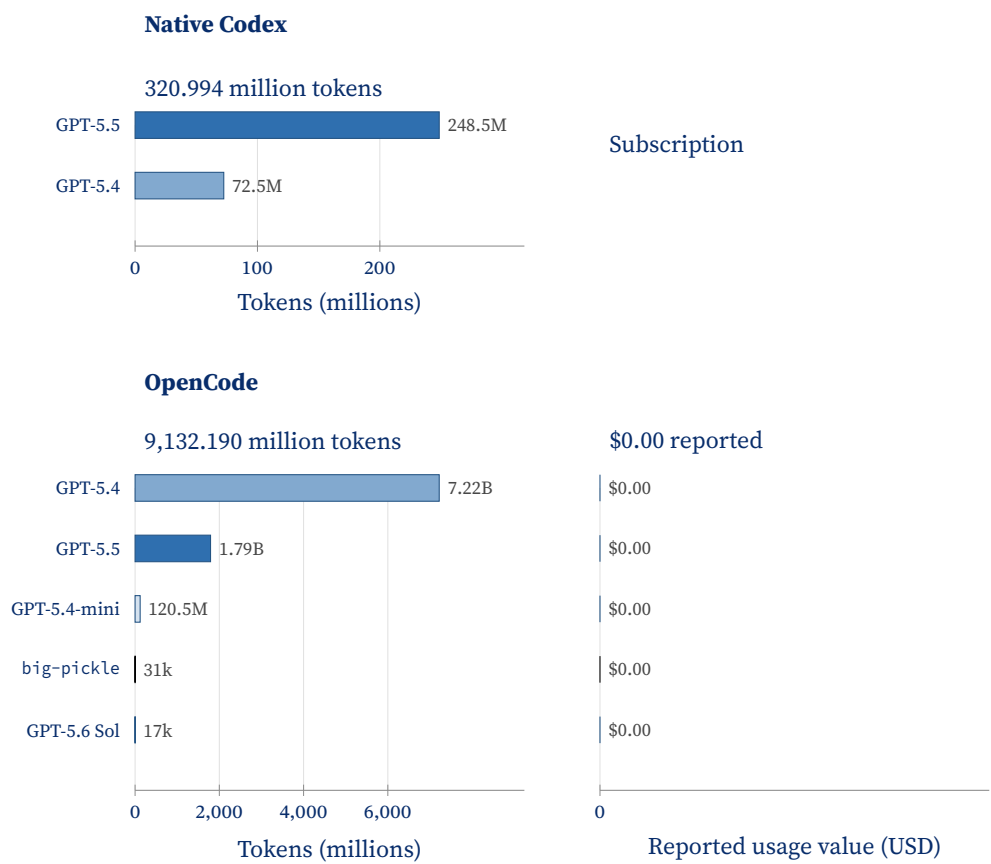}
  \begin{tikzpicture}
    \begin{axis}[
      name=nativeTokens0,
      width=5.15cm, height=1.6cm, scale only axis,
      xbar, bar width=10pt,
      title={320.994 million tokens},
      title style={font=\small, at={(0,1.03)}, anchor=south west},
      y dir=reverse, enlarge y limits=0.06,
      xmin=0, xmax=318.115, xtick={0,100,200},
      symbolic y coords={m0,m1,mpad}, ymax=mpad,
      ytick={m0,m1}, yticklabels={{GPT-5.5},{GPT-5.4}},
      axis line style={thin, cUsageAxis},
      xtick style={color=cUsageAxis},
      ytick style={color=cUsageAxis},
      xticklabel style={font=\scriptsize},
      yticklabel style={font=\scriptsize},
      xlabel style={font=\small},
      ylabel style={font=\small},
      grid style={line width=0.3pt, cUsageGrid},
      axis x line=bottom,
      axis y line=left,
      x axis line style={-},
      y axis line style={draw=none},
      ytick style={draw=none},
      yticklabel style={align=right},
      xmajorgrids=true,
      scaled x ticks=false,
      clip=false,
      xlabel={Tokens (millions)},
      point meta=explicit symbolic, nodes near coords,
      nodes near coords align={horizontal},
      every node near coord/.append style={anchor=west, font=\scriptsize, text=cUsageLabel, inner xsep=3pt},
    ]
      \addplot+[forget plot, bar shift=0pt, mark=none,
        fill=cGPT55, draw=cOAIStroke]
        coordinates {(248.527164,m0) [{248.5M}]};
      \addplot+[forget plot, bar shift=0pt, mark=none,
        fill=cGPT54, draw=cOAIStroke]
        coordinates {(72.466966,m1) [{72.5M}]};
    \end{axis}
    \node[anchor=south west, font=\small\bfseries, yshift=11mm]
      at (nativeTokens0.north west) {Native Codex};
    \node[anchor=north west, font=\small, xshift=10mm]
      at (nativeTokens0.north east) {Subscription};
    \begin{axis}[
      name=nativeTokens1, at={(nativeTokens0.south west)}, anchor=north west, yshift=-32mm,
      width=5.15cm, height=4.0cm, scale only axis,
      xbar, bar width=10pt,
      title={9,132.190 million tokens},
      title style={font=\small, at={(0,1.03)}, anchor=south west},
      y dir=reverse, enlarge y limits=0.06,
      xmin=0, xmax=9241.87, xtick={0,2000,4000,6000},
      symbolic y coords={m0,m1,m2,m3,m4,mpad}, ymax=mpad,
      ytick={m0,m1,m2,m3,m4}, yticklabels={{GPT-5.4},{GPT-5.5},{GPT-5.4-mini},{\code{big-pickle}},{GPT-5.6 Sol}},
      axis line style={thin, cUsageAxis},
      xtick style={color=cUsageAxis},
      ytick style={color=cUsageAxis},
      xticklabel style={font=\scriptsize},
      yticklabel style={font=\scriptsize},
      xlabel style={font=\small},
      ylabel style={font=\small},
      grid style={line width=0.3pt, cUsageGrid},
      axis x line=bottom,
      axis y line=left,
      x axis line style={-},
      y axis line style={draw=none},
      ytick style={draw=none},
      yticklabel style={align=right},
      xmajorgrids=true,
      scaled x ticks=false,
      clip=false,
      xlabel={Tokens (millions)},
      point meta=explicit symbolic, nodes near coords,
      nodes near coords align={horizontal},
      every node near coord/.append style={anchor=west, font=\scriptsize, text=cUsageLabel, inner xsep=3pt},
    ]
      \addplot+[forget plot, bar shift=0pt, mark=none,
        fill=cGPT54, draw=cOAIStroke]
        coordinates {(7220.208271,m0) [{7.22B}]};
      \addplot+[forget plot, bar shift=0pt, mark=none,
        fill=cGPT55, draw=cOAIStroke]
        coordinates {(1791.450717,m1) [{1.79B}]};
      \addplot+[forget plot, bar shift=0pt, mark=none,
        fill=cGPT54mini, draw=cOAIStroke]
        coordinates {(120.482352,m2) [{120.5M}]};
      \addplot+[forget plot, bar shift=0pt, mark=none,
        fill=cUnknown, draw=cNeutralStroke]
        coordinates {(0.0314,m3) [{31k}]};
      \addplot+[forget plot, bar shift=0pt, mark=none,
        fill=cGPT56, draw=cOAIStroke]
        coordinates {(0.016843,m4) [{17k}]};
    \end{axis}
    \node[anchor=south west, font=\small\bfseries, yshift=11mm]
      at (nativeTokens1.north west) {OpenCode};
    \begin{axis}[
      name=nativeCost1, at={(nativeTokens1.south east)}, anchor=south west, xshift=10mm,
      width=5.15cm, height=4.0cm, scale only axis,
      xbar, bar width=10pt,
      title={\$0.00 reported},
      title style={font=\small, at={(0,1.03)}, anchor=south west},
      y dir=reverse, enlarge y limits=0.06,
      xmin=0, xmax=0.001, xtick={0},
      symbolic y coords={m0,m1,m2,m3,m4,mpad}, ymax=mpad,
      ytick={m0,m1,m2,m3,m4}, yticklabels=\empty,
      axis line style={thin, cUsageAxis},
      xtick style={color=cUsageAxis},
      ytick style={color=cUsageAxis},
      xticklabel style={font=\scriptsize},
      yticklabel style={font=\scriptsize},
      xlabel style={font=\small},
      ylabel style={font=\small},
      grid style={line width=0.3pt, cUsageGrid},
      axis x line=bottom,
      axis y line=left,
      x axis line style={-},
      y axis line style={draw=none},
      ytick style={draw=none},
      yticklabel style={align=right},
      xmajorgrids=true,
      scaled x ticks=false,
      clip=false,
      xlabel={Reported usage value (USD)},
      point meta=explicit symbolic, nodes near coords,
      nodes near coords align={horizontal},
      every node near coord/.append style={anchor=west, font=\scriptsize, text=cUsageLabel, inner xsep=3pt},
    ]
      \addplot+[forget plot, bar shift=0pt, mark=none,
        fill=cGPT54, draw=cOAIStroke]
        coordinates {(0,m0) [{\$0.00}]};
      \addplot+[forget plot, bar shift=0pt, mark=none,
        fill=cGPT55, draw=cOAIStroke]
        coordinates {(0,m1) [{\$0.00}]};
      \addplot+[forget plot, bar shift=0pt, mark=none,
        fill=cGPT54mini, draw=cOAIStroke]
        coordinates {(0,m2) [{\$0.00}]};
      \addplot+[forget plot, bar shift=0pt, mark=none,
        fill=cUnknown, draw=cNeutralStroke]
        coordinates {(0,m3) [{\$0.00}]};
      \addplot+[forget plot, bar shift=0pt, mark=none,
        fill=cGPT56, draw=cOAIStroke]
        coordinates {(0,m4) [{\$0.00}]};
    \end{axis}

  \end{tikzpicture}
  \endgroup
  \caption{Recorded native model usage in the retained project sessions.
    Token bars show input plus output; cached input is counted once.
    OpenCode output includes recorded reasoning tokens.
    Native Codex was used through a subscription.
    Identified Codex sessions from TeXRA are excluded.
    Model colours match \cref{fig:token-usage-harnesses}; token counts here
    are in millions.
    Dollar bars show reported usage values.}
  \label{fig:token-usage-native}
\end{figure}
\begin{table}[htbp]
  \centering
  \caption{Recorded token usage in the MIPStarRE workspaces and the older
    workspace that enclosed them, combining retained TeXRA machine exports
    and recovered native Codex and OpenCode records.
    Token counts are in millions. Cached input is part of input;
    share is the fraction of all input and output tokens.}
  \label{tab:token-usage-model}
  \begin{tblr}{
    colspec = {X[3,l] X[1,r] X[1.35,r] X[0.7,r] X[1,r]},
    row{odd} = {bg=gray!5},
    row{1} = {bg=gray!25, font=\bfseries},
    rowsep = 3pt,
  }
    \toprule
    Model & Input & Cached input & Output & Share (\%) \\
    \midrule
    GPT-5.4 & 14,721.154 & 12,713.549 & 44.090 & 49.020 \\
    GPT-5.5 & 9,262.304 & 8,761.890 & 20.089 & 30.817 \\
    DeepSeek-pro & 2,162.273 & 2,143.377 & 6.847 & 7.201 \\
    Claude Opus 4.6 & 1,407.632 & 1,326.941 & 2.424 & 4.681 \\
    GPT-5.6 Sol & 793.349 & 705.352 & 2.481 & 2.642 \\
    TeXRA (model unrecorded) & 436.082 & 416.408 & 1.324 & 1.452 \\
    Claude Sonnet 4.6 & 432.370 & 420.189 & 2.351 & 1.443 \\
    Claude Opus 4.7 & 214.177 & 182.612 & 0.798 & 0.714 \\
    Kimi K3 & 208.834 & 204.026 & 1.706 & 0.699 \\
    GPT-5.4-mini & 171.567 & 155.870 & 1.667 & 0.575 \\
    Codex (model unrecorded) & 159.255 & 151.258 & 0.803 & 0.531 \\
    Claude Opus 4.8 & 44.753 & 43.214 & 0.393 & 0.150 \\
    Gemini 3.1 Pro & 15.710 & 13.613 & 0.366 & 0.053 \\
    Claude SDK (model unrecorded) & 2.723 & 2.519 & 0.044 & 0.009 \\
    GLM-4.7 & 1.579 & 1.512 & 0.013 & 0.005 \\
    DeepSeek-flash & 0.837 & 0.751 & 0.005 & 0.003 \\
    GPT-5.2 & 0.237 & 0.076 & 0.145 & 0.001 \\
    Claude Opus 5 & 0.274 & 0.000 & 0.065 & 0.001 \\
    \code{gpt54pro} & 0.189 & 0.000 & 0.003 & 0.001 \\
    GPT-5.6 Terra & 0.094 & 0.014 & 0.064 & 0.001 \\
    \code{big-pickle} & 0.031 & 0.015 & 0.000 & 0.000 \\
    \midrule
    Total & 30,035.425 & 27,243.188 & 85.676 & 100.000 \\
    \bottomrule
  \end{tblr}
\end{table}

\begin{figure}[tbp]
  \centering
  \begingroup
  \input{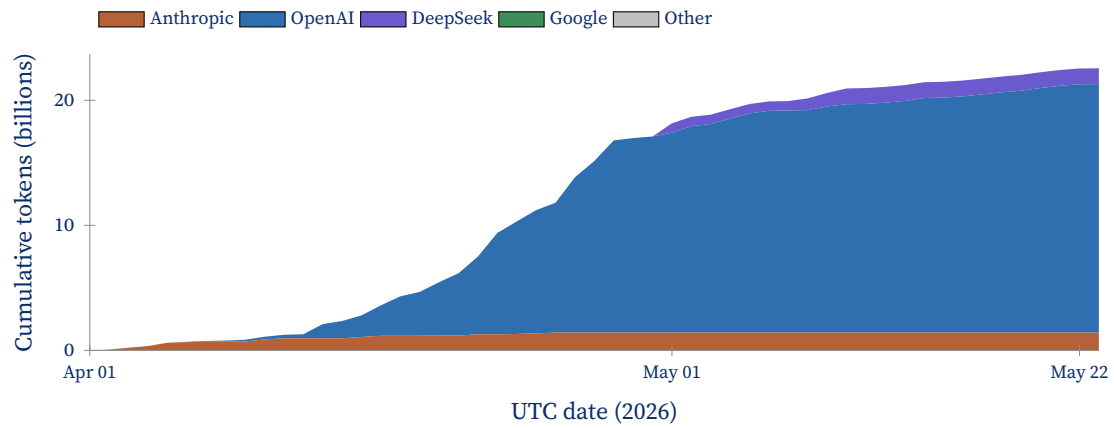}
  \begin{tikzpicture}
    \begin{axis}[
      width=0.9\linewidth, height=5.5cm,
      stack plots=y, area style, axis x line=bottom, axis y line=left,
      axis line style={thin, cUsageAxis},
      xtick style={color=cUsageAxis},
      ytick style={color=cUsageAxis},
      xticklabel style={font=\scriptsize},
      yticklabel style={font=\scriptsize},
      xlabel style={font=\small},
      ylabel style={font=\small},
      grid style={line width=0.3pt, cUsageGrid},
      x axis line style={-}, y axis line style={-},
      scaled y ticks=false, ylabel={Cumulative tokens (billions)},
      xlabel={UTC date (2026)},
      xmin=0, xmax=52, ymin=0,
      xtick={0,30,51},
      xticklabels={Apr 01,May 01,May 22},
      enlarge x limits=false, enlarge y limits={upper,value=0.05},
      legend entries={Anthropic,OpenAI,DeepSeek,Google,Other},
      legend columns=-1, legend style={at={(0,1.05)},anchor=south west,draw=none,fill=none,font=\scriptsize}
    ]
      \addplot[draw=none,fill=antbase] coordinates {(0,0.000000000) (1,0.062035995) (2,0.215931429) (3,0.346562397) (4,0.611742212) (5,0.656348159) (6,0.693828272) (7,0.693828272) (8,0.713538608) (9,0.877617444) (10,0.963943400) (11,0.963943400) (12,0.965578940) (13,0.965578940) (14,1.073682590) (15,1.170610440) (16,1.170610440) (17,1.170610440) (18,1.185089074) (19,1.188942283) (20,1.284349073) (21,1.284349073) (22,1.314369365) (23,1.361250191) (24,1.421206489) (25,1.421206489) (26,1.421206489) (27,1.421206489) (28,1.421206489) (29,1.421206489) (30,1.426708147) (31,1.426708147) (32,1.426708147) (33,1.426708147) (34,1.426708147) (35,1.427106868) (36,1.427106868) (37,1.427106868) (38,1.427106868) (39,1.427106868) (40,1.427106868) (41,1.427106868) (42,1.427106868) (43,1.427106868) (44,1.427106868) (45,1.427106868) (46,1.427106868) (47,1.427106868) (48,1.427106868) (49,1.427106868) (50,1.427106868) (51,1.427106868) (52,1.427106868)} \closedcycle;
      \addplot[draw=none,fill=oaibase] coordinates {(0,0.000000000) (1,0.000000000) (2,0.000000000) (3,0.000000000) (4,0.000000000) (5,0.032222840) (6,0.059401521) (7,0.088667353) (8,0.136282806) (9,0.212737969) (10,0.285260907) (11,0.324026392) (12,1.133759177) (13,1.383875963) (14,1.722521710) (15,2.429984768) (16,3.151105143) (17,3.504535564) (18,4.268381621) (19,4.979875297) (20,6.218442681) (21,8.104030499) (22,8.997755771) (23,9.853256067) (24,10.387802015) (25,12.423106379) (26,13.725772097) (27,15.370551204) (28,15.559933251) (29,15.683756985) (30,15.986299346) (31,16.504377872) (32,16.676016310) (33,17.113067992) (34,17.526920858) (35,17.733564168) (36,17.761245195) (37,17.789258655) (38,18.082698141) (39,18.257706206) (40,18.286437641) (41,18.385969161) (42,18.524706028) (43,18.754941486) (44,18.789341421) (45,18.891013423) (46,19.047927213) (47,19.210799998) (48,19.346702862) (49,19.555372609) (50,19.733676455) (51,19.852197589) (52,19.868515038)} \closedcycle;
      \addplot[draw=none,fill=dsbase] coordinates {(0,0.000000000) (1,0.000000000) (2,0.000000000) (3,0.000000000) (4,0.000000000) (5,0.000000000) (6,0.000000000) (7,0.000000000) (8,0.000000000) (9,0.000000000) (10,0.000000000) (11,0.000000000) (12,0.000000000) (13,0.000000000) (14,0.000000000) (15,0.000000000) (16,0.000000000) (17,0.000000000) (18,0.000000000) (19,0.000000000) (20,0.000000000) (21,0.000000000) (22,0.000000000) (23,0.000000000) (24,0.000000000) (25,0.000000000) (26,0.000000000) (27,0.000000000) (28,0.000000000) (29,0.000000000) (30,0.749964958) (31,0.749964958) (32,0.749964958) (33,0.749964958) (34,0.749964958) (35,0.749964958) (36,0.749964958) (37,0.939695496) (38,1.078792757) (39,1.264089445) (40,1.264089445) (41,1.264089445) (42,1.264089445) (43,1.264089445) (44,1.264089445) (45,1.264089445) (46,1.264089445) (47,1.264089445) (48,1.264089445) (49,1.264089445) (50,1.264089445) (51,1.264089445) (52,1.264089445)} \closedcycle;
      \addplot[draw=none,fill=gobase] coordinates {(0,0.000000000) (1,0.000000000) (2,0.000000000) (3,0.000000000) (4,0.000000000) (5,0.000000000) (6,0.000000000) (7,0.000000000) (8,0.000000000) (9,0.000000000) (10,0.000000000) (11,0.000000000) (12,0.000000000) (13,0.000000000) (14,0.000000000) (15,0.000000000) (16,0.000000000) (17,0.000000000) (18,0.000000000) (19,0.000000000) (20,0.000000000) (21,0.000000000) (22,0.000000000) (23,0.000000000) (24,0.000000000) (25,0.000000000) (26,0.000000000) (27,0.000000000) (28,0.000000000) (29,0.000000000) (30,0.000000000) (31,0.000000000) (32,0.000000000) (33,0.000041848) (34,0.000041848) (35,0.000041848) (36,0.000041848) (37,0.000041848) (38,0.000041848) (39,0.000041848) (40,0.000041848) (41,0.000041848) (42,0.000041848) (43,0.000041848) (44,0.000041848) (45,0.000041848) (46,0.000041848) (47,0.000041848) (48,0.000041848) (49,0.000041848) (50,0.000041848) (51,0.000041848) (52,0.000041848)} \closedcycle;
      \addplot[draw=none,fill=cUnknown] coordinates {(0,0.000000000) (1,0.000000000) (2,0.000000000) (3,0.000000000) (4,0.000000000) (5,0.000000000) (6,0.000000000) (7,0.000000000) (8,0.000000000) (9,0.000000000) (10,0.000000000) (11,0.000000000) (12,0.000000000) (13,0.000000000) (14,0.000000000) (15,0.000000000) (16,0.000000000) (17,0.000000000) (18,0.000000000) (19,0.000000000) (20,0.000000000) (21,0.000000000) (22,0.000000000) (23,0.000000000) (24,0.000000000) (25,0.000000000) (26,0.000000000) (27,0.000000000) (28,0.000000000) (29,0.000000000) (30,0.000000000) (31,0.000000000) (32,0.000031400) (33,0.000031400) (34,0.000031400) (35,0.000031400) (36,0.000031400) (37,0.000031400) (38,0.000031400) (39,0.000031400) (40,0.000031400) (41,0.000031400) (42,0.000031400) (43,0.000031400) (44,0.000031400) (45,0.000031400) (46,0.000031400) (47,0.000031400) (48,0.000031400) (49,0.000031400) (50,0.000031400) (51,0.000031400) (52,0.000031400)} \closedcycle;
    \end{axis}
  \end{tikzpicture}
  \endgroup
  \caption{Recorded usage during formalization.
  Calls are selected through 4 June 2026, as in the proof-status figure.
  Retained call dates cover 1 April to 22 May 2026.
  The bands accumulate 22.56 billion input and output tokens
  from 54,916 records, assigned to an explicit call or message creation timestamp
  of each call (UTC).
  A further 6.49 billion recorded tokens have no retained
  call date and are omitted; calls dated outside the formalization period
  are also excluded.}
  \label{fig:token-usage-timeline}
\end{figure}
\input{figures/fig_token_usage_roles}
\begin{figure}[tbp]
  \centering
  \begingroup
  \input{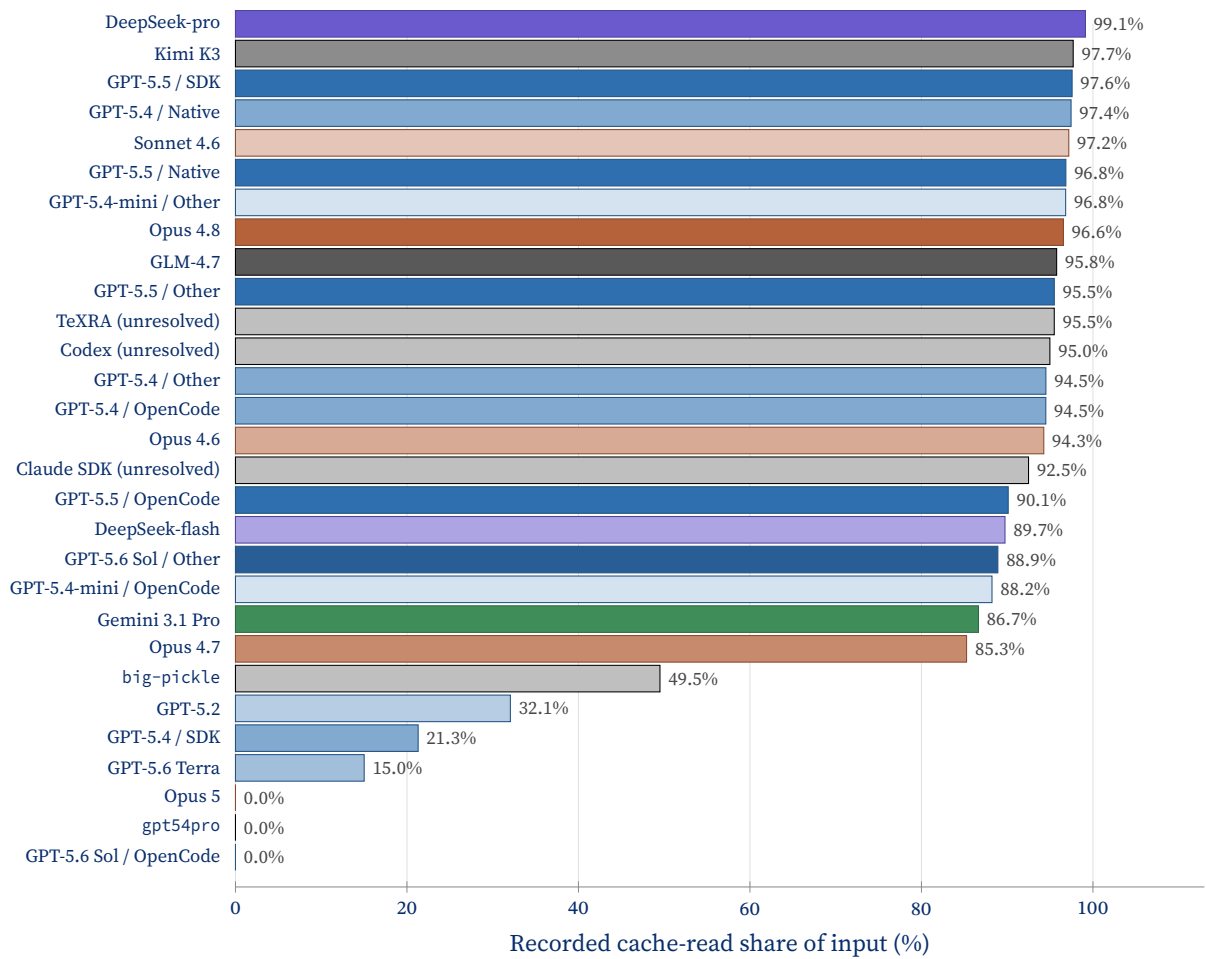}
  \begin{tikzpicture}
    \begin{axis}[
      width=14.4cm, height=13.0cm, xbar,
      bar width=10pt,
      y dir=reverse, enlarge y limits=0.035,
      xmin=0, xmax=113, xtick={0,20,40,60,80,100},
axis line style={thin, cUsageAxis},
      xtick style={color=cUsageAxis},
      ytick style={color=cUsageAxis},
      xticklabel style={font=\scriptsize},
      yticklabel style={font=\scriptsize},
      xlabel style={font=\small},
      ylabel style={font=\small},
      grid style={line width=0.3pt, cUsageGrid},
      axis x line=bottom,
      axis y line=left,
      x axis line style={-},
      y axis line style={draw=none},
      ytick style={draw=none},
      yticklabel style={align=right},
      xmajorgrids=true,
      scaled x ticks=false,
      clip=false,
      xlabel={Recorded cache-read share of input (\%)},
      symbolic y coords={m0,m1,m2,m3,m4,m5,m6,m7,m8,m9,m10,m11,m12,m13,m14,m15,m16,m17,m18,m19,m20,m21,m22,m23,m24,m25,m26,m27,m28,mpad},
      ymax=mpad,
      ytick={m0,m1,m2,m3,m4,m5,m6,m7,m8,m9,m10,m11,m12,m13,m14,m15,m16,m17,m18,m19,m20,m21,m22,m23,m24,m25,m26,m27,m28},
      yticklabels={{DeepSeek-pro},{Kimi K3},{GPT-5.5 / SDK},{GPT-5.4 / Native},{Sonnet 4.6},{GPT-5.5 / Native},{GPT-5.4-mini / Other},{Opus 4.8},{GLM-4.7},{GPT-5.5 / Other},{TeXRA (unresolved)},{Codex (unresolved)},{GPT-5.4 / Other},{GPT-5.4 / OpenCode},{Opus 4.6},{Claude SDK (unresolved)},{GPT-5.5 / OpenCode},{DeepSeek-flash},{GPT-5.6 Sol / Other},{GPT-5.4-mini / OpenCode},{Gemini 3.1 Pro},{Opus 4.7},{\code{big-pickle}},{GPT-5.2},{GPT-5.4 / SDK},{GPT-5.6 Terra},{Opus 5},{\code{gpt54pro}},{GPT-5.6 Sol / OpenCode}},
      point meta=explicit symbolic, nodes near coords,
      nodes near coords align={horizontal},
      every node near coord/.append style={anchor=west, font=\scriptsize, text=cUsageLabel, inner xsep=3pt},
    ]
      \addplot+[forget plot, bar shift=0pt, mark=none,
        fill=cDSpro, draw=cDSStroke]
        coordinates {(99.126104722,m0) [{99.1\%}]};
      \addplot+[forget plot, bar shift=0pt, mark=none,
        fill=cKimi3, draw=cNeutralStroke]
        coordinates {(97.697903740,m1) [{97.7\%}]};
      \addplot+[forget plot, bar shift=0pt, mark=none,
        fill=cGPT55, draw=cOAIStroke]
        coordinates {(97.570953858,m2) [{97.6\%}]};
      \addplot+[forget plot, bar shift=0pt, mark=none,
        fill=cGPT54, draw=cOAIStroke]
        coordinates {(97.448898736,m3) [{97.4\%}]};
      \addplot+[forget plot, bar shift=0pt, mark=none,
        fill=cSonnet46, draw=cAntStroke]
        coordinates {(97.182813142,m4) [{97.2\%}]};
      \addplot+[forget plot, bar shift=0pt, mark=none,
        fill=cGPT55, draw=cOAIStroke]
        coordinates {(96.841339192,m5) [{96.8\%}]};
      \addplot+[forget plot, bar shift=0pt, mark=none,
        fill=cGPT54mini, draw=cOAIStroke]
        coordinates {(96.816879519,m6) [{96.8\%}]};
      \addplot+[forget plot, bar shift=0pt, mark=none,
        fill=cOpus48, draw=cAntStroke]
        coordinates {(96.561651726,m7) [{96.6\%}]};
      \addplot+[forget plot, bar shift=0pt, mark=none,
        fill=cGLM47, draw=cNeutralStroke]
        coordinates {(95.767685425,m8) [{95.8\%}]};
      \addplot+[forget plot, bar shift=0pt, mark=none,
        fill=cGPT55, draw=cOAIStroke]
        coordinates {(95.493123046,m9) [{95.5\%}]};
      \addplot+[forget plot, bar shift=0pt, mark=none,
        fill=cUnknown, draw=cNeutralStroke]
        coordinates {(95.488564539,m10) [{95.5\%}]};
      \addplot+[forget plot, bar shift=0pt, mark=none,
        fill=cUnknown, draw=cNeutralStroke]
        coordinates {(94.978365894,m11) [{95.0\%}]};
      \addplot+[forget plot, bar shift=0pt, mark=none,
        fill=cGPT54, draw=cOAIStroke]
        coordinates {(94.512319547,m12) [{94.5\%}]};
      \addplot+[forget plot, bar shift=0pt, mark=none,
        fill=cGPT54, draw=cOAIStroke]
        coordinates {(94.502391181,m13) [{94.5\%}]};
      \addplot+[forget plot, bar shift=0pt, mark=none,
        fill=cOpus46, draw=cAntStroke]
        coordinates {(94.267576274,m14) [{94.3\%}]};
      \addplot+[forget plot, bar shift=0pt, mark=none,
        fill=cUnknown, draw=cNeutralStroke]
        coordinates {(92.488954777,m15) [{92.5\%}]};
      \addplot+[forget plot, bar shift=0pt, mark=none,
        fill=cGPT55, draw=cOAIStroke]
        coordinates {(90.119127745,m16) [{90.1\%}]};
      \addplot+[forget plot, bar shift=0pt, mark=none,
        fill=cDSflash, draw=cDSStroke]
        coordinates {(89.744055182,m17) [{89.7\%}]};
      \addplot+[forget plot, bar shift=0pt, mark=none,
        fill=cGPT56, draw=cOAIStroke]
        coordinates {(88.910026829,m18) [{88.9\%}]};
      \addplot+[forget plot, bar shift=0pt, mark=none,
        fill=cGPT54mini, draw=cOAIStroke]
        coordinates {(88.227780408,m19) [{88.2\%}]};
      \addplot+[forget plot, bar shift=0pt, mark=none,
        fill=cGem31Pro, draw=cGOStroke]
        coordinates {(86.650166153,m20) [{86.7\%}]};
      \addplot+[forget plot, bar shift=0pt, mark=none,
        fill=cOpus47, draw=cAntStroke]
        coordinates {(85.262194107,m21) [{85.3\%}]};
      \addplot+[forget plot, bar shift=0pt, mark=none,
        fill=cUnknown, draw=cNeutralStroke]
        coordinates {(49.505811146,m22) [{49.5\%}]};
      \addplot+[forget plot, bar shift=0pt, mark=none,
        fill=cGPT52, draw=cOAIStroke]
        coordinates {(32.060045118,m23) [{32.1\%}]};
      \addplot+[forget plot, bar shift=0pt, mark=none,
        fill=cGPT54, draw=cOAIStroke]
        coordinates {(21.305247237,m24) [{21.3\%}]};
      \addplot+[forget plot, bar shift=0pt, mark=none,
        fill=cGPT56mini, draw=cOAIStroke]
        coordinates {(15.011781263,m25) [{15.0\%}]};
      \addplot+[forget plot, bar shift=0pt, mark=none,
        fill=cOpus5, draw=cAntStroke]
        coordinates {(0.000000000,m26) [{0.0\%}]};
      \addplot+[forget plot, bar shift=0pt, mark=none,
        fill=cUnknown, draw=cNeutralStroke]
        coordinates {(0.000000000,m27) [{0.0\%}]};
      \addplot+[forget plot, bar shift=0pt, mark=none,
        fill=cGPT56, draw=cOAIStroke]
        coordinates {(0.000000000,m28) [{0.0\%}]};
    \end{axis}
  \end{tikzpicture}
  \endgroup
  \caption{Prompt caching in the combined workspace records.
    Each rate is the sum of recorded cache-read tokens divided by the
    sum of input tokens for that model and route.
    We separate recorded routes when a model appears in multiple
    harnesses, including OpenCode.
    These labels identify the route, not an additional model.
    Colours match \cref{fig:token-usage-harnesses}.
    }
  \label{fig:token-cache-rates}
\end{figure}

\cref{fig:token-usage-timeline} plots daily token volume by model provider for
dated records between 7 March and 4 June 2026; the workspace totals in
\cref{tab:token-usage-model} also include undated records and sessions outside
that date window.
\cref{fig:token-usage-roles} partitions token volume and metered cost by agent
role identifier and native client.
Across all providers, prompt cache hits supplied
\RecordedUsageCachePercent{}\% of all input tokens.
\cref{fig:token-cache-rates} compares the input-weighted cache hit rate across
models and execution harnesses.
\FloatBarrier
\endgroup

The QPBT run (\cref{sec:second-theorem}) used 13.7 billion tokens, including 2.7
billion in the orchestrating GPT-6 Astra sessions and, in subagent sessions, 3.1 billion
on GPT-6 Astra, 5.6 billion on GPT-5.6 Sol, and 2.2 billion on Claude Opus 5.

\suppappendix{The review dataset}
\label{app:review-census}

\subsection{Population}
\label{sec:census-population}

The review corpus analyzed in the main text comprises the 21,651 GitHub review
reports and discussion comments retrieved on 27 July 2026 across the 1,904
closed pull requests and repository issues in \code{MIPStarRE}.
Using the GitHub REST API on 27 July 2026, we retrieved all review reports
across the 1,904 closed pull requests (numbered 1 to 2,615, opened between 20
March and 17 July 2026, with 1,816 merged), together with repository-wide
listings of inline review comments, pull-request discussion comments, and issue
comments; every pull request carrying discussion comments belongs to this
closed set.
Each record entered the inventory under a unique identifier combining its
record type (\cref{tab:census-population}) with its GitHub database ID,
retaining its body text, author login, GitHub account category, timestamp, and
target thread.
Inline review comments are partitioned into thread-opening root comments and
subsequent replies.
The inventory also includes 1,089 review-report records with submission state
``commented'' and an empty body field; because these entries contain no comment
text, we include them in the population totals and separately evaluate the
sensitivity of the category distribution when all 1,089 entries are routed to
Repository process and evidence.

\begin{table}[htb]
  \centering
  \caption{The 21,651 review comments and reports retrieved on 27 July 2026,
    by type and by the account type that GitHub assigns to the author.}
  \label{tab:census-population}
  \begin{tblr}{
    colspec = {l r r r},
    row{odd} = {bg=gray!5},
    row{1} = {bg=gray!25, font=\bfseries},
    rowsep = 3pt,
  }
    \toprule
    Record type & Count & Bot accounts & User accounts \\
    \midrule
    Review reports & 7,656 & 6,859 & 797 \\
    Inline review comments, roots & 6,426 & 6,426 & 0 \\
    Inline review comments, replies & 790 & 8 & 782 \\
    Pull-request discussion comments & 2,675 & 1,208 & 1,467 \\
    Issue comments outside pull requests & 4,104 & 2,531 & 1,573 \\
    \midrule
    Total & 21,651 & 17,032 & 4,619 \\
    \bottomrule
  \end{tblr}
\end{table}

Because the inventory counts each review action separately, a single
mathematical issue typically spans multiple entries across its thread: an
inline root comment, an author or repair-agent reply, a review report, and a
subsequent approval.
Automated review agents operating under dedicated bot identities (Claude Code,
GitHub Copilot, Cursor Bugbot, and the Codex connector) authored all 6,426
inline root comments.
Conversely, the 782 inline replies recorded under user accounts were posted by
automated proving and repair agents running under developer GitHub credentials
(\cref{sec:division-of-labor}); human authors wrote no Lean code and used issue
and pull-request comments for high-level steering and review.

\subsection{Reading and classification}
\label{sec:census-reading}

We classified the 21,651 review records in two stages: a structured reading
pass with GPT-5.4, followed by a deterministic term-matching classifier.
Because the Lean kernel verifies type correctness without checking whether
definitions and theorem signatures match the intended statements in the paper,
pull-request review concentrated on checking formal declarations and
intermediate proof obligations against the published proof.
To preserve conversational context across threads, we grouped the 21,651
records chronologically within each pull request or issue into 2,733 reading
units of at most 20 records each.

For each record in a reading unit, GPT-5.4 produced a claim summary, a
descriptive mechanism label and definition, and categorical annotations for
signal class, defect support, severity, follow-up resolution status, and
thematic relevance.
To verify that each annotation was grounded in the source text, the prompt
required exact reading receipts consisting of verbatim substrings extracted from
the comment body and from any supporting resolution reply.
An automated validation harness checked every response against the input
records and retried generations that omitted records, produced out-of-schema
labels, or failed exact-substring checks, yielding 15,773 distinct validated
mechanism labels across the archived dataset.

A deterministic classifier then mapped each annotated record into the three
primary categories and nine subcategories of the main-text table.
The classifier scores normalized text across the model's structured output
fields against curated term lists for each subcategory.
Each record is assigned to the subcategory with the highest match score,
breaking ties in favor of the first-listed subcategory in the taxonomy; the 117
records with zero domain matches default to Repository process and evidence.
Running the versioned classification script on the annotated ledger reproduces
all twelve category and subcategory counts in the main-text table.

To evaluate sensitivity to tie-breaking and empty-body review reports, we
recomputed the primary category shares under three alternative rules.
Top-score ties across primary categories occurred in 2,574 of the 21,651
records.
Mathematics and agreement with the paper receives 55.2\% of all records under
the default taxonomy ordering, 45.2\% when tie-breaking priority is reversed,
51.3\% when all 2,574 cross-category ties are excluded, and 52.2\% when all
1,089 textless review reports are routed to Repository process and evidence.
Across all four classification rules, Mathematics and agreement with the paper
remains the largest primary category.

We also compared the term-matching classifier against an independent
multi-model pilot on the inline root comments, in which GPT-5.6 Terra and Sol
classified the comments in thread context and Claude Opus 4.8 adjudicated
disagreements.
When projected onto the three primary categories, the pilot labels agreed with
the term classifier on 62.7\% of the mapped root comments (4,023 of 6,412).
The pilot assigned 37.2\% of root comments to Mathematics and agreement with
the paper versus 57.0\% under the term classifier, reflecting the broader
keyword lists in the deterministic script.

Finally, the follow-up status annotations connect review findings to subsequent
code repairs within the same discussion thread.
Across the 21,651 records, 3,953 records identified a concrete defect claim
(2,827) or a plausible risk claim (1,126).
Of these 3,953 findings, 713 (18.0\% of defect and risk claims, or 3.3\% of all
21,651 review events) were annotated as having an observed repository repair in
the same discussion thread, whereas 3,038 (76.9\%) were accepted without an
observed in-thread code change, deferred to follow-up issues, or left without a
reply; the remaining records consisted primarily of routine status bookkeeping
(6,240) and positive confirmations (5,564).

\subsection{Examples of review findings}
\label{sec:census-examples}

\cref{tab:census-examples-math}, \cref{tab:census-examples-design}, and
\cref{tab:census-examples-infra} present one to three representative findings
for each of the nine subcategories, selected from review records that the
reading pass annotated as confirmed defects with an observed repository change.
For each entry, the first line gives the specification required by the paper,
blueprint, or module interface, the second gives the defective form in the
reviewed draft, and the third gives the repaired form after review, with
\mathdefect{\text{defect highlighting}} marking the defect and
\mathrepair{\text{repair highlighting}} marking the repair.

\begin{table}
  \centering
  \caption{Review findings under ``Mathematics and agreement with the
    paper''. Paper: the required form; Draft: the reviewed form; Repair:
    the merged form.}
  \label{tab:census-examples-math}
  \begin{tblr}{
    colspec = {X[1.8,l] X[9,l]},
    row{odd} = {bg=gray!5},
    row{1} = {bg=gray!25, font=\bfseries},
    rowsep = 3pt,
  }
    \toprule
    Subcategory & Finding \\
    \midrule
    Mathematical content &
      Paper (Prop.~4.9, $A$ and $B$ projective):
      $\mathrm{sdd}(\psi;A,B)=2\,\mathrm{cons}(\psi;A,B)$, so
      $\approx_{2\delta}\Leftrightarrow\simeq_{\delta}$.\newline
      Draft: \mathdefect{\mathrm{cons}\le\mathrm{sdd}}, although the same
      calculation had established the equality, giving only
      $\approx_{\delta}\Rightarrow\simeq_{\delta}$ and doubling the error
      at every use.\newline
      Repair: \mathrepair{2\,\mathrm{cons}=\mathrm{sdd}} and the implication
      restated as $\approx_{2\delta}\Rightarrow\simeq_{\delta}$
      (\mippr{515}). \\
    Mathematical content &
      Paper (commutativity of points):
      $A^u_a\otimes I\simeq_{\gamma m}I\otimes L^{\ell}_{[f(u)=a]}$ on
      average over a uniformly random point $u$ and a uniformly random
      line $\ell$ through it.\newline
      Draft: the question distribution was
      \mathdefect{\mathcal{D}=\delta_{(\ell_0,t_0)}}, a point mass on one
      default question, so the averaged relation constrained a single
      $(\ell,t)$ and a strategy failing almost everywhere still satisfied
      the bound.\newline
      Repair: \mathrepair{\mathcal{D}=\mathrm{Unif}(\mathcal{L}\times
      \F_q)}, and $\mathrm{Unif}(\mathcal{L}\times\F_q^2)$ for the
      two-point relation, with the line set $\mathcal{L}$ finite through
      $\ell\mapsto(\mathrm{base},\mathrm{direction})$ (\mippr{121}). \\
    Mathematical content &
      Paper (local variance of points): six steps with errors
      $2\delta,2\varepsilon,\tfrac{md}{q},\tfrac{md}{q},2\varepsilon,
      2\delta$ and the $k$-step triangle inequality give
      $\approx_{6(4\varepsilon+4\delta+2md/q)}$, relaxed to
      $24(\varepsilon+\delta+\tfrac{md}{q})$.\newline
      Draft obligation: \mathdefect{\le 4\varepsilon+4\delta+
      2\tfrac{md}{q}}, the factor $k=6$ dropped, six times stronger than
      the chain proves.\newline
      Repair: \mathrepair{\le 6\,(4\varepsilon+4\delta+2\tfrac{md}{q})},
      with $6(\cdots)\le24(\varepsilon+\delta+\tfrac{md}{q})$ as a
      separate lemma (\mippr{780}). \\
    Source and blueprint correspondence &
      Blueprint (pasted sum to a polynomial in $G$): for a bipartite state
      $\psi_{\mathrm{bi}}$,
      $\mathbb{E}_{x}\sum_{|\tau|\ge d+1}\sum_{g}
      \langle\psi_{\mathrm{bi}}|\widehat{H}^{x}_{g}\otimes I
      |\psi_{\mathrm{bi}}\rangle\approx_{\nu_8}
      \sum_{r=d+1}^{k}\binom{k}{r}\langle\psi_{\mathrm{bi}}|
      G^{r}(I-G)^{k-r}\otimes I|\psi_{\mathrm{bi}}\rangle$.\newline
      Draft: a free state parameter with the added premise
      \mathdefect{\psi_{\mathrm{bi}}=\psi}, absent from the blueprint and
      never used in the proof.\newline
      Repair: \mathrepair{\text{parameter and premise removed}}; the lemma
      is stated at the strategy's state $\psi$, the identification the
      paper makes (\mippr{446}). \\
    Source and blueprint correspondence &
      Paper (cascade bound): $\nu\le 10000\,k^2m^2\bigl(\varepsilon^{1/1024}
      +(d/q)^{1/1024}\bigr)\Rightarrow\sigma\le 10000\,k^2m^4E$.\newline
      Draft blueprint, tagged complete:
      \mathdefect{0\le\nu\ \wedge\ \nu\le 1000\,k^2m^2(\ldots)}, a conjunct
      Lean never assumes and a coefficient ten times too small.\newline
      Repair: \mathrepair{\nu\le 10000\,k^2m^2(\ldots)}, matching Lean
      symbol for symbol (\mippr{507}). \\
    Semantic and API invariants &
      Paper: a sub-measurement is $\{A^a\}$ with $A^a\succeq0$ and
      $\sum_a A^a=T\preceq I$.\newline
      Draft: the three invariants became structure fields with
      \mathdefect{\text{default proofs left unproved}}, so any caller
      obtained them for free.\newline
      Repair: \mathrepair{\text{no defaults}}; every construction site
      proves $A^a\succeq0$, $\sum_a A^a=T$, and $T\preceq I$
      (\mippr{124}). \\
    \bottomrule
  \end{tblr}
\end{table}

\begin{table}
  \centering
  \caption{Review findings under ``Exposition and library design''.}
  \label{tab:census-examples-design}
  \begin{tblr}{
    colspec = {X[1.8,l] X[9,l]},
    row{odd} = {bg=gray!5},
    row{1} = {bg=gray!25, font=\bfseries},
    rowsep = 3pt,
  }
    \toprule
    Subcategory & Finding \\
    \midrule
    Mathematical exposition &
      Interface (base case $m=1$): the distinguished outcomes are the
      zero polynomial, $a_A=a_B=0$.\newline
      Draft: the docstring said so; the code took
      \codedefect{Classical.choice (inferInstance: Nonempty _)}, an
      arbitrary element not provably equal to $0$.\newline
      Repair: \mathrepair{a_A=a_B=0} built explicitly, with the proof
      $\deg_{x_i}0\le d$ (\mippr{1044}). \\
    Mathematical exposition &
      Definition: normalization is $\tau(\rho)=1$ for the density
      operator $\rho$ under the normalized trace, with no purity
      assumption.\newline
      Draft docstring: \mathdefect{\langle\psi|\psi\rangle=1}, ``as is
      standard for pure strategies''.\newline
      Repair: \mathrepair{\tau(\rho)=1}, coinciding with
      $\langle\psi|\psi\rangle=1$ for pure states; the corrected reading
      gave $\tau(\rho)=1\Rightarrow\dim\mathcal{H}\ge1$ and removed a
      redundant instance argument (\mippr{443}). \\
    Library architecture and API &
      Blueprint: the public successor-step wrapper returns
      $G\in\mathrm{PolyMeas}(m+1,q,d)$ ``without exposing the intermediate
      bookkeeping packages''.\newline
      Draft: its hypotheses mentioned a \codedefect{private} helper for
      the restriction package $\mathcal{R}$ five times, a name no caller
      outside the module can write.\newline
      Repair: \mathrepair{\mathcal{R}:=\mathrm{ofRestrictedProbabilities}
      (\ldots)} bound locally inside each hypothesis (\mippr{649}). \\
    Library architecture and API &
      Paper (restricted probabilities):
      $\mathbb{E}_x[\tfrac{m}{m+1}\varepsilon_x]\le\varepsilon$ and
      $\mathbb{E}_x[\tfrac{m}{m+1}\gamma_x]\le\gamma$, one transverse
      weight for both branches.\newline
      Draft: \mathdefect{\mathbb{E}_x[w_{\mathrm{diag}}\gamma_x]\le\gamma}
      with $w_{\mathrm{diag}}$ deleted earlier and undefined, so the module
      did not elaborate.\newline
      Repair: \mathrepair{\mathbb{E}_x[\tfrac{m}{m+1}\gamma_x]\le\gamma}
      (\mippr{660}). \\
    Reuse and maintainability &
      The review certified the six-step chain and
      $6(4\varepsilon+4\delta+2\tfrac{md}{q})\le24(\varepsilon+\delta+
      \tfrac{md}{q})$ against the paper.\newline
      Draft: the line-chart identity
      $\ell_{u,i,t_0}(t)=u[i\mapsto(u_i-t_0)+t]$ was proved twice,
      \mathdefect{\text{once as a helper nothing called}} and once
      inline.\newline
      Repair: \mathrepair{\text{helper deleted}}, the identity stated once
      (\mippr{898}). \\
    Reuse and maintainability &
      The final stage needs $\sqrt{100m}\le10m$, $\sqrt{10m}\le4m$,
      $\sqrt{400m}\le20m$, and $\sqrt{960m}\le31m$ for $m\ge1$.\newline
      Draft: the four bounds were \codedefect{private} and unused in the
      helper module and \mathdefect{\text{re-proved verbatim}} in the
      module that needed them.\newline
      Repair: \mathrepair{\text{public, stated once}}, the duplicates
      deleted (\mippr{1295}). \\
    \bottomrule
  \end{tblr}
\end{table}

\begin{table}
  \centering
  \caption{Review findings under ``Audit and execution infrastructure''.}
  \label{tab:census-examples-infra}
  \begin{tblr}{
    colspec = {X[1.8,l] X[9,l]},
    row{odd} = {bg=gray!5},
    row{1} = {bg=gray!25, font=\bfseries},
    rowsep = 3pt,
  }
    \toprule
    Subcategory & Finding \\
    \midrule
    Build, CI and review automation &
      Invariant: review runs skip commits produced by automation prefixes
      (Claude and Codex, automatic and review-driven).\newline
      Draft guard: \codedefect{^\[claude-(auto|review)-fix\]}, allowing
      Codex prefixes to escape and re-trigger review loops.\newline
      Repair: \coderepair{^\[(claude|codex)-(auto|review)-fix\]}
      (\mippr{1392}). \\
    Build, CI and review automation &
      Coverage test: required scripts in workflow $w$ must be included in path
      filters for each trigger $e$.\newline
      Draft: evaluated path filters across the union of all YAML blocks,
      masking missing triggers.\newline
      Repair: enforced path filter coverage independently for each trigger
      block (\mippr{960}). \\
    Reproducibility, security and environment &
      The documented per-file check must terminate.\newline
      Draft: one simplification step with 22 arguments made the file's
      check exceed 120~s and caused the CI build to time out.\newline
      Repair: reverted the step to an explicit tracked obligation naming
      the missing reindexing, allowing the build to pass (\mippr{662}). \\
    Reproducibility, security and environment &
      Toolchain pin audit: ensure toolchain version strings match between
      README and lakefile.\newline
      Draft: string mismatch between `4.28.0' and `v4.28.0'.\newline
      Repair: normalized version strings on both sides before comparison
      (\mippr{762}). \\
    Repository process and evidence &
      Idempotent notices: automated merge notices include a deduplication
      marker to prevent re-posting upon webhook redelivery.\newline
      Draft: notice body inserted text inside the marker, causing re-posting
      on redelivered merge events.\newline
      Repair: moved deduplication marker to a strict body prefix, plus a
      regression test (\mippr{2595}). \\
    \bottomrule
  \end{tblr}
\end{table}

\suppappendix{Prompt for auditing and repairing the formalization}
\label{app:repair-prompt}
The prompt below directed the repository-wide statement audit and repair pass
during May 2026 (\hyperref[sec:methods]{Methods}).
Run in goal mode with automated turn-end progress checks, it instructs the agent
to audit paper-facing Lean declarations and non-green or missing blueprint
nodes against the paper, classify any statement discrepancy, and either restore
the intended statement with tracked obligations or isolate conditional internal
helpers without linking them as completed paper theorems.
\begin{promptbox}{Audit-and-repair prompt}
\begin{lstlisting}[style=agentprompt]
Audit and repair the LDT formalization so that paper-facing Lean and blueprint statements match the source paper, and so that remaining red or unfinished dependency-graph nodes are classified by their real mathematical status.

Work in ~/Local/agentFormalization/MIPStarRE.  The source of truth is references/ldt-paper/, then blueprint/src/, then MIPStarRE/.  Also check the GitHub Pages branch in a separate worktree and inspect the generated blueprint dependency graph, including dep_graph_document.html, so that the audit reflects the public non-green nodes.

This is not a renaming task.  Mechanical renames of Bridge, Package, Residual, Repair, Producer, Input, or Hypotheses do not solve the problem.  The task is to determine what mathematical assertion is missing or incorrectly represented, and then either prove it, state it faithfully with a tracked sorry, or remove the misleading paper-facing link.

Main invariants:

1. A theorem, lemma, or proposition advertised as a paper result must match the cited statement in references/ldt-paper/ up to faithful formal encoding.  Do not add a bridge, residual, repair, package, producer, input, generic hypotheses bundle, or arbitrary implication hypothesis to make the theorem compile.

2. If a proof step is missing, keep the paper-facing theorem visible and source-faithful.  It may contain a tracked sorry during paper-realignment mode.  The missing step should become a named proof obligation or construction theorem, not an extra assumption on the paper theorem.

3. Conditional helpers may remain only when they preserve useful mathematics.  They must have names and docstrings that say they are internal obligations, not source hypotheses.  They must not be linked by \leanok to the source-labelled blueprint theorem.

4. Some boundary hypotheses may be faithful formal encodings: positivity needed for division, nonemptiness, decidability, field-model instances, or finite-type structure.  These must be distinguished from load-bearing invented proof assumptions.

5. Prefer larger useful repair batches.  Each PR should discharge several related audit items when possible, but avoid unrelated refactoring.

Audit procedure:

A. Inspect the public dependency graph from the GitHub Pages branch and list the non-green or missing nodes.  For each node, compare:

- the paper statement in references/ldt-paper/;

- the blueprint statement and its \lean{} / \leanok status;

- the Lean declaration, if any;

- whether the Lean proof contains sorry, axiom, or proof-evasion scaffolding;

- whether the Lean statement has extra hypotheses or a weakened conclusion.

B. Scan the Lean and blueprint sources for proof-debt vocabulary and conditional scaffolding: Bridge, bridge, Residual, residual, Repair, repair, Package, package, Producer, producer, Input, input, Hypotheses, hypotheses, Assumptions, assumptions, sorryAx, obstruction, conditional, ofBridge, ofObligations.

Classification for every item:

- Missing statement: no Lean declaration yet corresponds to the paper statement.

- Stated with proof hole: Lean declaration is source-faithful but contains sorry.

- Unlinked statement: Lean declaration exists and is faithful, but the blueprint does not point to it.

- Unfaithful statement: Lean declaration has extra non-paper hypotheses, changed quantifiers, weakened conclusions, or packaged conclusions.

- Conditional helper: useful internal theorem, but not the paper theorem.

- Boundary condition: extra Lean hypothesis appears mathematically necessary for a faithful formal encoding.

- Obsolete scaffolding: conditional object has no mathematical value and should be removed or replaced by a sorry in the source-faithful theorem.

Repair policy:

1. First recover actual mathematics from existing scaffolding. If a bridge or package contains a real construction or inequality, extract a self-contained theorem stated from paper hypotheses.

2. If the scaffold merely assumes the missing step, do not preserve it as a paper theorem.  Restore the paper-aligned statement and leave the missing proof as a tracked sorry or named obligation.

3. Update the blueprint only when the Lean statement is faithful.  Remove or avoid \leanok for conditional helpers.

4. Add concise docstrings for proof obligations explaining the paper label, the missing mathematical step, and the intended discharge.

5. Update tracking issues.  Use native GitHub subissues to connect the main bridge-debt tracking issue to specific repair tasks.  One PR may address multiple subissues.

6.  Treat statement drift and definition drift as high priority, because incorrect definitions propagate through downstream theorems.

7. Assign yourself when you do it

Deliverables:

- A table or issue comment classifying the non-green dependency graph nodes and remaining sorry sites by mathematical status.

- One or more PRs that make real mathematical progress: source-faithful statements restored, useful obligations proved or isolated, misleading conditional paper links removed, and local checks improved where cheap.

- For each PR, include a statement integrity audit: paper assumptions, Lean assumptions, paper conclusion, Lean conclusion, verdict.

- Validation commands in the PR body.  Prefer local single-file checks first; run lake build only when the batch touches shared declarations.
\end{lstlisting}
\end{promptbox}

\suppappendix{Tools distilled for new formalizations}
\label{app:starter-tools}
We extracted the project-agnostic infrastructure developed during the LIDT
formalization into four standalone packages that initialize \sysname{} for new
Lean projects.
During the LIDT proof, recurring statement discrepancies and unproved
obligations prompted iterative revisions to our build checks, blueprint schema,
and agent instructions.
The resulting packages organize a new formalization around three functions:
configuring the Lean build and continuous-integration environment, maintaining
the blueprint and proof-gap records, and standardizing agent proving procedures
across sessions.

\paragraph*{Preparing the project.}
The repository template
\href{https://github.com/LionSR/oh-my-formalization}{\code{oh-my-formalization}}
and the GitHub Action
\href{https://github.com/texra-ai/lean-env-action}{\code{lean-env-action}}
configure the Lean package structure and automated build pipeline for a new
formalization.
The template initializes a Lean workspace with a pinned Mathlib dependency, a
blueprint directory, paper-gap note templates, and publishing workflows that
deploy the blueprint and gap notes alongside the formal library.
Within continuous integration, \code{lean-env-action} provisions the pinned Lean
toolchain and retrieves compiled Mathlib caches before running \code{lake build}
and the repository's verification scripts.
A new formalization supplies the paper, definitions, and target statements in
place of the template's sample mathematics.

\paragraph*{Maintaining the mathematical account.}
The package
\href{https://github.com/LionSR/texra-blueprint}{\code{texra-blueprint}}
renders the interactive blueprint and enforces the proof-gap protocol that links
formal declarations to the paper.
It builds LaTeX blueprint nodes with Lean declaration links, dependency
references, and bibliographies, while checking the output for specified
rendering errors.
When kernel verification exposes a missing hypothesis or a modified error bound
in the paper, its paper-gap notes record the cited assertion, the mathematical
discrepancy, the proposed correction, and the review status of that decision.
Automated checks require registered source identifiers, reject references to
missing notes, and compile the notes into individual PDFs with a shared index.
This infrastructure establishes the proof-gap protocol from the start of a
project so that suspected mathematical departures identified by agents or
reviewers are recorded and indexed for human adjudication.

\paragraph*{Carrying the procedures across sessions.}
The repository
\href{https://github.com/texra-ai/texra-lean-skills}{\code{texra-lean-skills}}
collects reusable operational instructions for agents working on Lean proofs.
These instruction modules specify how agents search the existing library,
develop proofs, synchronize the blueprint, record paper gaps, and simplify
completed proofs without altering theorem statements.
They also define shared documentation and code-review conventions.
While the project repository supplies the mathematical context and open proof
obligations, \code{texra-lean-skills} gives successive proving sessions a
consistent procedure that can be refined as further proof attempts and reviews
expose new failure modes.

Together, these four packages provide a reusable verification harness for
applying \sysname{} to a new mathematical theorem.
The target theorem and its dependency graph define the proof tasks, and the
blueprint and paper-gap notes preserve the correspondence between the formal
declarations and the paper as those tasks are resolved.
Continuous-integration checks and periodic audits, following the procedures in
Methods and \cref{app:repair-prompt}, then verify that the completed library
proves the intended mathematical statements.

\stopcontents[si]

\clearpage
\bibliographystyle{unsrtnat}
\bibliography{paper_refs}

\end{document}